\documentclass[12pt]{article}
\usepackage[dvipdfmx]{graphicx}
\usepackage{amsfonts,amsmath,amssymb,epsf}
\usepackage{amsmath,braket}
\usepackage{graphicx,hyperref,color}
\hypersetup{
    unicode=false,          
    pdftoolbar=true,        
    pdfmenubar=true,       
    pdffitwindow=false,     
    pdfstartview={FitH},    
    pdfnewwindow=true,      
    colorlinks=true,       
    linkcolor=blue,          
    citecolor=blue,        
    filecolor=blue,      
    urlcolor=blue           
}
\newcommand{\de}{\partial}
\newcommand{\be}{\begin{equation}}
\newcommand{\ba}{\begin{eqnarray}}
\newcommand{\ea}{\end{eqnarray}}
\newcommand{\ee}{\end{equation}}

\newcommand{\f}{\frac}
\newcommand{\s}{\sqrt}
\newcommand{\vp}{\varphi}

\newcommand{\ti}{\tilde}
\newcommand{\ap}{\alpha}

\newcommand{\ddd}{\cdot\cdot\cdot}
\newcommand{\no}{\nonumber \\}
\newcommand{\la}{\langle}
\newcommand{\lb}{\rangle}
\newcommand{\bea}{\begin{eqnarray}}
\newcommand{\eea}{\end{eqnarray}}
\newcommand{\bes}{\begin{equation*}}
\newcommand{\beas}{\begin{eqnarray*}}
\newcommand{\eeas}{\end{eqnarray*}}
\newcommand{\bas}{\begin{array*}}
\newcommand{\eas}{\end{array*}}
\newcommand{\ees}{\end{equation*}}

\newcommand{\ep}{\epsilon}

\newcommand{\fr}{\frac}

\newcommand{\pa}[1]{\left(#1 \right)}

\newcommand{\abs}[1]{\left|#1\right|}

\newcommand{\ar}[1]{\xrightarrow[#1]{}}

\usepackage{cite}

\usepackage{tcolorbox}

\begin{document}

\begin{titlepage}
\thispagestyle{empty}

\begin{flushright}
YITP-20-91
\\
IPMU20-0079
\\
\end{flushright}

\bigskip

\begin{center}
\noindent{{\Large \textbf{Codimension two holography for wedges}}}\\
\vspace{2cm}

Ibrahim Akal$^{a}$,
Yuya Kusuki$^{a}$,
Tadashi Takayanagi$^{a,b,c}$,
and
Zixia Wei$^{a}$
\vspace{1cm}\\

{\it $^a$Center for Gravitational Physics,\\
Yukawa Institute for Theoretical Physics,
Kyoto University, \\
Kitashirakawa Oiwakecho, Sakyo-ku, Kyoto 606-8502, Japan}\\

{\it $^b$Inamori Research Institute for Science,\\
620 Suiginya-cho, Shimogyo-ku,
Kyoto 600-8411 Japan}\\

{\it $^{c}$Kavli Institute for the Physics and Mathematics
 of the Universe (WPI),\\
University of Tokyo, Kashiwa, Chiba 277-8582, Japan}

\end{center}

\begin{abstract}
We propose a codimension two holography between a gravitational theory on a $d+1$ dimensional wedge spacetime and a $d-1$ dimensional CFT which lives on the corner of the wedge. 
Formulating this as a generalization of AdS/CFT,  we explain how to compute the free energy, entanglement entropy and correlation functions of the dual CFTs from gravity. 
In this wedge holography, the holographic entanglement entropy is computed by a double minimization procedure.    
Especially, for a four dimensional gravity ($d=3$), we obtain a two dimensional CFT and the holographic entanglement entropy perfectly reproduces the known result expected from the holographic conformal anomaly. We also discuss a lower dimensional example ($d=2$) and find that a universal quantity naturally arises from gravity, which is  analogous to the boundary entropy.  
Moreover, we consider a gravity on a wedge region in Lorentzian AdS, which is expected to be dual to a CFT with a space-like boundary.  We formulate this new holography and compute the holographic entanglement entropy via a Wick rotation of the AdS/BCFT construction.  Via a conformal map, this wedge spacetime is mapped into a geometry where a bubble-of-nothing expands under time evolution.
We reproduce the holographic entanglement entropy for this gravity dual via CFT calculations. 
\end{abstract}

\end{titlepage}

\newpage
\tableofcontents

\section{Introduction}

The AdS/CFT correspondence \cite{Ma,Gubser:1998bc,Witten:1998qj} provides a non-perturbative definition of quantum gravity on an anti-de Sitter space (AdS) in terms of a lower dimensional conformal field theory (CFT).  Since one of the final goals in string theory is to understand how our universe is created,  it is important to extend the idea of holography \cite{tHooft:1993dmi,Susskind:1994vu} to more realistic spacetimes.

Till now, there have been  several progresses in this direction by generalizing or deforming the AdS/CFT correspondence. For example, this includes the brane world holography \cite{Randall:1999ee,Randall:1999vf,Karch:2000ct} which proposes a holographic duality between AdS with a finite geometric cutoff in the radial direction and a CFT coupled to dynamical gravity. By further developing the brane world approach, a holographic description of de Sitter gravity was proposed in \cite{Alishahiha:2004md,Alishahiha:2005dj,Dong:2018cuv}. For holographic duals of de Sitter gravity, there have also been worked out more direct approaches called dS/CFT by explicitly changing the sign of the cosmological constant  and by regarding the future/past infinity as the holographic dual boundary where the dual CFT looks non-unitary 
\cite{Strominger:2001pn,Maldacena:2002vr,Anninos:2011ui}. There also exists a general approach called surface/state correspondence, which is applicable to holography in any spacetime
\cite{Miyaji:2015yva,Takayanagi:2018pml}, regarding  codimension two and codimension one surface as a quantum state and  a quantum circuit, respectively. 

Another approach to extend the standard AdS/CFT is to introduce boundaries on the manifold where the CFT lives. When a part of conformal symmetry is preserved, such a theory is called a boundary conformal field theory (BCFT). Holographic duals of  BCFTs, called AdS/BCFT, can be constructed by cutting the AdS along an end-of-the-world brane with backreactions taken into account \cite{Karch:2000gx,AdSBCFT,AdSBCFT2,AdSBCFT3}. One characteristic feature of AdS/BCFT is that the holographic entanglement entropy \cite{RT,RT2,HRT,Faulkner:2013ana,Engelhardt:2014gca} can have a contribution from minimal surfaces ending on the end-of-the-world brane, which has recently been called an Island. Indeed, by combining the AdS/BCFT with the brane world holography, the Island phenomena \cite{Penington:2019npb,Almheiri:2019psf}
in effective gravity theories have been derived holographically in \cite{Almheiri:2019hni} 
(also see \cite{Rozali:2019day,Chen:2019uhq,Almheiri:2019psy,Kusuki:2019hcg,Balasubramanian:2020hfs,Sully:2020pza,Geng:2020qvw,Chen:2020uac} for more progresses on the applications of brane world holography and AdS/BCFT to this subject).

\begin{figure}
  \centering
  \includegraphics[width=8cm]{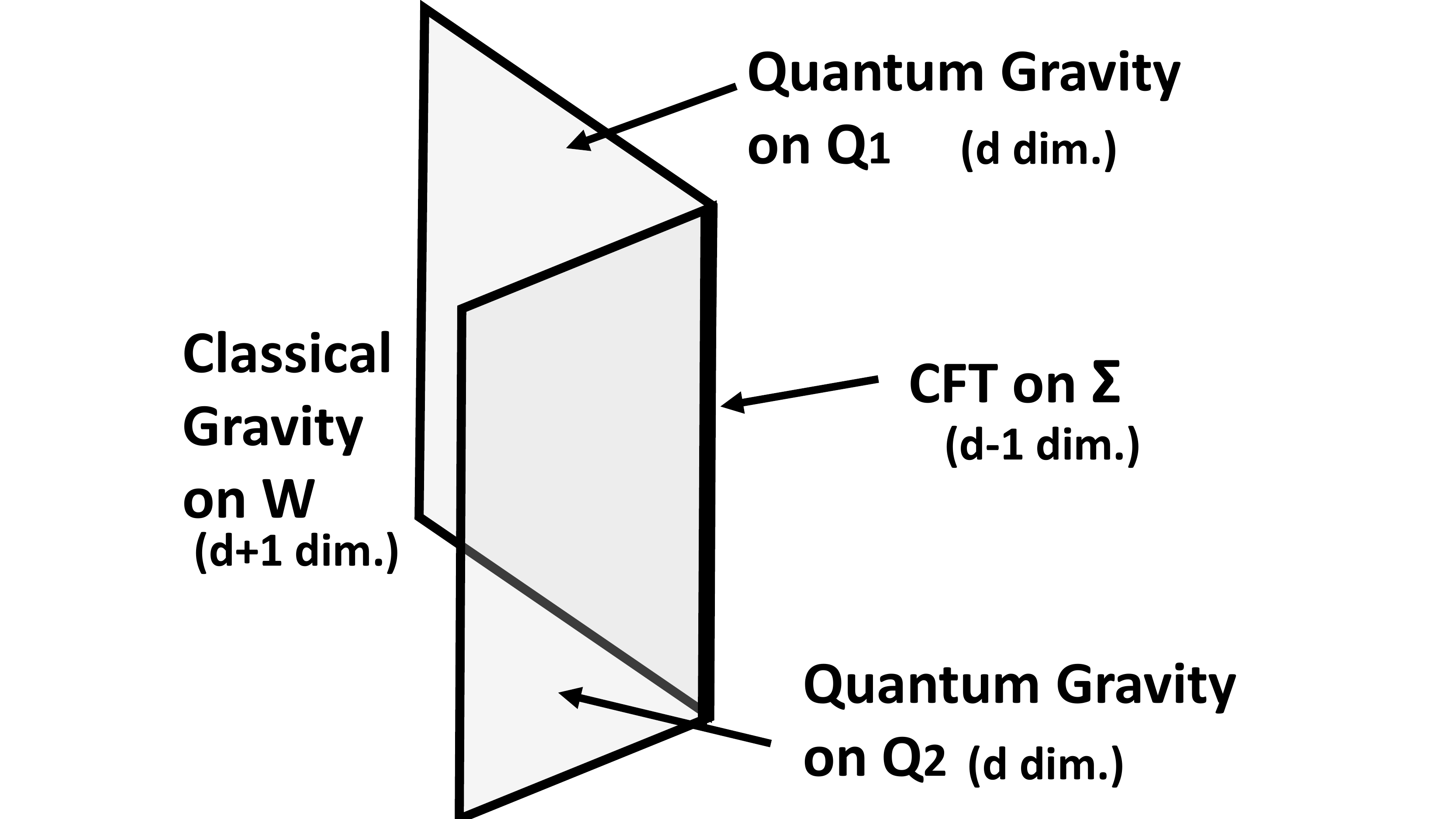}
  \caption{A sketch of wedge holography.}
\label{wedgesebtupfig}
\end{figure}

The main aim of the present paper is to combine these ideas to explore a novel holographic duality --- we call this wedge holography --- between 
a $d+1$ dimensional wedge space $W$ and a $d-1$ dimensional CFT on $\Sigma$ as sketched in Fig.~\ref{wedgesebtupfig}. Similar ideas of higher codimension holography were already discussed in a brane world setup \cite{Bousso:2001cf},  the dS/dS correspondence \cite{Alishahiha:2004md,Alishahiha:2005dj}, a dimensional reduction of de-Sitter space 
\cite{Arias:2019zug} and also  in the context of gravity edge modes\cite{Takayanagi:2019tvn}.

After we show that our wedge holography can be understood as a certain limit of AdS/BCFT, we will study this wedge holography by calculating the free energy, the conformal anomaly, the entanglement entropy and correlation functions. The results of these computations confirm the proposed holographic relation. In particular, in the $d=3$ case, we show that the conformal anomaly computed from the gravity on-shell action agrees with that computed from the holographic entanglement entropy.
We also analyze Lorentzian wedge spacetimes which is dual to a CFT on a manifold with a space-like boundary via a Wick rotation of an Euclidean AdS/BCFT setup. We see that this opens up an interesting possibility of nucleations of bubbles-of-nothing in AdS and its CFT dual.

This paper is organized as follows: In section \ref{sec:wh-general}, we present a general formulation of wedge holography, including computations of the free energy, entanglement entropy and correlation functions. In section \ref{sec:wh-3d}, we study the details by focusing on $d=3$. In section \ref{whdt}, we present details in the case of $d=2$. In section \ref{sec:wh-joinCFTs}, we give a holographic connection between wedge holography and the gravity dual of joining two CFTs. 
In section \ref{sec:wh-btz}, we analyze wedge holography for the BTZ spacetime.  In section \ref{sec:space}, we discuss Lorentzian wedge spacetimes whose CFT duals have space-like boundaries. In section \ref{sec:concs}, we summarize our conclusions and discuss future problems. In appendix \ref{sec:schac}, we present a short review of the Schwarzian action.
In appendix \ref{sec:another}, we present gravity calculations with another choice of UV cutoff.

Note added: When we finished computations and were writing up this paper, we got aware of the recent paper \cite{Bousso:2020kmy}, where a similar codimension two holography and a calculation of holographic entanglement entropy via  double minimizations were independently considered.

\section{Wedge holography in general dimensions}
\label{sec:wh-general}

Here, we would like to formulate our codimension two holography, called wedge holography, in any dimensions by keeping only a wedge region in the standard AdS/CFT correspondence. Consider a Poincare AdS$_{d+1}$ with the AdS radius $L$, whose metric can be written in the following two different choices of the coordinates:
\ba
ds^2&=&L^2\left(\frac{dz^2-dt^2+dx^2+\sum_{i=1}^{d-2}d\xi_i^2}{z^2}\right) \label{AdsP} \\
&=&d\rho^2+L^2 \cosh^2\frac{\rho}{L}\left(\frac{dy^2-dt^2+\sum_{i=1}^{d-2} d\xi_i^2}{y^2}\right), \label{adsg}
\ea 
where the coordinates $(z,x)$ are related to $(\rho ,y)$ via 
\ba
z=\frac{y}{\cosh\frac{\rho}{L}},\ \ \ \ x=y\tanh\frac{\rho}{L}.
\ea
The Euclidean solution is simply obtained by setting $t=-it_E$.

\begin{figure}
  \centering
  \includegraphics[width=8cm]{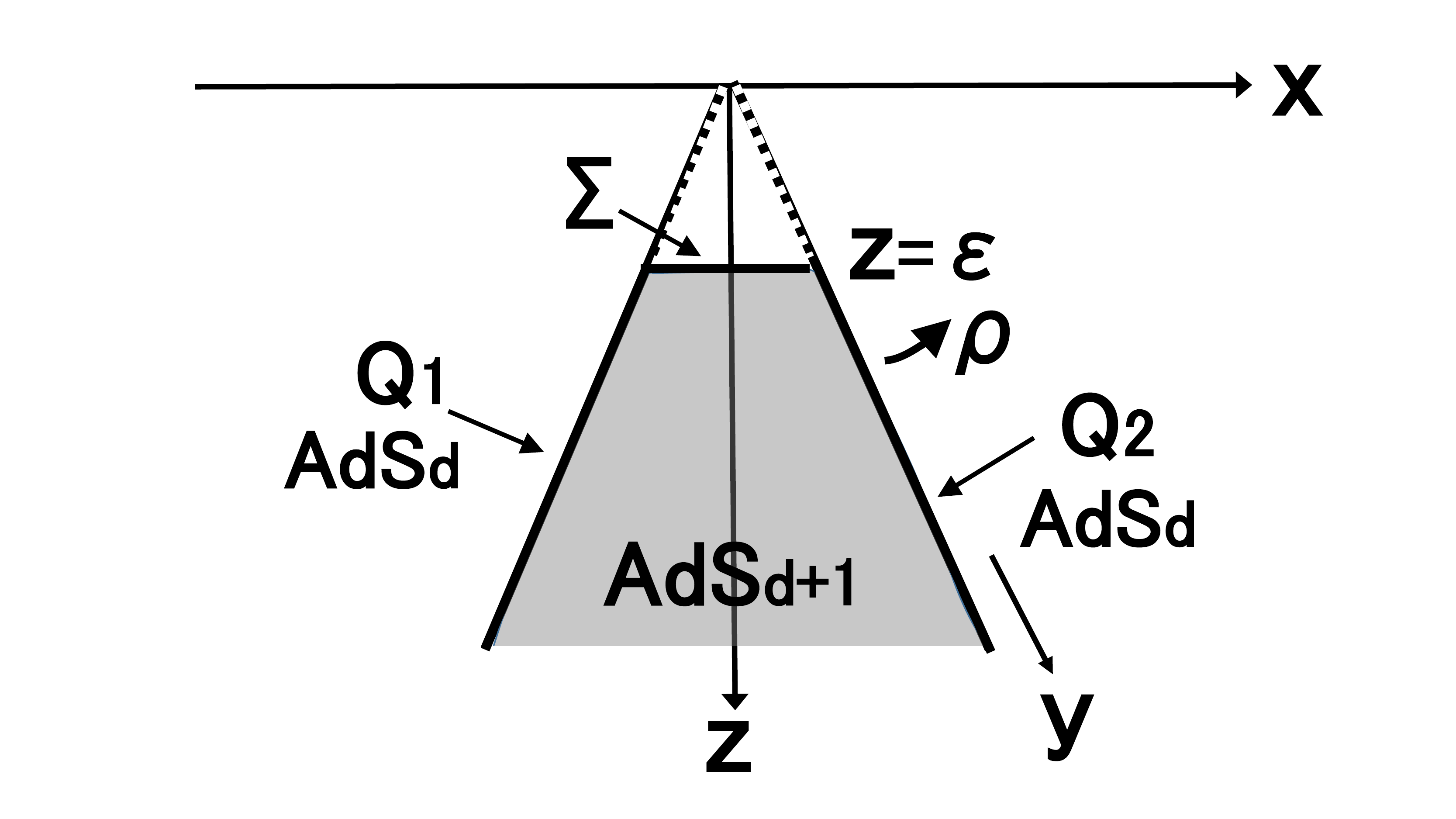}
  \caption{A basic setup of wedge holography with UV regularization.}
\label{wedgesetupfig}
\end{figure}

Now, we consider a $d+1$ dimensional wedge geometry, called $W_{d+1}$, in this background defined by the following limited region in the $\rho$ coordinate introduced in \eqref{adsg}: 
\ba
-\rho_*\leq \rho\leq \rho_*,
\ea
where we call the two boundaries of the wedge i.e. $\rho=-\rho_*$ and $\rho=\rho_*$, the $d$ dimensional 
suface $Q_1$ and $Q_2$, respectively. As usual in AdS/CFT, we regard the UV cutoff  of CFT as the geometric cutoff $z\geq \ep$.  This generates an extra $d$ dimensional boundary surface $\Sigma$
given by $z=\ep$ and $|x|\leq \ep\sinh\frac{\rho_*}{L}$.
This setup of the wedge geometry is depicted in 
Fig.~\ref{wedgesetupfig}.

We define the classical gravity on this wedge $W_{d+1}$ by imposing the Neumann boundary condition
in AdS/BCFT \cite{AdSBCFT,AdSBCFT2}:
\ba
K_{ab}-Kh_{ab}=-Th_{ab},  \label{bdycond}
\ea
on $Q_1$ and $Q_2$ where $h_{ab}$ and $K_{ab}$ are the induced metric and the extrinsic curvature on the surfaces $Q_{1,2}$, while $T$ is the tension parameter. Since we have
\ba
&& K_{ab}=\frac{1}{L}\tanh\frac{\rho_*}{L}\cdot h_{ab},\no
&& K\equiv h^{ab}K_{ab}=\frac{d}{L}\tanh\frac{\rho_*}{L},
\ea
we choose the tension to be 
\ba
T=\frac{d-1}{L}\tanh\frac{\rho_*}{L}.
\ea
On the asymptotically AdS surface $\Sigma$, we impose the standard  Dirichlet boundary condition which fixes the metric on $\Sigma$, denoted by $\gamma_{\ap\beta}$, to be given by \eqref{AdsP}.

\subsection{Wedge holography}

Thinking of the above setup, we now would like to argue for the following codimension two holographic correspondence.

A classical gravity on the $d+1$ dimensional wedge $W_{d+1}$, defined with the above boundary conditions, is dual to a (effectively) $d-1$ dimensional CFT located on $\Sigma$, as depicted 
in Fig.~\ref{wedgesetupfig}. Notice that since the width in $x$ direction of the surface $\Sigma$ is $\mathcal{O}(\ep)$, i.e. infinitesimally small, we can effectively regard $\Sigma$ as being $d-1$ dimensional, namely $R^{1,d-2}$, spanned by $(t,\xi_1,\ddd,\xi_{d-2})$, which we write as $\Sigma_{d-1}$. 
Our proposal is summarized as follows.\\

\begin{tcolorbox}

{\bf Wedge holography proposal}
\ba
\mbox{Classical gravity on wedge $W_{d+1}$} &\simeq& \mbox{(Quantum) gravity on two AdS}_d\ (=Q_1\cup Q_2)\no
&\simeq& \mbox{CFT on}\ \Sigma_{d-1} \label{duality}
\ea

\end{tcolorbox}

This holography can be understood by two steps. For simplicity, let us take the strict UV limit $\ep=0$ as in 
Fig.~\ref{wedgesebtupfig}. First, the classical gravity on the $d+1$ dimensional 
wedge $W_{d+1}$ is dual to a (quantum) gravity on its $d$ dimensional boundary $\de W_{d+1}=Q_1\cup Q_2$ via the brane world holography \cite{Randall:1999ee,Randall:1999vf,Karch:2000ct}.  Since $Q_1$ and $Q_2$ are $d$ dimensional AdS spacetimes, the quantum gravity theories on them are expected to be dual to $d-1$ dimensional CFTs which live on $\de Q_1=\de Q_2=\Sigma$. We combine these two CFTs on $\Sigma$ and treat them as a single $d-1$ dimensional CFT. This argument explains the previous proposal of the codimension two holography.

In the middle expression of \eqref{duality}, i.e. $d$ dimensional gravity, one may think there is also  a contribution from the CFT$_{d}$ on a small interval. However, we can neglect such a contribution. To see this, we can look, for example, at the anomaly for $d=3$ which we will discuss later. The anomaly is proportional to $\rho_*$ and this cannot be explained by the fields on the small interval, which should not depend on $\rho_*$. 

One may also understand our codimension two holography by decomposing the wedge geometry 
into an interval $[-\rho_*,\rho_*]$ 
in the $\rho$ direction and the transverse AdS$_{d}$. After a compactification 
on an interval we can apply the AdS$_{d}/$CFT$_{d-1}$ duality to find the codimension two holography.
However, our $d+1$ dimensional description of wedge holography has several advantages. First of all, as we will see below the nature of wedge holography is sensitive to the choice of UV cut off surface as a function of $\rho$, which is implicit after the compactification.  Indeed below we choose the UV cut off surface with a non-trivial $\rho$ dependence. Another advantage is that we can employ the full symmetry of the AdS$_{d+1}$ to have analytical controls over computations of physical quantities, even when the bulk fields have nontrivial profiles in the $\rho$ direction. Moreover, the structure of the wedge geometry leaves the $\rho$ direction explicit in the bulk and we expect this helps us to understand nonlocal physics such as entanglement on the compact manifold.

\subsection{Derivation from AdS/BCFT}

\begin{figure}
  \centering
  \includegraphics[width=6cm]{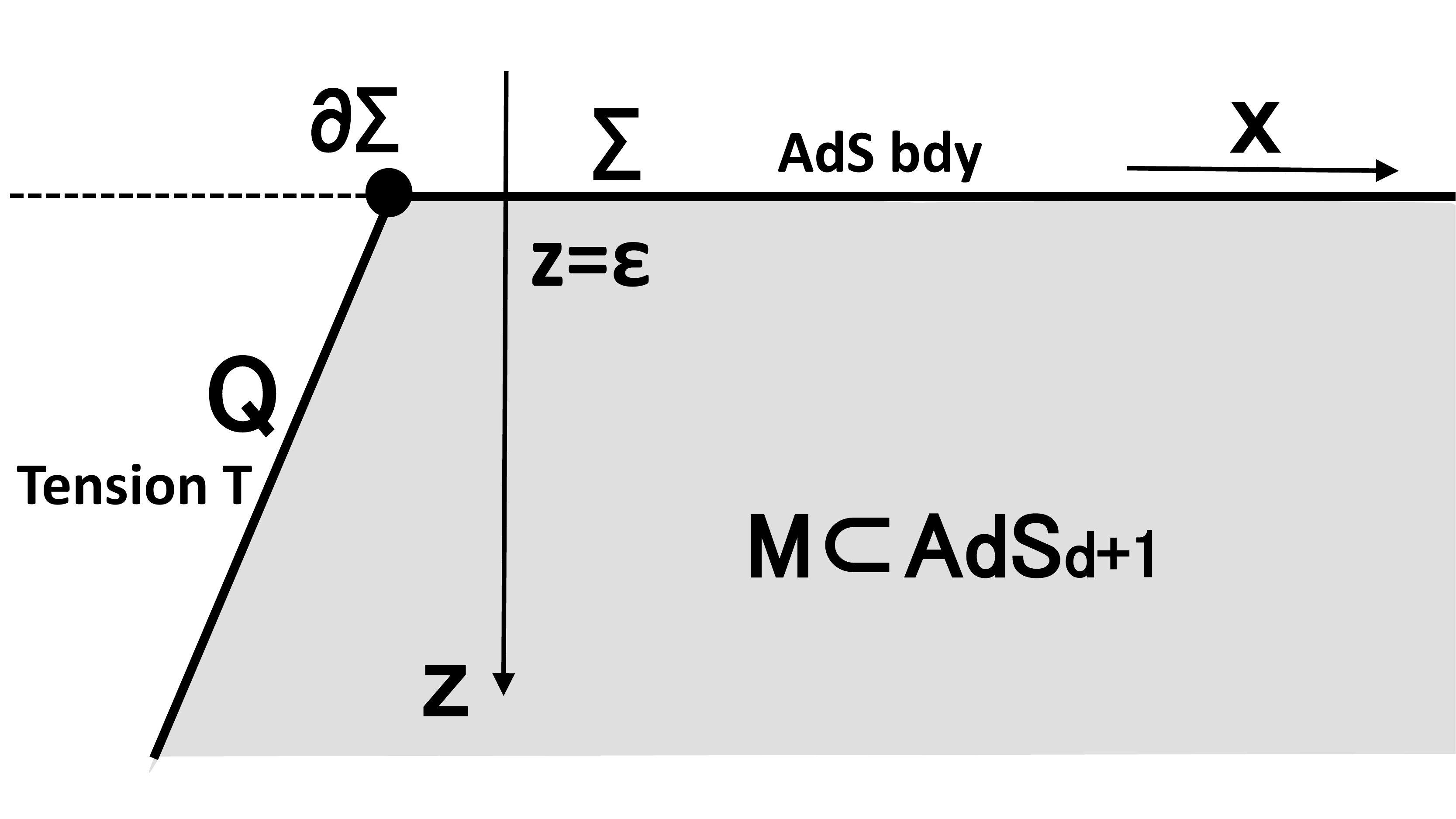}
\hspace{1cm}
  \includegraphics[width=6cm]{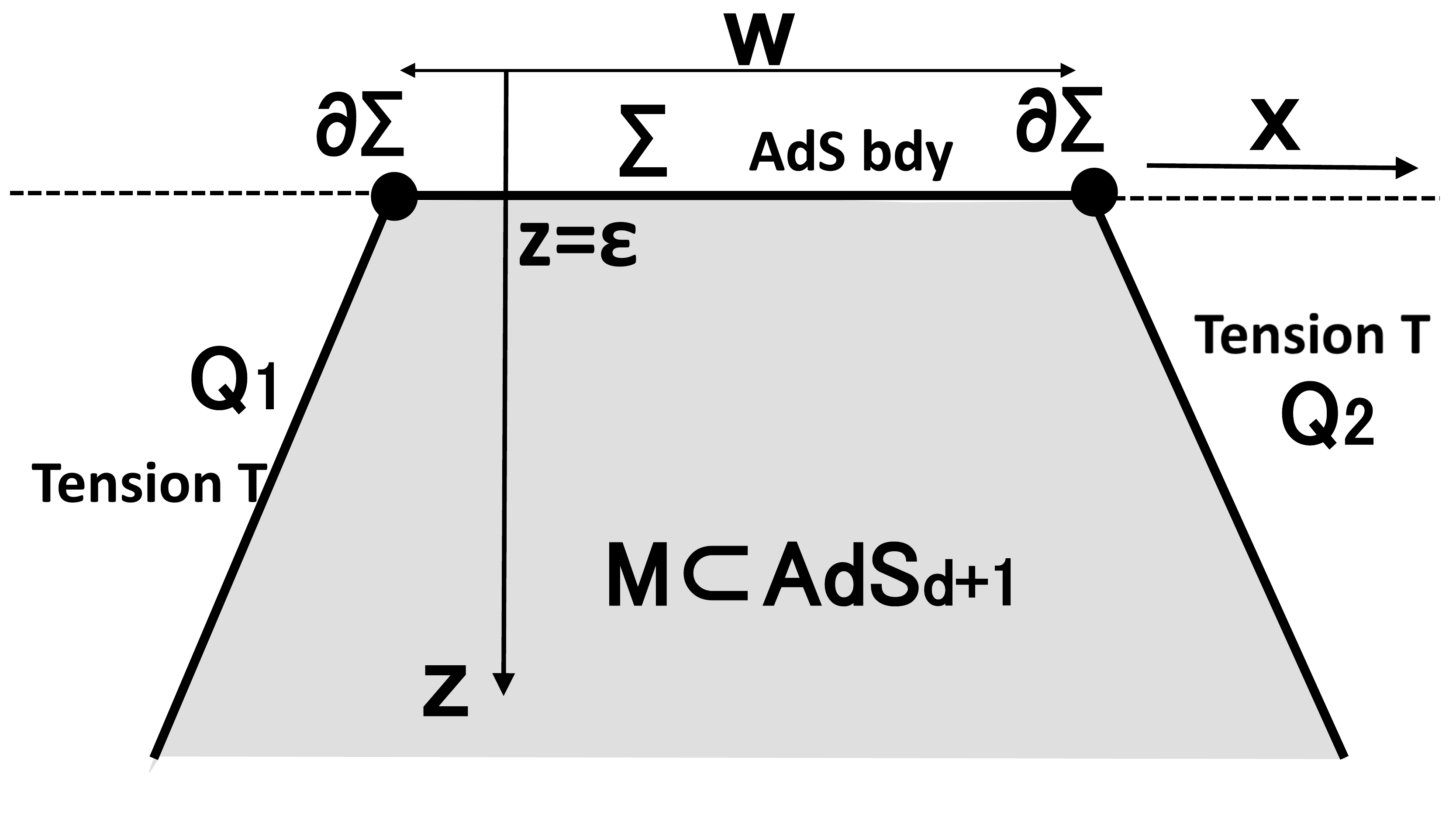}
  \caption{A sketch of derivation of wedge holography from AdS/BCFT. The left picture describes the gravity dual of a CFT on a semi-infinite plane. The right one shows the gravity dual of a CFT on an interval.}
\label{AdSBCFTfig}
\end{figure}

We can derive the previous codimension two holography by taking a suitable limit of the AdS/BCFT.
Consider a gravity dual of a $d$ dimensional CFT on a manifold $\Sigma$ with boundaries $\de\Sigma$.  The AdS/BCFT construction \cite{AdSBCFT,AdSBCFT2,AdSBCFT3} argues that the gravity dual of the CFT on $\Sigma$ is given by a gravity on a $d+1$ dimensional manifold $M$ with extra $d$ dimensional boundaries $Q$ (so called end-of-the-world brane), in addition to the standard AdS boundary $\Sigma$ such that $\de M=Q\cup \Sigma$. This manifold $M$ is determined by solving the Einstein equation and by imposing the Neumann-type boundary condition \eqref{bdycond}, parameterized by the tension $T$, on the surface $Q$. When the boundary $\Sigma$ is a semi-infinite plane $x>0$ in \eqref{AdsP}, the corresponding construction is sketched in the left panel of Fig.~\ref{AdSBCFTfig}. 

When we consider a gravity dual of a $d$ dimensional CFT on a strip $0\leq x\leq w$, there are two possibilities in general, which may be called as the confined solution and deconfined solution, as in the Hawking-Page transition. In the confined phase, a mass gap is generated from the strong interactions in the holographic CFT and the $d$ dimensional surface $Q$ is connected in the bulk as explicitly examined in \cite{AdSBCFT,AdSBCFT2,AdSBCFT3}. However, since we are interested only in gravity duals with scale invariance in this paper, we will not consider this confined case below. We would like to focus on the deconfined phase. For example, only this phase is allowed when we impose an appropriate  supersymmetric boundary condition (analogous to the R-sector of open strings) on $\de \Sigma$ which helps the system to be conformally invariant even in the presence of the boundary. In this deconfined case, the surface $Q$ consists of two disconnected planes $Q_1$ and $Q_2$, both of which  satisfy the boundary condition \eqref{bdycond}. Note that $Q_1$ and $Q_2$ are parallel with those in Fig.~\ref{wedgesetupfig}. 

If we take the limit $w\to 0$ of the vanishing width of strip $\Sigma$, we can regard $\Sigma$ as a $d-1$ dimensional space $R^{d-1}$. The gravity dual of AdS/BCFT is now reduced to the wedge geometry in 
Fig.~\ref{wedgesebtupfig} or its regularized version shown in Fig.~\ref{wedgesetupfig}. This explains our proposal of the codimension two holography. Note that we expect the original CFT$_{d}$ on $\Sigma$ is now reduced 
to a $d-1$ dimensional CFT in the zero width limit $w\to 0$. This phenomenon is quite usual. For example,
a massless free scalar on a $d$ dimensional strip $[0,w]\times R^{d-1}$ is reduced to a $d-1$ dimensional massless free scalar on $R^{d-1}$ in the limit $w\to 0$.

\subsection{Free energy in non-compact space}

As a fundamental quantity which characterizes the holographic correspondence, 
we would like to evaluate the free energy first. This is evaluated from the on-shell 
gravity action on the Euclidean wedge geometry of Fig.~\ref{wedgesetupfig}:
\ba
I_W=-\frac{1}{16\pi G_N}\int_{W} \s{g}(R-2\Lambda)-\frac{1}{8\pi G_N}\int_{Q_1\cup Q_2}\s{h}(K-T)-\frac{1}{8\pi G_N}\int_\Sigma \s{\gamma} K.
\ea
We calculate the explicit values using the metric \eqref{adsg} and \eqref{AdsP}:
\ba
&&R=-\frac{(d+1)d}{L^2},\ \ \ \ \ \Lambda=-\frac{d(d-1)}{2L^2},\no
&&K|_Q=\frac{d}{L}\tanh\frac{\rho_*}{L},\ \ \ \ \ T=\frac{d-1}{L}\tanh\frac{\rho_*}{L},\ \ \ \ \ 
K|_\Sigma=\frac{d}{L}.
\ea
The gravity action is evaluated as follows
\ba
&& I_W=\frac{d}{8\pi G_N L^2}\int_W\s{g}-\frac{\tanh\frac{\rho_*}{L}}{8\pi G_N L}\int_{Q_1\cup Q_2} \s{h}
-\frac{d}{8\pi G_N L}\int_\Sigma \s{\gamma}\no
&& =\frac{d\cdot V_{d-1}L^{d-2}}{4\pi G_N }\int^{\rho_*}_0 d\rho \left(\cosh\frac{\rho}{L}\right)^{d}
\int^\infty_{\ep\cosh\frac{\rho}{L}}\frac{dy}{y^d}\no
&&\ \ \ \ -\frac{L^{d-1}V_{d-1}\tanh\frac{\rho_*}{L}}{4\pi G_N }\left(\cosh\frac{\rho_*}{L}\right)^d
\int^\infty_{\ep\cosh\frac{\rho_*}{L}}\frac{dy}{y^d}\no
&&\ \ \ \ -\frac{d\cdot L^{d-1} V_{d-1}\sinh\frac{\rho_*}{L}}{4\pi G_N \ep^{d-1}} \no
&& =\frac{V_{d-1}L^{d-1}}{4\pi G_N \ep^{d-1}}\left[\frac{d}{(d-1)L}\int^{\rho_*}_0
d\rho \cosh\frac{\rho}{L}-\frac{1}{d-1}\sinh\frac{\rho_*}{L}-d\sinh\frac{\rho_*}{L}\right]\no
&&= (1-d)\cdot \frac{V_{d-1}L^{d-1}}{4\pi G_N \ep^{d-1}}\sinh\frac{\rho_*}{L}, \label{noncpt}
\ea
where we defined $V_{d-1}=\int dt_E d\xi_1\ddd d\xi_{d-2}$. 
Indeed, the scaling $I_G\propto V_{d-1}\ep^{1-d}$ is consistent with the vacuum energy of a $d-1$ dimensional CFT (CFT$_{d-1}$).  Also we find that the degrees of freedom of the CFT$_{d-1}$ 
are estimated as
\ba
\sim \frac{L^{d-1}}{G_N}\sinh\frac{\rho_*}{L}.
\ea

However, when there are cusp like codimension two singularities  on the boundaries, we need to add the Hayward term 
\cite{Hayward:1993my,Brill:1994mb} to the standard gravity action $I_W$ in \eqref{gravwed}. For each cusp at $\ti{\Sigma}$, the corresponding Hayward term $I_H$ is written as 
\ba
I_H=\frac{1}{8\pi G_N}\int_{\ti{\Sigma}} (\Theta-\pi)\s{\gamma}, \label{Hay}
\ea
where $\gamma$ is the induced metric on $\ti{\Sigma}$ and $\Theta$ is the cusp angle between two surfaces.
In our setup shown in the right panel of Fig.~\ref{AdSBCFTfig}, $\ti{\Sigma}$ and $\Theta$ are the $d-1$ dimensional intersection and the angle between $Q_{1,2}$ and $\Sigma$. This angle is given by $\theta+\frac{\pi}{2}$, where $\theta$ is 
\ba
\tan\theta=\sinh\frac{\rho_*}{L}.
\ea

The reason why we add the Hayward term is because we need to have a sensible variation on $\Sigma$
\ba
\delta (I_W+I_H)=\frac{1}{8\pi G_N}\int_{\ti{\Sigma}}(\Theta-\pi)\delta\s{\gamma},
\ea
when we impose the on-shell conditions, i.e. the Einstein equation and the boundary condition \eqref{bdycond}
on $Q_{1,2}$. The Dirichlet boundary condition on $\Sigma$ (and thus on $\ti{\Sigma}$) leads to $\delta\s{\gamma}=0$ and we therefore have 
$\delta (I_W+I_H)=0$.
In our setup, this is evaluated as follows
\ba
I_H=(2\theta-\pi)\frac{V_{d-1}L^{d-1}}{8\pi G_N\ep^{d-1}}.  \label{ihij}
\ea

Finally, the total gravity action is given by 
\ba
I_W+I_H=(1-d)\frac{V_{d-1}L^{d-1}}{4\pi G_N \ep^{d-1}}\sinh\frac{\rho_*}{L}+(2\theta-\pi)\frac{V_{d-1}L^{d-1}}{8\pi G_N\ep^{d-1}}.\label{noncpta}
\ea
Though one may think that it is strange that the total  action does not vanish even when the wedge region gets squeezed to zero size, $\rho_*=0$, the Hayward term contribution $I_H$  is important to appropriately take into account the gravity edge mode degrees of freedom \cite{Takayanagi:2019tvn}.

It is useful to compare this \eqref{noncpta} with the standard gravity partition function on the Poincare AdS$_d$ with the cutoff $z=\ep$, which is dual to a $d-1$ dimensional CFT on $R^{d-1}$ (we impose the Dirichlet boundary condition on $z=\ep$):
\ba
&& I_P=-\frac{1}{16\pi G^{(d)}_N}\int_{AdS_d} \s{g}(R-2\Lambda)-\frac{1}{8\pi G^{(d)}_N}
\int_{R^{d-1}}\s{h} K\no
&& =\frac{d-1}{8\pi G^{(d)}_NL^2}\int_{AdS_d} \s{g}-\frac{d-1}{8\pi G^{(d)}_N L}\int_{R^{d-1}} \s{h}\no
&& =(2-d)\frac{V_{d-1}L^{d-2}}{8\pi G^{(d)}_N\ep^{d-1}}.\label{padsp}
\ea
This takes the same form as in \eqref{noncpta}.

\subsection{Free energy in compact space}

It is also useful to calculate the free energy in a setup dual to a  CFT$_{d-1}$ on $S^{d-1}$.
In particular, this is useful to determine the conformal anomaly in the next section.
For this, we employ the following metric expression of Euclidean AdS$_{d+1}$:
\ba
ds^2=dr^2+L^2\sinh^2\frac{r}{L}(d\theta^2+\cos^2\theta d\Omega_{d-1}^2),
\ea
where $d\Omega_{d-1}$ describes the area element of $S^{d-1}$.
Let us introduce another coordinate system $(\rho,\eta)$ instead of $(r,\theta)$ via
\ba
&& \cosh\frac{r}{L}=\cosh\eta\cosh\frac{\rho}{L},\no
&& \sinh\frac{r}{L}\sin\theta=\sinh\frac{\rho}{L}.
\ea
This leads to the equivalent expression of the metric
\ba
ds^2=d\rho^2+L^2\cosh^2\frac{\rho}{L}\left(d\eta^2+\sinh^2\eta d\Omega_{d-1}^2 \right).
\label{adsrte}
\ea

\begin{figure}
  \centering
  \includegraphics[width=8cm]{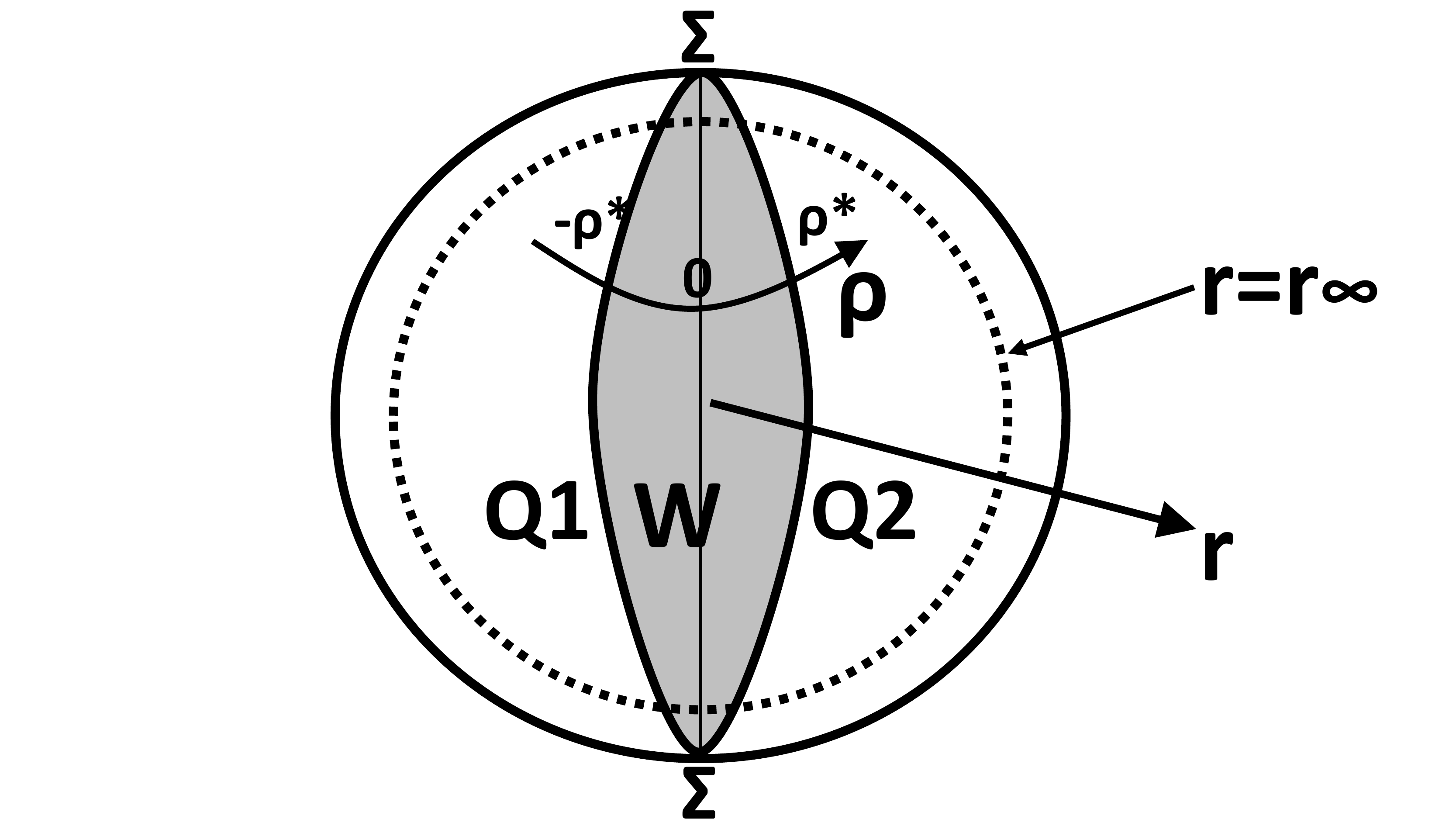}
  \caption{A sketch of compactified setup of wedge holography.}
\label{compactfig}
\end{figure}

We define the wedge geometry $W_{d+1}$ as the $d+1$ dimensional subspace of the Euclidean AdS$_{d+1}$ \eqref{adsrte} given by $|\rho|\leq \rho_*$, as depicted in Fig.~\ref{compactfig}. Therefore, $Q_1$ and $Q_2$ are defined by $\rho=\rho_*$ and $\rho=-\rho_*$, respectively. In the UV limit, the surface $\Sigma$, where the dual CFT$_{d-1}$ lives, is the $d-1$ dimensional surface given by $(r,\theta)=(\infty,0)$. To regulate the UV divergences, we introduce the cutoff such that $r<r_\infty(\to \infty)$ and the surface $\Sigma$ is identified with $r=r_\infty$ and $|\theta|\leq \theta_*$, where
\ba
\sinh\frac{\rho_*}{L}=\sin\theta_*\sinh\frac{r_\infty}{L}.
\ea
In the limit $r_\infty\to \infty$, we obtain
\ba
\theta_*\simeq 2\sinh\frac{\rho_*}{L}e^{-\frac{r_\infty}{L}}+2\left(\sinh\frac{\rho_*}{L}
+\frac{2}{3}\sinh^3\frac{\rho_*}{L}\right)e^{-\frac{3r_\infty}{L}}.
\ea
Note also that the extrinsic curvature of the surface $r=r_\infty$ is found as 
\ba
K|_\Sigma=\frac{d}{L\tanh\frac{r_\infty}{L}}.
\ea

Under this UV regularization, the $\eta$ coordinate in \eqref{adsrte} follows the regularization near the AdS boundary as 
$\eta< \eta_\infty$ such that
\ba
\cosh\eta_\infty \cosh\frac{\rho}{L}=\cosh\frac{r_\infty}{L} .
\ea
This is solved as 
\ba
e^{\eta_\infty}\simeq \frac{1}{\cosh\frac{\rho}{L}}e^{\frac{r_{\infty}}{L}}
+\left(\frac{1}{\cosh\frac{\rho}{L}}-\cosh\frac{\rho}{L}\right)e^{-\frac{r_{\infty}}{L}}.
\ea

The gravity action is calculated as follows
\ba
&& I_W=\frac{d}{8\pi G_N L^2}\int_W\s{g}-\frac{\tanh\frac{\rho_*}{L}}{8\pi G_N L}\int_{Q_1\cup Q_2} \s{h}
-\frac{d}{8\pi G_N L\tanh\frac{\rho_*}{L}}\int_\Sigma \s{\gamma}\no
&&=\frac{dL^{d-2}\cdot \mbox{Vol}(S^{d-1})}{4\pi G_N}\int^{\rho_*}_0 d\rho \cosh^d\frac{\rho}{L}
\int^{\eta_\infty(r)}_0 \sinh^{d-1}\eta\no
&&\ \ \ -\frac{L^{d-1}\cdot \mbox{Vol}(S^{d-1})\cdot \tanh\frac{\rho_*}{L}}{4\pi G_N}\cosh^d \frac{\rho_*}{L}\cdot \int^{\eta^*_\infty}_0 d\eta \sinh^{d-1}\eta\no
&&\ \ \ -\frac{d\cdot L^{d-1}\cdot \mbox{Vol}(S^{d-1})}{4\pi G_N\tanh\frac{r_\infty}{L}}
\sinh^d \frac{r_\infty}{L}\cdot \int^{\theta_*}_0 d\theta \cos^{d-1}\theta. 
\ea
Here $\mbox{Vol}(S^{d-1})$ is the volume of a unit radius of $d-1$ dimensional sphere given by 
\ba
\mbox{Vol}(S^{d-1})=\frac{2\pi^{\frac{d}{2}}}{\Gamma\left(\frac{d}{2}\right)}.
\ea 

In the limit $r_\infty\to \infty$, we can expand the action as
\ba
I_W=\frac{L^{d-1}}{4\pi G_N}\cdot \left(\frac{2\pi^{d/2}}{\Gamma(d/2)}\right)\cdot 2^{1-d}\cdot J_W,
\label{ijw}
\ea
where $J_W$ is given by 
\ba
&& J_W=(1-d)s_* e^{(d-1)r_\infty/L}+\left[\left(d^2-4d-1-\frac{4}{d-3}\right)s_*
+\frac{2}{3}(d-1)(d-2)s_*^3\right]e^{(d-3)r_\infty/L}\no
&&\ \ \ \ \ \ \ \ +O(e^{(d-5)r_\infty/L}),   \label{jw}
\ea
where we set $s_*=\sinh\frac{\rho_*}{L}$ for simplicity. Note that the leading contribution $O(e^{(d-1)r_\infty/L})$ agrees with the non-compact result \eqref{noncpt} by
identifying $e^{r_\infty/L}\sim \frac{1}{\ep}$.

In addition, there is the Hayward term contribution (\ref{Hay}), as we have evaluated in the previous example of the non-compact CFT dual. Since this calculation is straightforward and does not contribute to what we are interested in for the upcoming arguments, we simply omit this here.

\subsection{\texorpdfstring{$d$}{d} dimensional gravity viewpoint}
\label{dgrav}

If we regard the $d+1$ dimensional gravity on $W_{d+1}$ as a $d$ dimensional gravity on AdS$_d$ via the compactification along $\rho$ direction as in the brane world holography \cite{Randall:1999ee,Randall:1999vf,Karch:2000ct}, we find the effective $d$ dimensional Newton constant by reducing the original $d+1$ dimensional Einstein-Hilbert action in terms of the $d$ dimensional one:
\ba
&& \frac{1}{16\pi G_N}\int_W \s{g}R  \no
&& =\frac{1}{16\pi G_N}\int^{\rho_*}_{-\rho_*} d\rho
\left(\cosh\frac{\rho}{L}\right)^{d-3}\int_{\mbox{AdS}_d}\s{g^{(d)}}R^{(d)}
\equiv \frac{2}{16\pi G^{(d)}_N}\int_{\mbox{AdS}_d}\s{g^{(d)}}R^{(d)},\no
\ea
where $g^{(d)}$ is the metric of AdS$_d$ given by $ds^2=L^2y^{-2}(dy^2-dt^2+\sum_{i=1}^{d-2} d\xi_i^2)$ and $R^{(d)}$ is its scalar curvature. We inserted the factor $2$ in the final expression because there are two surfaces $Q_1$ and $Q_2$ where both are identical to AdS$_d$.

This leads to the relation
\ba
\frac{1}{G^{(d)}_N}=\frac{1}{G_N}\int^{\rho_*}_0 d\rho \left(\cosh\frac{\rho}{L}\right)^{d-2}. \label{newg}
\ea
In the small wedge limit $\rho_*\ll L$, we find 
\ba
\frac{G_N}{LG^{(d)}_N}\simeq \frac{\rho_*}{L}+\frac{d-2}{6}\left(\frac{\rho_*}{L}\right)^3+\ddd. \label{nep}
\ea

In the opposite limit $\rho_*\gg L$, where the wedge gets larger and gets closer to the full AdS$_{d+1}$, 
the $d$ dimensional Newton constant \eqref{newg} in terms of the CFT metric, behaves as follows\footnote{
Notice that here we normalized $d$ dimensional metric and the corresponding Newton constant $G^{(d)}_N$ 
in terms of the CFT metric which is obtained from the bulk metric via $ds^2_{bulk}=\frac{L^2}{y^2}(dy^2+ds^2_{CFT})$.}
\ba
\frac{G_N}{LG^{(d)}_N}=\frac{2^{2-d}}{d-2}e^{\frac{d-2}{L}\rho_*}\gg 1.  \label{newddd}
\ea

\subsection{Holographic entanglement entropy}
\label{sec:hee}

\begin{figure}
  \centering
  \includegraphics[width=6cm]{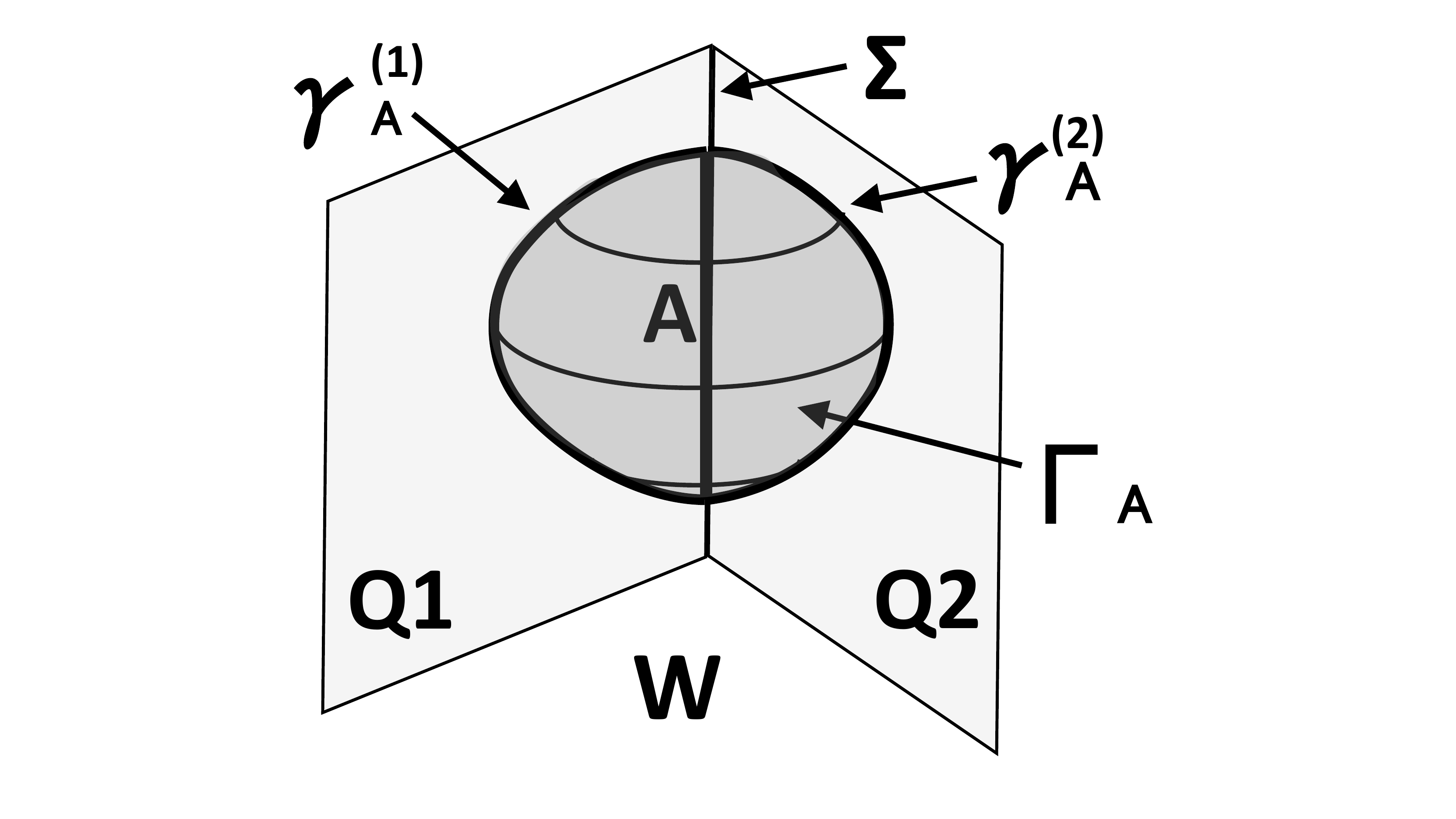}
\hspace{1cm}
 \includegraphics[width=6cm]{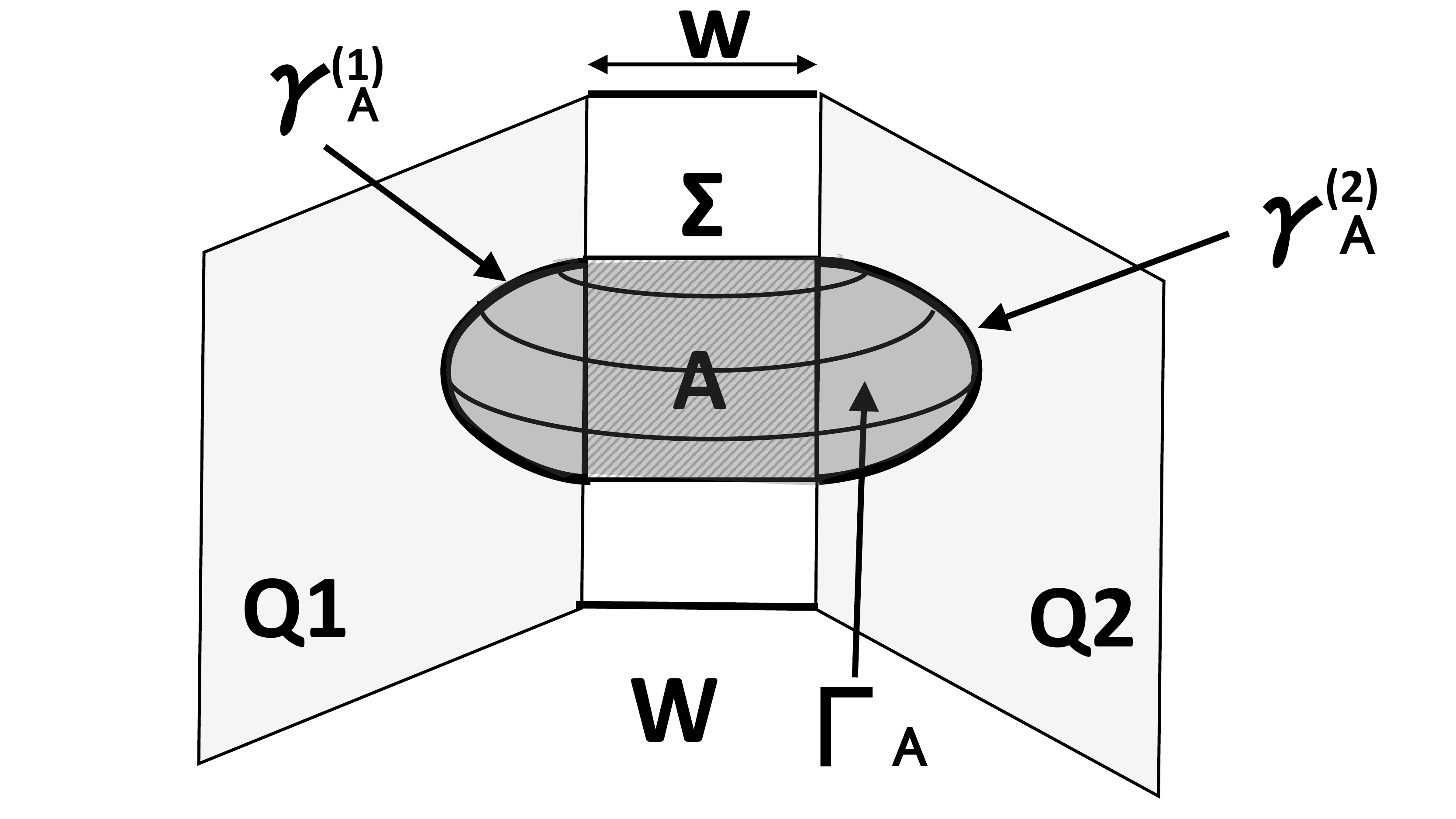}
  \caption{A sketch of calculation of holographic entanglement entropy in codimension two 
holography (left) and in AdS/BCFT (right).}
\label{HEEfig}
\end{figure}

When a gravity dual is given, a useful and universal probe of the geometry is the holographic entanglement entropy \cite{RT,RT2,HRT}. Let us consider how we can calculate the holographic entanglement entropy 
in our wedge holography. We choose a $d-2$ dimensional subsystem $A$ in the $d-1$ dimensional 
space $\Sigma_{d-1}$ at a time slice, chosen to be $t=0$ (remember the setup of Fig.~\ref{wedgesetupfig}).
 We would like to calculate the holographic entanglement entropy which is equal to the entanglement entropy $S_A$ in the dual CFT$_{d-1}$. We argue for the following holographic formula which involves the two step minimization:
\ba
S_A=\mbox{Min}_{\gamma_A\ s.t.\ \de\gamma^{(1)}_A= \de\gamma^{(2)}_A
=\de A}\left[\mbox{Min}_{\Gamma_A\ s.t.\ \de\Gamma_A=\gamma^{(1)}_A\cup \gamma^{(2)}_A}\left[ \frac{A(\Gamma_A)}{4G_N}\right]\right]. \label{formulaee}
\ea
As in the left picture of Fig.~\ref{HEEfig}, 
we choose  $d-2$ dimensional surfaces $\gamma^{(1)}_A$ on $Q_1$ and 
 $\gamma^{(2)}_A$ on $Q_2$ so that $\de\gamma^{(1,2)}_A=\de A$. Then we extend the $d-2$ dimensional 
compact surface $\gamma^{(1)}_A\cup \gamma^{(2)}_A$ to the bulk wedge $W_{d+1}$ as a $d-1$ dimensional surface $\Gamma_A$ such that $\de \Gamma_A=\gamma^{(1)}_A\cup \gamma^{(2)}_A$. First we fix 
the shape of $\gamma^{(1,2)}_A$ and minimize the area of $\Gamma_A$, denoted by $A(\Gamma_A)$.
Next we minimize this area by changing the shape of $\gamma^{(1,2)}_A$. This procedure selects a single minimal surface and this area gives the holographic entanglement entropy dual to $S_A$ in the $d-1$ dimensional CFT. This is what the formula (\ref{formulaee}) means. Notice that this formula guarantees
the property of strong subadditivity as in the usual holographic entanglement entropy \cite{Headrick:2007km}.
In the Lorentzian time dependent backgrounds we just need to replace the minimizations with the extremalizations as in \cite{HRT}. 

We can derive the formula \eqref{formulaee} by taking the zero width limit $w\to 0$ of the AdS/BCFT with the two boundaries (refer to the right picture of Fig.~\ref{AdSBCFTfig}). This is depicted in the right picture of Fig.~\ref{HEEfig}. Remember that in the AdS/BCFT \cite{AdSBCFT,AdSBCFT2}, we calculate the holographic entanglement entropy by minimizing the area among surfaces $\Gamma_A$ which end on the boundary surface $Q$ as well as the boundary of the subsystem $A$.

Consider the holographic entanglement entropy when the subsystem $A$ is given by $d-2$ dimensional 
round disk $\sum_{i=1}^{d-2}\xi_i^2\leq l^2$ with the radius $l$. To find the correct minimal surface
$\Gamma_A$, we need to first solve the partial differential equation for arbitrary choices of $\gamma^{(1,2)}_A$, which requires a lot of numerical computations. 

Let us first focus on the case where the wedge is very small i.e. $\rho_*\ll L$. In this case, the wedge geometry is approximated by 
the direct product of AdS$_d$, whose coordinate is given by $(y,t,\xi_1,\ddd,\xi_{d-2})$, and an interval $|\rho|\leq \rho_*$ as the warp factor stays almost constant $\cosh\frac{\rho}{L}\simeq 1$. Therefore, it is clear that both  $\gamma^{(1)}_A$ and  $\gamma^{(2)}_A$  should also be minimal surfaces on $Q_1$ and $Q_2$, i.e. the AdS$_d$. Thus we can identify $\Gamma_A$ with the wedge part of the sphere
\ba
x^2+z^2+\sum_{i=1}^{d-2}(\xi_i)^2=l^2,  
\ea
which is equivalent to 
\ba
y^2+\sum_{i=1}^{d-2}(\xi_i)^2=l^2,\label{minsph}
\ea
where the $\rho$ dependence drops out.

Furthermore, we would like to argue that the surface (\ref{minsph}) is the correct minimal surface defined by  (\ref{formulaee}) for any $\rho_*$. Though it is well-known that this surface solves the minimal surface equation in the Poincare AdS, we need to confirm that the area of the minimal surface with various choices of boundaries $\gamma^{(1,2)}$  is minimized when  $\gamma^{(1,2)}$ are both the same semi circle.
To see this, let us set $d=3$ just for notational simplicity and consider a perturbation around this semi-sphere solution
\ba
y=\s{l^2-\xi^2}+f(\xi,\rho),
\ea
where we impose $f(\pm l,\rho)=0$ as the surface should end on the boundary of subsystem $A$ in CFT$_{d-1}$. 

We find that under this perturbation the area of the surface looks like
\ba
&& A=\int d\xi d\rho \s{G}= -\left[\frac{L\cosh\frac{\rho}{L}}{l(l^2-\xi^2)^{\f{1}{2}}}\xi f(\xi)\right]^{\xi=l}_{\xi=-l}+
\int d\rho d\xi \Biggl[\frac{Ll\cosh\frac{\rho}{L}}{l^2-\xi^2}\no
&& +\! \frac{\!L\cosh\!\frac{\rho}{L}}{2l^3(l^2\!-\!\xi^2)^2}\!\left(\!2l^4f^2\!+\!L^2l^2\left(\!\cosh^2\!\frac{\rho}{L}\!\right)\!
(l^2\!-\!\xi^2)\!
(\de_\rho f)^2\!+\!2l^2\xi (l^2\!-\!\xi^2)f\de_\xi f\! +\!(l^2\!-\!\xi^2)^3(\de_\xi f)^2\!\right)\!\Biggr]. \no
\ea
The first term represents the linear perturbation and this takes the form of total derivative as the profile 
$f=0$ satisfies the minimal surface equation. This surface contribution vanishes due to the boundary condition.  From the form of quadratic terms of $f$, the local minimum is clearly identical to the solution without $\rho$ dependence, i.e. the minimal surface (or geodesic) in AdS$_3$. Thus we can confirm that the surface $y=\s{l^2-\xi^2}$ should be the exact minimal surface we want even when $\rho_*$ is finite.

Now, using this minimal surface solution, we can evaluate the minimal area $A(\Gamma_A)$ and the holographic entanglement entropy reads 
\ba
S_A=\frac{A(\Gamma_A)}{4G_N}=\frac{L^{d-2}}{2G_N}\cdot \mbox{Vol}(S^{d-3})\int^{\rho_*}_{0}d\rho  \left(\cosh\frac{\rho}{L}\right)^{d-2}\cdot 
\int^l_{\ep\cosh\frac{\rho}{L}} dy\frac{l(l^2-y^2)^{\frac{d}{2}-2}}{y^{d-2}},  \label{heegen}
\ea
where $\mbox{Vol}(S^{d-3})$ is the volume of a unit radius $d-3$ dimensional sphere given by 
\ba
\mbox{Vol}(S^{d-3})=\frac{2\pi^{\frac{d-2}{2}}}{\Gamma\left(\frac{d-2}{2}\right)}.
\ea 
In the limit $\rho_*/L\ll 1$, the holographic entanglement entropy looks like
\ba
S_A=\frac{A(\Gamma_A)}{4G_N}\simeq \frac{\rho_*}{4G_N}\cdot \left[A(\gamma^{(1)}_A)
+A(\gamma^{(2)}_A)\right],
\ea
where $A(\gamma^{(1,2)}_A)$ is the minimal surface area in $Q_{1,2}$, given by
\ba
A(\gamma^{(1)}_A)=A(\gamma^{(2)}_A)=L^{d-2}\cdot \mbox{Vol}(S^{d-1})\cdot 
\int^l_{\ep} dy\frac{l(l^2-y^2)^{\frac{d}{2}-2}}{y^{d-2}}.
\ea
This agrees with the (standard) holographic entanglement entropy in CFT$_{d-1}$ calculated from the minimal surface in Poincare AdS$_d$,  by relating the $d$ dimensional Newton constant to $G_N$ in $d+1$ dimension
as 
\ba
\frac{\rho_*}{G_N}\simeq\frac{1}{G^{(d)}_N},  \label{newtone}
\ea  
when $\rho_*\ll L$. This agrees with \eqref{nep}.

Now, let us go further to perform the full computation in (\ref{heegen}) for finite $\rho_*$ using a power expansion with respect to $l/\ep$. As a first step, we get
\begin{align}
    &\int^l_{\ep\cosh\frac{\rho}{L}} dy\frac{l(l^2-y^2)^{\frac{d}{2}-2}}{y^{d-2}} \nonumber\\
    =& {p_{1}}{\left(\cosh\frac{\rho}{L}\right)^{-d+3}}\left(\frac{l}{\ep}\right)^{d-3}+{p_{3}}{\left(\cosh\frac{\rho}{L}\right)^{-d+5}}\left(\frac{l}{\ep}\right)^{d-5}+\cdots \nonumber\\
    & \ldots+\left\{\begin{array}{ll} 
    p_{d-3}\left(\cosh\frac{\rho}{L}\right)^{-1}\left(\frac{l}{\ep}\right)+p_{d-2}+\mathcal{O}\left(\frac{\ep}{l}\right) & d: \text { even } \\
    p_{d-4}\left(\cosh\frac{\rho}{L}\right)^{-2}\left(\frac{l}{\ep}\right)^2+q \log \left(\frac{l}{\ep}\right)+\mathcal{O}(1) & d: \text { odd } 
\end{array}\right. \\
    =& 
    \left\{\begin{array}{ll} 
    \sum_{k=1}^{(d-2)/2} p_{2k-1} \left(\cosh\frac{\rho}{L}\right)^{-d+2k+1}\left(\frac{\ep}{l}\right)^{d-2k-1} +p_{d-2}+\mathcal{O}\left(\frac{\ep}{l}\right)
     & d: \text { even } \\
     \sum_{k=1}^{(d-3)/2} p_{2k-1} \left(\cosh\frac{\rho}{L}\right)^{-d+2k+1}\left(\frac{\ep}{l}\right)^{d-2k-1} + q \log \left(\frac{l}{\ep}\right)+\mathcal{O}(1), & d: \text { odd } 
\end{array}\right.
\end{align}
where the coefficients are as follows:
\begin{align}
    &p_1 = (d-3)^{-1}, p_3 = -(d-4)/[2(d-5)], \cdots \nonumber\\
    &p_{d-2} = (2 \sqrt{\pi})^{-1} \Gamma((d-2) / 2) \Gamma((3-d) / 2) \quad(\text{if } d\text{ is even}), \nonumber\\
    &q =\frac{\sqrt{\pi}}{\Gamma((4-d)/2)\Gamma((d-1)/2)} \quad(\text{if } d\text{ is odd}). 
\end{align}
Accordingly,
\begin{align}
    S_A&=\frac{A(\Gamma_A)}{4G_N}=\frac{L^{d-2}}{2G_N}\cdot \mbox{Vol}(S^{d-3})\int^{\rho_*}_{0}d\rho  \left(\cosh\frac{\rho}{L}\right)^{d-2}\cdot 
    \int^l_{\ep\cosh\frac{\rho}{L}} dy\frac{l(l^2-y^2)^{\frac{d}{2}-2}}{y^{d-2}}, \nonumber\\
    &=\frac{\pi ^{\frac{d-2}{2}}L^{d-2}}{G_N\Gamma\left(\frac{d-2}{2}\right)}
    \left(\int^{\rho_*}_{0}d\rho  \left(\cosh\frac{\rho}{L}\right) 
    {p_{1}}\left(\frac{l}{\ep}\right)^{d-3}+\int^{\rho_*}_{0}d\rho  \left(\cosh\frac{\rho}{L}\right)^3
    {p_{3}}\left(\frac{l}{\ep}\right)^{d-5}+\cdots\right)\nonumber\\
    &=\frac{\pi ^{\frac{d-2}{2}}L^{d-1}}{G_N\Gamma\left(\frac{d-2}{2}\right)}
    \Bigg(\frac{\sinh\frac{\rho_*}{L}}{d-3}\left(\frac{l}{\ep}\right)^{d-3}
    -\frac{(d-4)}{2(d-5)}\left(\frac{\sinh\frac{3\rho_*}{L}}{12}+\frac{3\sinh\frac{\rho_*}{L}}{4}\right)\left(\frac{l}{\ep}\right)^{d-5}+
    \cdots \nonumber\\
    &~~~~~~~~~~~~~~~~~~~~~~\cdots+\left\{\begin{array}{ll} 
    p_{d-2}\int^{\rho_*/L}_{0}d\eta  \left(\cosh\eta\right)^{d-2}+\mathcal{O}\left(\frac{\ep}{l}\right),
     & d: \text { even } \\
      q \int^{\rho_*/L}_{0}d\eta  \left(\cosh\eta\right)^{d-2} \cdot \log \left(\frac{l}{\ep}\right)+\mathcal{O}(1), & d: \text { odd } 
\end{array}\right. \Bigg).
\end{align}
First, notice that the above scaling profile of holographic entanglement entropy agrees with general results \cite{RT2, Casini:2011kv} in $d-1$ dimensional CFTs.
Moreover, by comparing the constant terms in even $d$ cases (or the log terms in odd $d$ cases) with those in the conventional holographic entanglement entropy from ${\rm AdS}_{d}/{\rm CFT}_{d-1}$, we can obtain the same relationship as ($\ref{newg}$) between the two Newton constants in ${\rm AdS}_{d+1}$ and ${\rm AdS}_{d}$, respectively. This observation gives a cross check to (\ref{newg}).

Some low dimensional results are shown as follows. For $d=3$, i.e. the AdS$_4$/CFT$_2$ case, 
\begin{align}
    S_A =  \frac{L^2 }{G_N^{(4)}}\sinh\frac{\rho_*
    }{L}\log\frac{l}{\epsilon} + \mathcal{O}(1),
\end{align}
whose details will be discussed later in subsection \ref{sec:heethrd}.
For $d=4$, i.e the AdS$_5$/CFT$_3$ case, we have
\begin{align}
    S_A = \left(\frac{\pi L^3}{G_N^{(5)}}\sinh\frac{\rho_*}{L}\right)\frac{l}{\epsilon} - \frac{\pi L^3}{G^{(5)}_N}\left(\frac{1}{2}\frac{\rho_*}{L} +\frac{1}{4}\sinh\frac{2\rho_*}{L}\right) + \mathcal{O}\left(\frac{\epsilon}{l}\right).
\end{align}
For $d=5$, i.e the AdS$_6$/CFT$_4$ case, we obtain
\begin{align}
    S_A = \left(\frac{\pi L^4}{G_N^{(6)}}\sinh\frac{\rho_*}{L}\right)\left(\frac{l}{\epsilon}\right)^2 - \frac{\pi L^4}{G^{(6)}_N}\left(\frac{\sinh\frac{3\rho_*}{L}}{12}+\frac{3\sinh\frac{\rho_*}{L}}{4}\right)\log\frac{l}{\epsilon} + \mathcal{O}\left(1\right).
\end{align}

\subsection{Bulk scalar field and spectrum}

\subsubsection{Dimensional reduction}
Consider the following metric in AdS$_{d+1}$ (we set $L=1$)
\ba
ds^2 = d \rho^2 + \cosh^2 \rho \frac{dy^2 + dx^i dx_i}{y^2},
\ea
describing the wedge bulk spacetime characterized by $-\rho_* \leq \rho \leq \rho_*$. We consider a massive Klein-Gordon field $\phi \equiv \phi(\rho,y,x^i)$ with mass $m$. For the latter, we aim at solving the corresponding wave equation,
\ba
\frac{1}{\sqrt{g}} \partial_\mu  (\sqrt{g} g^{\mu \nu} \partial_\nu \phi ) - m^2 \phi = 0
\ea
near the end-of-the-world branes $Q_1$ and $Q_2$ enclosing the AdS$_{d+1}$ wedge. Remember that
in the AdS/CFT correspondence, the mass $m$ of the scalar field is dual to a scalar operator with the conformal dimension
$\Delta = d/2 + \sqrt{(d/2)^2 + m^2}$ in the CFT$_d$.

Employing the coordinates above, the equation of motion takes the explicit form
\ba
\frac{1}{\cosh^d(\rho)} \partial_\rho (\cosh^d(\rho) \partial_\rho \phi) + \frac{y^2}{\cosh^2(\rho)} (\partial^2_y \phi + \sum_{i=1}^{d-1}\partial^2_i \phi) - \frac{d-2}{\cosh^2 (\rho)} y\partial_y \phi - m^2 \phi = 0.
\label{eq:waveeqnwedge}
\ea
Next, we impose a dimensional reduction in the $\rho$ direction, so that we have the wave equation on AdS$_d$ described by $ds^2 = \frac{dy^2 + dx^i dx_i}{y^2}$,
\ba
y^2 (\partial^2_y \phi + \sum_{i=1}^{d-1} \partial^2_i \phi) -(d-2)y\partial_y \phi - M^2 \phi = 0,
\label{eq:Meqn}
\ea
where $M$ corresponds to the Kaluza-Klein mass. We can now use \eqref{eq:Meqn} in order to rewrite \eqref{eq:waveeqnwedge} as 
\ba
\frac{1}{\cosh^d(\rho)} \partial_\rho (\cosh^d(\rho) \partial_\rho \phi) + \frac{M^2 \phi}{\cosh^2(\rho)} - m^2 \phi = 0.
\label{eq:waveeqnwedge2}
\ea
We want to solve this equation. For doing so, we consider the following standard ansatz
\ba
\phi = \tilde \phi(\rho) \varphi(y,x^i).
\label{eq:ansatz}
\ea
For accuracy, now we restrict our discussion to $d=2$. Results in higher dimensional setups can be obtained in a similar way. Note that the $d$s appearing below in the section are set to be $d=2$.

For the dimensionally reduced system on AdS$_d$, by following the described procedure, we would end up with a modified conformal dimension in CFT$_{d-1}$ of the form 
\ba
\tilde\Delta = \frac{d-1}{2} + \sqrt{\frac{(d-1)^2}{4} + M^2}.  \label{confddl}
\ea
The latter would then determine the correlation functions.

Having said this, let us consider the Dirichlet boundary condition on the surfaces $Q_1$ and $Q_2$:
\ba
\phi( \rho_*) = \phi( - \rho_*) = 0.
\label{eq:bdycond}
\ea
Imposing the first equality in \eqref{eq:bdycond} on the wave function $\phi$, we find the following $\rho$ dependent part 
\ba
\tilde \phi_+(\rho) = \left[ P^\theta_{\eta}(\tilde \rho)Q^\theta_{\eta}(\tilde \rho_*) - P^\theta_{\eta}(\tilde \rho_*)Q^\theta_{\eta}(\tilde \rho) \right] \times \frac{\sqrt{1 - {\tilde \rho}^2}}{ Q^\theta_{\eta}(\tilde \rho_*) },
\label{eq:phiPlusDiri}
\ea
where, for simplicity, we have defined
\begin{align}
\begin{split}
\tilde \rho &= \tanh (\rho),\quad \tilde \rho_* = \tanh (\rho_*),\\
\theta &= \sqrt{m^2 + 1},\quad \eta = \frac{\Pi - 1}{2},\quad \Pi = \sqrt{ 4M^2 + 1}.
\label{eq:var-defs}
\end{split}
\end{align}
The function $P_n^m$ denotes the associated Legendre function of the first kind, whereas $Q_n^m$ corresponds to the associated Legendre function of the second kind. It is useful to note that 
when 
\ba
\theta = \eta + 1,
\label{eq:N-D-conda}
\ea
the expression in \eqref{eq:phiPlusDiri} does vanish. 

In order to satisfy the second boundary condition in \eqref{eq:bdycond}, we should solve the following equation 
\ba
\tilde \phi_+ (\rho = - \rho_*) = 0
\ea
for the variable $\eta$. The final result, rewritten as $M$ using \eqref{eq:var-defs}, will depend on the AdS$_{d+1}$ mass $m$ and brane location $\rho_*$.  

We may alternatively impose the Neumann boundary condition, i.e. 
\ba
\phi^\prime( \rho_*) = \phi^\prime( - \rho_*) = 0.
\label{eq:bdycond-Neu}
\ea
For this case, the exact solution for the $\rho$ dependent part satisfying the first condition in \eqref{eq:bdycond-Neu} takes the following, more complicated form
\begin{align}
\begin{split}
\tilde \phi_+(\rho) 
&= \bigg[ 
(2 \theta - \Pi - 1)
\left[
P_{\eta}^\theta (\tilde \rho) Q_{1+\eta}^\theta(\tilde \rho_*) - P_{1+\eta}^\theta (\tilde \rho_*) Q_{\eta}^\theta(\tilde \rho)
\right]\\
&+ (\Pi \tilde \rho_* - \tilde \rho_*)  \left[ P_\eta^\theta(\tilde \rho) Q_\eta^\theta(\tilde \rho_*)
-  P_\eta^\theta(\tilde \rho_*) Q_\eta^\theta(\tilde \rho) \right]
\bigg]\\
&\times \frac{\sqrt{1 - {\tilde \rho}^2}}{ (2 \theta - \Pi - 1) Q_{1 + \eta}^\theta (\tilde \rho_*) + (\Pi \tilde \rho_* - \tilde \rho_*) Q_\eta^\theta(\tilde \rho_*)   }.
\label{eq:phiPlusNeu}
\end{split}
\end{align}
It can be seen that when 
\ba
2 \theta - \Pi = 1 \quad \Leftrightarrow \quad  \theta = \eta + 1,
\label{eq:N-D-cond}
\ea
the solution \eqref{eq:phiPlusNeu} does vanish as before. 
In the following, we discuss the mass spectrum in more detail when $\rho_*$ is taken to be very small.

Before doing so, let us note that for both boundary conditions, we can find the mass spectrum in the dimensionally reduced system 
 $M_n$ as a function of $m$ and $\rho_*$. The two point function of the dual  scalar operator $O(x)$ in the CFT$_{d-1}$ then behaves as 
\ba
\la O(x_1)O(x_2)\lb=\frac{1}{|x_1-x_2|^{2\tilde{\Delta}}},
\ea
where the conformal dimension is given by \eqref{confddl}.
Instead of the full numerical analysis, we below give the mass spectrum in the small $\rho_*$ limit by bringing the problem into a Schr{\"o}dinger-like form.
As we will see, for the Neumann boundary condition, the lowest mass in AdS$_d$ is given by $M_0=m$, while in the Dirichlet case, the lowest mass is heavy, i.e. $M_1=\mathcal{O}(1/\rho_*)$.

\subsubsection{\texorpdfstring{Schr{\"o}dinger}{Schrodinger} analysis}

By introducing a rescaled function $\psi(\rho)$, such that 
\ba
\psi(\rho)=\cosh(\rho) \phi(\rho),
\ea
we can transform the equation of motion \eqref{eq:waveeqnwedge2} into the following Schr{\"o}dinger-like form 
\ba
-\de^2_\rho\psi+V(\rho)\psi=0,
\ea
where
\ba
V(\rho)=1+m^2-\frac{M^2}{\cosh^2\rho}+\tanh^2\rho.
\label{eq:Vpot}
\ea

\paragraph{Dirichlet boundary condition}

When we impose the Dirichlet boundary condition $\psi(\rho_*)=\psi(-\rho_*)=0$ or \eqref{eq:bdycond}, 
the problem is similar to the quantum mechanics in a box. In particular, when $\rho_*$ is very small, we can approximate the potential \eqref{eq:Vpot} as 
\ba
V(\rho)\simeq 1+m^2-M^2.
\ea
Since in this small $\rho_*$ approximation, we can solve the boundary condition as 
\ba
\psi(\rho)\propto \sin\left[\frac{n\pi}{2}\left(\frac{\rho}{\rho_*}+1\right)\right] ,\  \ \ (n=1,2,3,\ldots),
\ea
the spectrum of $M$ can be found as follows
\ba
M_n\simeq \s{1+m^2+\frac{n^2\pi^2}{4\rho_*^2}},\  \ \ (n=1,2,3,\ldots).  \label{ewew}
\ea
Therefore, the lowest excitation gives $M_1= \s{1+m^2+\frac{\pi^2}{4\rho_*^2}}\simeq \frac{\pi}{2\rho_*}$.
In this case, we should be careful because of an artificial \textit{zero point} of $\phi(\rho)$ that arises when $\frac{1}{2}(\s{1+4M^2}-1)=\s{1+m^2}$. 
Note that the latter observation precisely agrees with the vanishing condition in \eqref{eq:N-D-conda} and \eqref{eq:N-D-cond}.

\paragraph{Neumann boundary condition}

If we impose the Neumann boundary condition \eqref{eq:bdycond-Neu}, 
then when $\rho_*$ is very small, we find the obvious lowest mode
\ba
\phi(\rho)=\mbox{const.}
\ea
This leads to the lowest spectrum $M_0=m$.
For excited modes, we find 
\ba
\psi(\rho)\propto \cos\left[\frac{n\pi}{2}\left(\frac{\rho}{\rho_*}+1\right)\right] ,\  \ \ (n=1,2,3,\ldots),
\ea
which leads to the same spectrum as in \eqref{ewew}.

\section{Wedge holography for \texorpdfstring{$d=3$}{d=3}}
\label{sec:wh-3d}

Here we focus on $d=3$ case of the wedge holography, where the dual CFT is two dimensional and study how this holographic duality works in detail.

\subsection{Conformal anomaly}

First we would like to study the behavior of free energy of the dual theory on $S^2$ and extract the value of 
central charge from the conformal anomaly as an extension  of holographic weyl anomaly \cite{Henningson:1998gx}. By setting $d=3$ in the free energy \eqref{ijw} with \eqref{jw}, 
we obtain the following behavior in the limit $r_\infty=-\log \ep\to \infty$:
\ba
I_W=-\frac{L^2}{2G_N\ep^2}\sinh\frac{\rho_*}{L}+\frac{L^2}{G_N}\sinh\frac{\rho_*}{L}\cdot \log \ep +O(1).
\ea
Note that we neglect the Hayward contribution $I_H$ as this does not contribute to the conformal anomaly. 
Since the Euclidean action of a two dimensional CFT include the conformal anomaly term as 
$I_{\mbox{2dCFT}}=\# \ep^{-2}+\frac{c}{6}\chi(\Sigma)\log\ep+O(1)$, where $c$ is the central charge of the two dimensional CFT and $\chi(\Sigma)$ is the Euler character of the two dimensional manifold $\Sigma$.
Since $\chi(S^2)=2$ we obtain
\ba
c=\frac{3L^2}{G_N}\sinh\frac{\rho_*}{L}. \label{centr}
\ea

\subsection{Holographic entanglement entropy}
\label{sec:heethrd}

Next we would like to analyze the holographic entanglement entropy in the $d=3$ case. We consider the 
non-compact setup in section \ref{sec:hee}. The dual two dimensional CFT is defined on R$^2$ and we  define the entanglement entropy $S_A$ by choosing the subsystem $A$ on $\Sigma$ as the interval $-l\leq w\leq l$ at a time $t=0$. By setting $d=3$ in the general result \eqref{heegen} we obtain the holographic entanglement entropy $S_A$ when $A$ is a length $2l$ interval:.
\ba
S_A&=&\frac{L}{2G_N}\int^{\rho_*}_{-\rho_*}d\rho \left(\cosh\frac{\rho}{L}\right)\int^{l}_{\ep\cosh\frac{\rho}{L}}dy
\frac{l}{y\s{l^2-y^2}}\no
&=&\frac{L^2}{G_N}\sinh\frac{\rho_*}{L}\log\frac{2l}{\ep}-4L^2\int^{\rho_*/L}_0 dw \cosh(w)\log\cosh(w),
\label{funeef}
\ea
which agrees with the well-known form \cite{Holzhey:1994we} in a two dimensional CFT 
$S_A=\frac{c}{3}\log\frac{2l}{\ep}+(\mbox{const.})$. This leads to the precisely same identification of the central charge as that of  \eqref{centr}. We can view this as a quantitative test of the wedge holography proposal.

It is also intriguing to consider more general choices of the subsystem $A$. When $A$ consists of two disjoint intervals,  we can again calculate the holographic entanglement entropy  (\ref{formulaee}) using two spherical minimal surfaces as in \eqref{minsph}. Since the resulting entanglement entropy computed from such spheres takes the form of the logarithmic function of $l$ as in (\ref{funeef}), the result of holographic entanglement entropy for the two intervals also simply reproduces the known results of holographic CFTs \cite{Headrick:2010zt,Hartman:2013mia}. In this case, we know that there is a phase transition phenomenon between the connected and disconnected minimal surfaces. The same is true when $A$ consists of  multiple disjoint intervals. In this way, the holographic entanglement entropy in wedge holography perfectly reproduces  know results in holographic CFTs.

\subsection{Three dimensional gravity viewpoint}

If we apply the Brown-Henneaux formula \cite{Brown:1986nw} to the $d=3$ dimensional gravity on $Q_1$
and that on $Q_2$ with the Newton constant $G^{(3)}_N$, we find the central charges of the dual two dimensional CFTs
\ba
c_1=c_2=\frac{3L}{2G^{(3)}_N}.
\ea
Since the relation \eqref{newg} at $d=3$ leads to
\ba
\frac{1}{G^{(3)}_N}=\frac{L}{G_N}\sinh\frac{\rho_*}{L}, \label{relationtd}
\ea
we can confirm that the central charge\eqref{centr} is the sum of the above two central charges $c=c_1+c_2$ 
as we expect.


\section{Wedge holography for \texorpdfstring{$d=2$}{d=2}}
\label{whdt}

Here we would like to focus on $d=2$ setup of the wedge holography depicted in Fig.~\ref{wedgesetupfig}. 
In this case, the dual CFT lives in $d-1=1$ dimension and therefore one may wonder whether this theory turns out to be a Schwarzian theory as in the JT gravity \cite{Maldacena:2016upp} or some other theory. Notice that since our setup is based on three dimensional gravity rather than two dimensional one, we can assume a standard Einstein gravity, which looks different from the situation in two dimension. 

Since the wedge $W_3$ is a part of the AdS$_3$ geometry, we have the advantage that all solutions to the Einstein equation with the Neumann boundary condition on $Q_1$ and $Q_2$ is locally expressed as the metric \eqref{AdsP} or \eqref{adsg}. In the Lorentzian signature the metric is written as  
\ba
ds^2&=&L^2\left(\frac{dz^2-dt^2+dx^2}{z^2}\right)=d\rho^2+L^2 \cosh^2\frac{\rho}{L}\left(\frac{-dt^2+dy^2}{y^2}\right). \label{adstf}
\ea 
The surfaces  $Q_1$ and $Q_2$ are again identified with $\rho=-\rho_*$ and $\rho=\rho_*$. We can change the shape of $\Sigma$ by deforming the choice of the UV cutoff.  We specify the form of $\Sigma$ by
\ba
z=g(t).
\ea

\subsection{Bulk on-shell action}
The total gravity action in the Lorentzian signature looks like
\ba
I_G=\frac{1}{16\pi G_N}\int_{W} 
\s{-g}(R-2\Lambda)+\frac{1}{8\pi G_N}\int_{Q_1\cup Q_2}\s{-\gamma}(K-T)
+\frac{1}{8\pi G_N}\int_{\Sigma}\s{-\gamma}K.\no \label{actiongg}
\ea
Note that in our setup we have
\ba
R=-\frac{6}{L^2},\ \ \ \Lambda=-\frac{1}{L^2},\ \ \ T=\frac{1}{L}\tanh\frac{\rho_*}{L}.
\ea
The induced metric on $\Sigma$ is 
\ba
ds^2=L^2\frac{-(1-\dot{g}^2)dt^2+dx^2}{g^2},
\ea
and its extrinsic curvature $K_{ab}=(\nabla_a N_b)_{\Sigma}$ for the normal vector $N^a$
\ba
(N^z,N^t,N^x)=\frac{g}{L\s{1-\dot{g}^2}}(-1,-\dot{g},0),
\ea
has the trace value
\ba
K|_\Sigma=\frac{2-2\dot{g}^2-g\ddot{g}}{L(1-\dot{g}^2)^{3/2}}.
\ea

Thus we find
\ba
&& I_W =-\frac{1}{4\pi G_NL^2}\int_W \s{-g}+\frac{\tanh\frac{\rho_*}{L}}{8\pi G_N L}\int_{Q_1\cup Q_2}\s{-h}
+\frac{1}{8\pi G_N}\int_{\Sigma}\s{-\gamma}K_\Sigma\no
&&=-\frac{1}{2\pi G_N}\int^{\rho_*}_0d\rho \cosh^2\frac{\rho}{L}\int dt \int^\infty_{g(t)\cosh\frac{\rho}{L}}
\frac{dy}{y^2}+\frac{L\sinh\frac{\rho_*}{L}\cosh\frac{\rho_*}{L}}{4\pi G_N}\int dt \int^\infty_{g(t)\cosh\frac{\rho_*}{L}}\frac{dy}{y^2}\no
&&\ +\frac{L}{4\pi G_N}\int dt \int^{g(t)\sinh\frac{\rho_*}{L}}_0 dx \left( \frac{2-2\dot{g}^2-g\ddot{g}}{g^2(1-\dot{g}^2)}\right)\no
&&=\frac{L\sinh\frac{\rho_*}{L}}{4\pi G_N}\int \frac{dt}{g}\left[1-\frac{g\ddot{g}}{1-\dot{g}^2}\right].
\ea

We introduce the Weyl scaling factor as follows:
\ba
e^{2\vp}=\frac{1-\dot{g}^2}{g^2}.
\ea
Assuming the usual UV cutoff property $g\ll 1$, we can expand 
\ba
g\simeq e^{-\vp}-\frac{1}{2}e^{-3\vp}\dot{\vp}^2+\ddd.
\ea
This leads to 
\ba
I_W&\simeq &\frac{L\sinh\frac{\rho_*}{L}}{4\pi G_N}\int \frac{dt}{g}=\frac{L\sinh\frac{\rho_*}{L}}{4\pi G_N}\int dt \left[e^{\vp}+\frac{1}{2}\dot{\vp}^2e^{-\vp}\right],  \label{imtwo}
\ea
where we performed a partial integration. The kinetic term of $\vp$ in \eqref{imtwo} is proportional to the Schwarzian action (refer to the appendix \ref{sec:schac} for a brief review of Schwarzian action). 

\subsection{Hayward term contribution}

However, it is still too early to conclude as there is another contribution from the Hayward term
\eqref{Hay} as we will see below. We write the angle between 
$Q_{1,2}$ and $\Sigma$ by $\eta+\frac{\pi}{2}$, where $\eta$  is given by
\ba
\sin\eta=N^{(1)}\cdot N^{(2)}=\frac{\tanh\frac{\rho_*}{L}}{\s{1-\dot{g}^2}}. \label{sineta}
\ea
$N^{(1)}$ and $N^{(2)}$ are the normal vector on $\Sigma$ and $Q_2$, explicitly given by 
\ba
&& (N^{(1)z},N^{(1)t},N^{(1)x})=\frac{g}{L\s{1-\dot{g}^2}}(-1,-\dot{g},0),\no
&& (N^{(2)z},N^{(2)t},N^{(2)x})=\frac{g}{L\cosh\frac{\rho_*}{L}}
\left(-\sinh\frac{\rho_*}{L},0,1\right).
\ea

We can evaluate the (Lorentzian) Hayward term as follows
\ba
I_H&=&\frac{L}{4\pi G_N}\int dt \frac{\s{1-\dot{g}^2}}{g}\left[\frac{\pi}{2}-\eta\right] \no
&=&\frac{L}{8G_N}\int dt e^{\vp}-\frac{L}{4\pi G_N}\int dt e^{\vp}\eta. \label{etaa}
\ea

By expanding \eqref{sineta} assuming $\dot{g}\ll 1$, we find 
\ba
\eta=\eta_0+\frac{1}{2}\sinh\frac{\rho_*}{L}\cdot \dot{g}^2+\ddd,
\ea
where $\eta_0$ is defined by $\sin\eta_0=\tanh\frac{\rho_*}{L}$.
Thus we can approximate \eqref{etaa} as follows
\ba
I_H\simeq \frac{L}{4\pi G_N}\int dt\left[(\frac{\pi}{2}-\eta_0)e^{\vp}-\frac{1}{2}\sinh\frac{\rho_*}{L}\cdot \dot{\vp}^2
e^{-\vp}\right].  \label{ihjt}
\ea

\subsection{Total gravity action}

By adding \eqref{imtwo} and \eqref{ihjt}, we obtain the total contribution to the free energy:
\ba
I_W+I_H\simeq \frac{L}{4\pi G_N}\int dt \left(\sinh\frac{\rho_*}{L}+\frac{\pi}{2}-\eta_0\right)e^{\vp}.
\ea
Thus we can conclude that there is no contribution which look like the Schwazian action.
One may worry that this cancellation can depend on the details of UV regularization. However, as we show in the appendix \ref{sec:another}, we again find no Schwarzian action term in another choice of the regularization where the Hayward contribution is vanishing. Also this is consistent with the fact that as opposed to the JT gravity, our three dimensional Einstein gravity manifestly preserves the conformal symmetry. In summary, we expect the dual CFT$_1$ is not a Schwarzian theory but a conformal quantum mechanics or a certain topological theory.

If we were to start with the Dirichlet boundary condition also on both $Q_1$ and $Q_2$, then we would not need to add the term $-\frac{T}{8\pi G_N}\int \s{-\gamma}$ in \eqref{actiongg}. In this case, the bulk contribution vanishes and we find the extra contribution in addition to \eqref{imtwo}:
\ba
\Delta I_W\simeq \frac{L\sinh\frac{\rho_*}{L}}
{4\pi G_N}\int dt\left[e^{\vp}+\frac{1}{2}\dot{\vp}^2\right]
=
\frac{L\sinh\frac{\rho_*}{L}}
{4\pi G_N}\int dt e^{\vp}\left(1- \mbox{Sch}(t,u)\right), \label{scdesf}
\ea
where we defined the coordinate $u$ by $e^{\vp}dt=du$ and $\mbox{Sch}(t,u)$ is the Schwarzian action.

This suggests the following interpretation. The latter Dirichlet setup of our wedge holography is dual to 
two CFTs on AdS$_2$ coupled to a quantum mechanics on the defect. This defect quantum mechanics is effectively described by the Schwarzian action \eqref{scdesf}. In the original Neumann setup, the CFTs on AdS$_2$ are coupled to quantum gravity. Therefore, it absorbs some degrees of freedom and this cancels the previous degrees of freedom of defect quantum mechanics given by the Schwarzian action.\\

\section{Wedge holography as interacting two CFTs}
\label{sec:wh-joinCFTs}

Here we would like to consider the limit where the size (i.e. width) of $\Sigma$ is strictly vanishing and its holographic interpretation. 
 In the Poincare AdS$_{d+1}$, it is straightforward to realize this limit by choosing the boundary surfaces $Q_1$ and 
$Q_2$ to be
\ba 
 Q_1:\ x=- \sinh\frac{\rho_*}{L}(z-\ep), \ \ \ \ \ Q_2:\ x= \sinh\frac{\rho_*}{L}(z-\ep).  \label{qab}
\ea

\begin{figure}
  \centering
  \includegraphics[width=8cm]{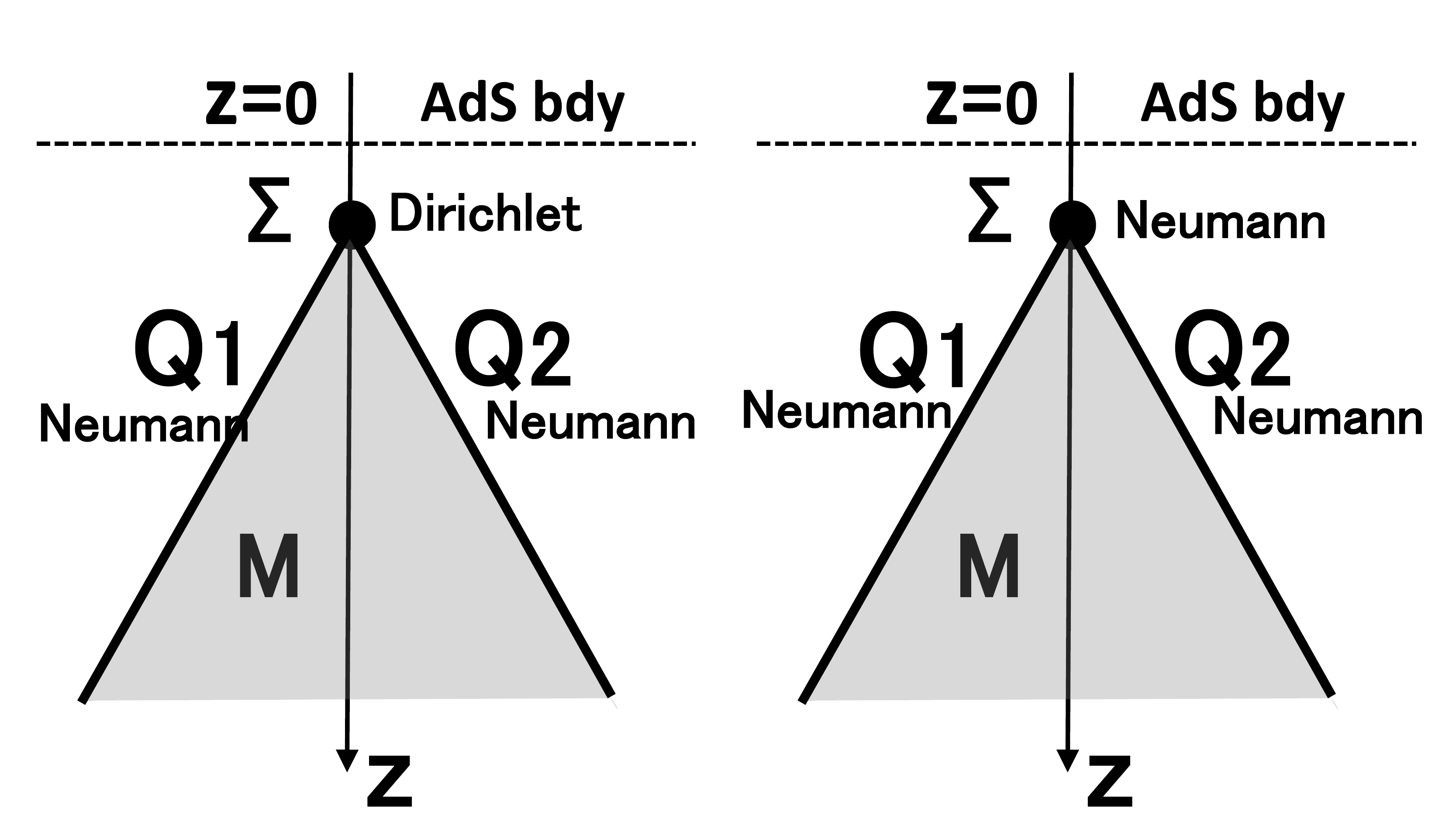}
\caption{Other two versions of wedge holography setups. The left (and right) setup imposes the Dirichlet (and Neuman) boundary condition on the tip $\Sigma$ of the wedge. }
\label{setupwegfig}
\end{figure}

 In this limit, we obtain the setup of the left picture in Fig.~\ref{setupwegfig}, where the surface $\Sigma$ is now $d-1$ dimensional. Here we impose the Neumann boundary condition on the $d$ dimensional surfaces $Q_1$ and $Q_2$ and the Dirichlet boundary condition on $\Sigma$. This corresponds to a limit of the wedge holography with the vanishing size of $\Sigma$.

On the other hand, as a different setup, we can also impose the Neumann boundary condition also on $\Sigma$ i.e. the right picture in Fig.~\ref{setupwegfig}, which is holographic dual to a gravity on a manifold defined by two AdS$_d$ glued along their boundaries. If we apply the brane-world holography \cite{Randall:1999ee,Randall:1999vf,Karch:2000ct}, the gravity on the wedge $W_{d+1}$ is dual to the (quantum) gravity on 
two copies of AdS$_d$ glued along their boundaries. Therefore, by applying the AdS/CFT holography once more, we expect that this theory is equivalent to two $d-1$ dimensional CFTs on $\Sigma_{d-1}$, interacting as the common metric is dynamical and fluctuating owing to the Neumann boundary condition on  $\Sigma_{d-1}$.

We will analyze these two setups with different boundary conditions below.

\subsection{Dirichlet boundary condition on \texorpdfstring{$\Sigma$}{Sigma}}

The gravity action for the region surrounded by \eqref{qab} can be computed as in our previous analysis.
Actually we find that the gravity action $I_W$ defined by
\ba
I_W=-\frac{1}{16\pi G_N}\int_{W} \s{g}(R-2\Lambda)-\frac{1}{8\pi G_N}\int_{Q_1\cup Q_2}\s{h}(K-T) \label{gravwed}
\ea
is vanishing in this wedge background:
\ba
I_W&=&\frac{1}{16\pi G_N}\int_W\frac{2d}{L^2}\s{g}-\frac{\tanh\frac{\rho_*}{L}}{8\pi G_N L}\int _{Q_1\cup Q_2}\s{\gamma}
\no
&=&\frac{dV_{d-1}L^{d-1}}{4\pi G_N}\int^\infty_{\ep\sinh\frac{\rho_*}{L}}dx\int^{\infty}_{\frac{x}{\sinh\frac{\rho_*}{L}}}
\frac{dz}{z^{d+1}}\no
&&-\frac{\tanh\frac{\rho_*}{L}V_{d-1}L^d}{4\pi G_N L}\left(\cosh\frac{\rho_*}{L}\right)^d\int^\infty_{\ep\cosh\frac{\rho_*}{L}}\frac{dy}{y^d}\no
&=&\frac{V_{d-1}L^{d-1}\sinh\frac{\rho_*}{L}}{4(d-1)\pi G_N \ep^{d-1}}-\frac{V_{d-1}L^{d-1}\sinh\frac{\rho_*}{L}}{4(d-1)\pi G_N \ep^{d-1}}=0.
\ea
This result differs from \eqref{noncpt} as the latter includes the contribution from the small but 
non-zero size of $\Sigma$.

Also the Hayward contribution \eqref{Hay}, where in the present setup we have 
\ba
I_{H}=\frac{1}{4\pi G_N}\int_{\Sigma} \left(\theta-\frac{\pi}{2}\right)\s{\gamma} \label{Hayy}
\ea 
is evaluated to be the same as  \eqref{ihij}.
Since $I_W=0$, the full gravity action in this Dirichlet case is given by \eqref{ihij}.

If we apply the brane world holography, we will find a copy of AdS$_d$ attached along their boundaries $\Sigma_{d-1}$. Performing another AdS/CFT, the setup is dual to two CFT$_{d-1}$s on $\Sigma_{d-1}$. 
Since we impose the Dirichlet boundary on $\Sigma$, there is no dynamical gravity coupled to them.
One may worry that they look decoupled, which may contradict with the fact that in the wedge geometry 
$W_{d+1}$, the two asymptotic regions are connected. However, the two CFTs are interacting through the large $N$ CFT$_{d}$ degrees of freedom, though they are not via a dynamical gravity on $\Sigma$.

\subsection{Neumann boundary condition on \texorpdfstring{$\Sigma$}{Sigma}}

To impose a Neumann boundary condition which fixes the value of $\theta$, we need to add to the Hayward term \eqref{Hayy} a cosmological constant term localized on $\Sigma$:
\ba
I_{C}=\frac{\kappa}{8\pi G_N}\int_{\Sigma} \s{\gamma}.
\ea
The vanishing variation condition $\delta(I_W+I_H+I_C)=0$ leads to the following condition (Neumann boundary condition 
on $\Sigma$):
\ba
2\theta-\pi+\kappa=0.  \label{nedocm}
\ea
Thus for any $\theta$ we want to realize, we can choose $\kappa$ so that the Neumann boundary condition is satisfied.
In this case with the Neumann boundary condition \eqref{nedocm}, we find that the total action vanishes
\ba
I_W+I_H+I_C=0.
\ea
This vanishing action is not surprising. We can easily see, for example, that the pure AdS with the cutoff $z\geq \ep$, where we impose the Neumann boundary condition  \eqref{bdycond} on the cutoff surface, the gravitational action vanishes, by adding the contribution $\frac{T}{8\pi G^{(d)}_N}\int_{R^{d-1}} \s{h}$ to the Dirichlet computation \eqref{padsp}.

\begin{figure}
  \centering
  \includegraphics[width=8cm]{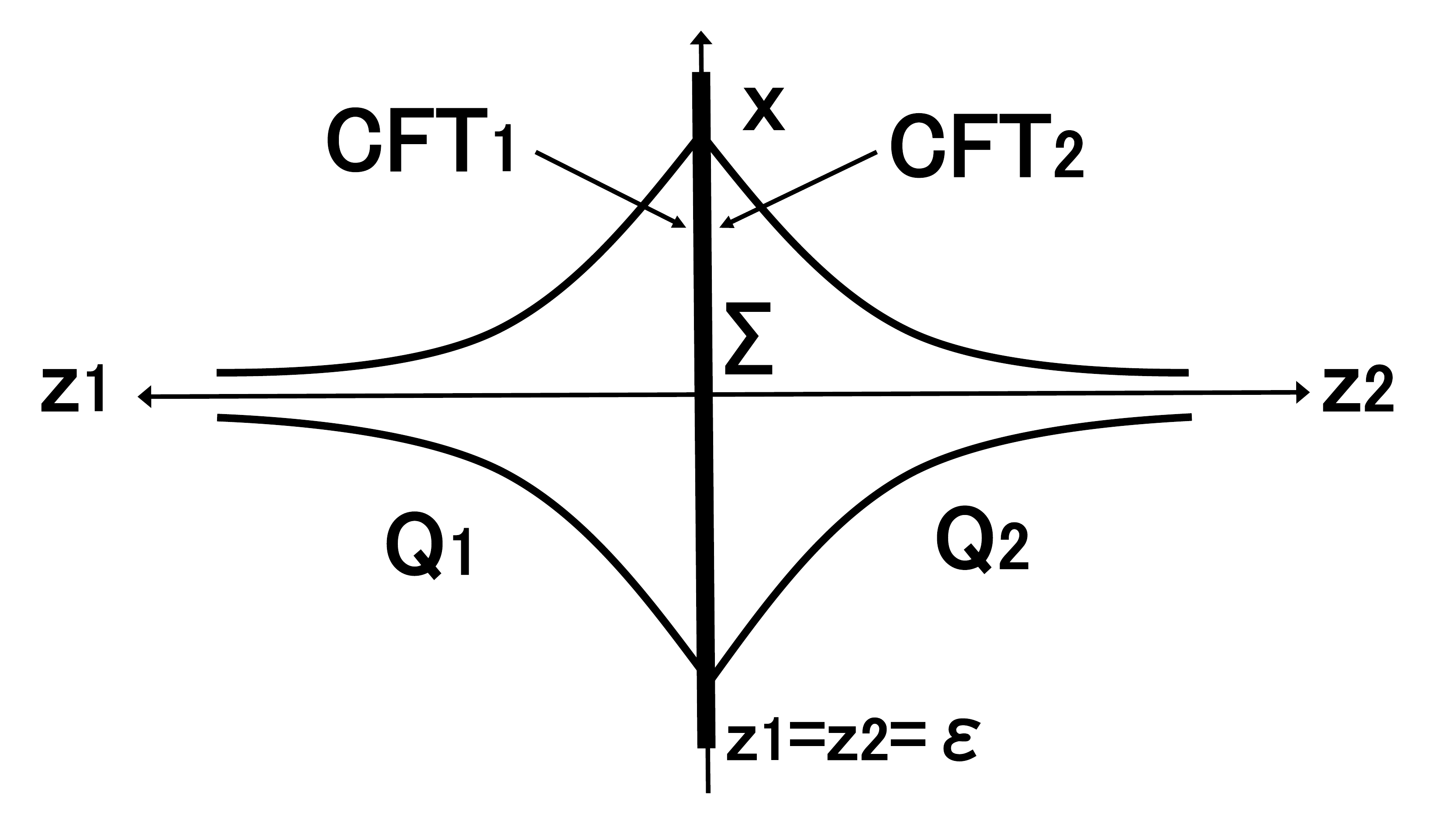}
\caption{Gluing two AdS backgrounds.}
\label{setupgluefig}
\end{figure}

In this Neumann case, the classical gravity on the $d+1$ dimensional wedge region is dual to a gravity on
the $d$ dimensional space $Q_1\cup Q_2$.  Since $Q_1$ and $Q_2$ are both AdS$_d$, this gravity lives on a space obtained by gluing two AdS$_d$ along their boundary $R^{d-1}$ as depicted in Fig.~\ref{setupgluefig}. 

In the limit $\theta\simeq \frac{\pi}{2}$ (i.e. $\rho_*\gg L$), we expect the gravity on AdS$_d$ gets weakly coupled in terms of the CFT metric. By approximating this $d$ dimensional gravity by the Einstein gravity, the bulk gravity action can be expressed as follows: 
\ba
I_{gravity}&=&-\frac{1}{16\pi G^{(d)}_N}\int_{Q_1} \s{g}(R-2\Lambda^{(d)})-\frac{1}{8\pi G_N}\int_{\Sigma}\s{h}(K^{(1)}-T)\no
&&-\frac{1}{16\pi G^{(d)}_N}\int_{Q_2} \s{g}(R-2\Lambda^{(d)})-\frac{1}{8\pi G_N}\int_{\Sigma}\s{h}(-K^{(2)}-T), \label{gravacaq}
\ea
where $\Lambda^{(d)}=-\frac{(d-1)(d-2)}{2L^2}$.
On their common boundary $\Sigma$, we impose the Neumann boundary condition which connect the two AdS$_d$ regions:
\ba
&& h^{(1)}_{\ap\beta}=h^{(2)}_{\ap\beta}(\equiv \ti{\gamma}_{ab}),\no
&& (K^{(1)}_{\ap\beta}-K^{(1)}\ti{\gamma}_{\ap\beta})-(K^{(2)}_{\ap\beta}-K^{(2)}\ti{\gamma}_{\ap\beta})
=-2\ti{T}\ti{\gamma}_{\ap\beta},  \label{bcondim}
\ea
where $h^{(1)}_{\ap\beta}$ and $K^{(1)}_{\ap\beta}$ are the metric and extrinsic curvature on $\Sigma$ in the left AdS$_d$ (i.e. Q$_1$), while   $h^{(2)}_{\ap\beta}$ and $K^{(2)}_{\ap\beta}$ are those in the right one $Q_2$. $\ti{T}$ is the tension of each boundary, which gives totally $\frac{2\ti{T}}{8\pi G^{(d)}_N}\int_\Sigma \s{\ti{\gamma}}$ contribution to the total $d$ dimensional gravity action.

As in section \ref{dgrav}, we relate the $d+1$ wedge geometry to AdS$_d$ via $ds^2_{wedge}=d\rho^2+\cosh^2\frac{\rho}{L}ds^2_{AdS_d}$. Then the extrinsic curvature on $\Sigma$ for the AdS$_d$ gravity is estimated as
\ba
K^{(1)}_{\ap\beta}=-K^{(2)}_{\ap\beta}=\frac{d-1}{L},
\ea
which leads to the value of the tension 
\ba
\ti{T}=\frac{d-2}{L},
\ea
by solving the boundary condition \eqref{bcondim}.
In this limit $\rho_*\gg L$ (or equally $\theta\simeq \frac{\pi}{2}$), we can indeed  confirm the boundary cosmological constant term in the $d$ dimensional gravity agrees with that in the $d+1$ dimensional one $I_C$:
\ba
\frac{2\ti{T}}{8\pi G^{(d)}_N}\int_\Sigma \s{\ti{\gamma}}\simeq \frac{\kappa}{8\pi G_N}\int_\Sigma \s{\gamma},
\ea
where we substituted  $\gamma_{\ap\beta}=\cosh\frac{\rho_*}{L}\cdot \ti{\gamma}_{\ap\beta}$, 
$\kappa=\pi-2\theta$,  $\cos\theta=\frac{1}{\cosh\frac{\rho_*}{L}}$ and  \eqref{newddd}. 

We can apply the AdS/CFT once more to the $d$ dimensional gravity on two AdS$_d$ regions glued each other. If we impose the Neumann boundary condition on $\Sigma$, which leads to dynamical gravity on $\Sigma$, the holographic dual is given by the CFT$_1$ and CFT$_2$, which are now interacting via the common metric field 
$\ti{\gamma}_{\ap\beta}$. This interacting two CFTs can be explicitly described by the total CFT action 
\ba
S_{tot}=S_{CFT}(\Phi^{(1)},\ti{\gamma})+S_{CFT}(\Phi^{(2)},\ti{\gamma})-\ti{T}\int_N \s{\ti{\gamma}},
\ea
where we path-integrate over the metric $\ti{\gamma}_{\ap\beta}$. Here $S_{CFT}(\Phi^{(1,2)},\ti{\gamma})$ are the CFT$_{d-1}$ actions in a curved background with the metric $\ti{\gamma}_{\ap\beta}$. In the linear level analysis, we expect a double trace interaction $\int \s{\ti{\gamma}}T^{(1)}_{\ap\beta}T^{(2)\ap\beta}$, where $T^{(1,2)}_{\ap\beta}$ are energy stress tensor in the two CFTs.
 It will be an intriguing future problem to study directly the $d$ dimensional gravity dual of this two interacting $d-1$ dimensional CFTs.

\section{BTZ and wedge holography}
\label{sec:wh-btz}

As an example of wedge holography at finite temperature, we would like to consider that in
the BTZ black hole spacetime. For this we first investigate the profiles of end of the world-branes in the BTZ background where we impose the Neumann boundary condition for the metric, namely solutions to 
(\ref{bdycond}). Since the BTZ geometry can be locally equivalent to the Poincare AdS$_3$ via a coordinate transformation, let us start with the Poincare AdS$_3$.

\subsection{End-of-the-world branes in Poincare \texorpdfstring{AdS$_3$}{AdS3}}

General solutions to the Neumann boundary condition (\ref{bdycond}) in the Poincare AdS$_3$ \eqref{adstf}
are given in the form
\ba
(z-\ap)^2+(x-p)^2-(t-q)^2=\beta^2,  \label{solab}
\ea
where $\ap,p,q$ and $\beta$ are arbitrary real valued constants (we take  $\beta\geq 0$).
In this solution, the tension $T$ is found to be 
\ba
T=\frac{\ap}{\beta L},  \label{tensionp}
\ea
assuming that we choose the inside of the region \eqref{solab} as a physical space for which we consider the AdS/BCFT holography. If instead we choose the outside region, the tension is given by (\ref{tensionp}) with 
$-1$ multiplied.

In particular, by taking the limits where $(\ap,p,q)$ goes to infinity, we can find
the plane solutions of the form
\ba
\ap z+ p x- q t=\mbox{const.}
\ea
This plane boundary surface $Q$ was already employed many times in this paper as our basic setup of wedge holography as in Fig.~\ref{wedgesetupfig}.

\subsection{End-of-the-world branes in BTZ}
\begin{figure}
  \centering
  \includegraphics[width=8cm]{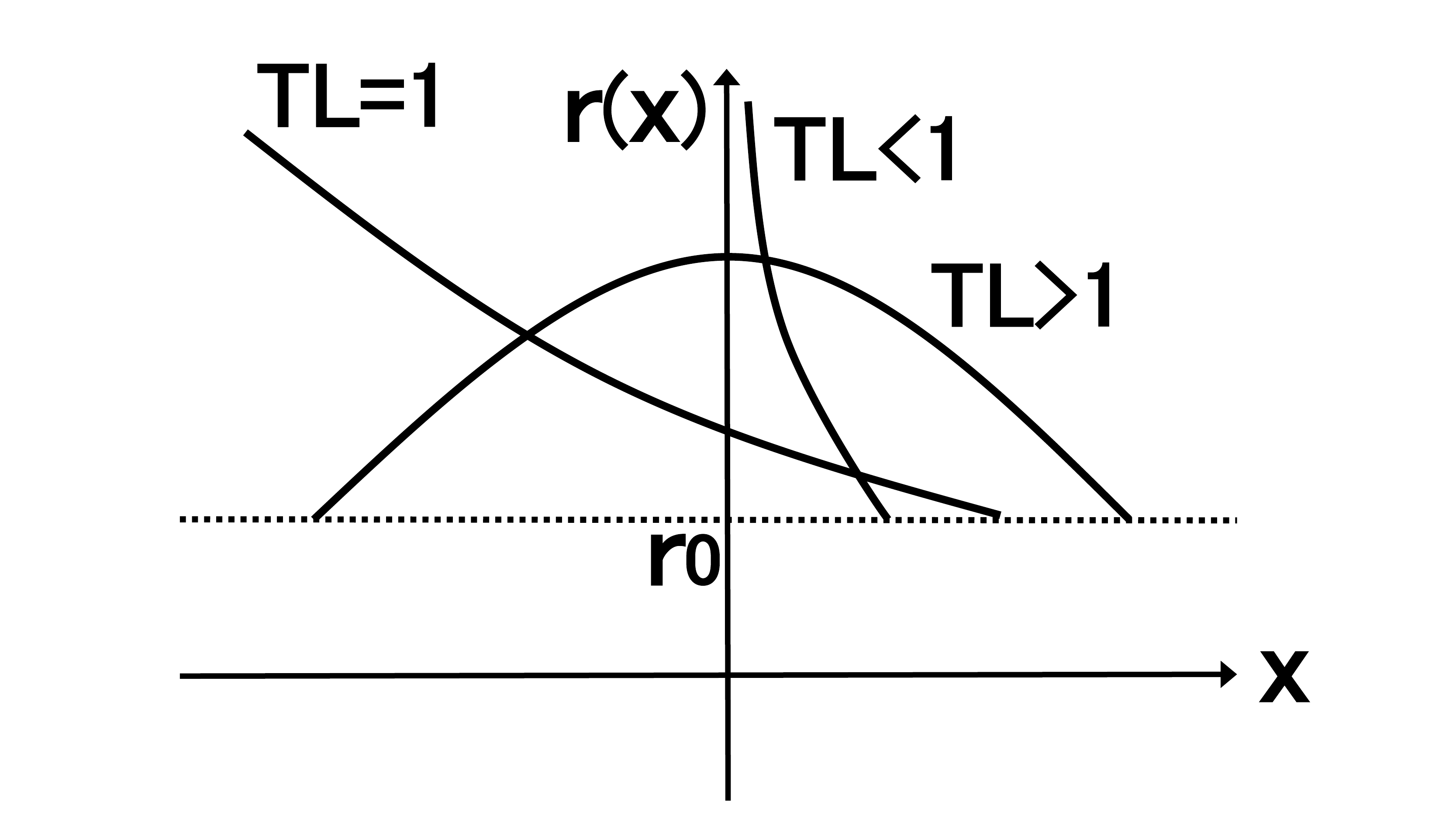}
  \caption{Profiles of end-of-the-world branes in BTZ.}
\label{btzbranefig}
\end{figure}

The BTZ metric
\ba
ds^2=-(r^2-r^2_0)d\tilde{T}^2+L^2\frac{dr^2}{r^2-r^2_0}+r^2dX^2,  \label{btzm}
\ea
can be mapped into the Poincare AdS$_3$ via the coordinate transformation
\ba
t-x=-e^{\frac{r_0}{L}(X-\tilde{T})}\s{1-\frac{r_0^2}{r^2}},\ \ \ \
 t+x=e^{\frac{r_0}{L}(X+\tilde{T})}\s{1-\frac{r_0^2}{r^2}}, \ \ \ \
z=\frac{r_0}{r}e^{\frac{r_0}{L}X}.
\ea

By mapping the solution \eqref{solab} at $p=q=0$ and then performing appropriate translation along $X$ direction in each case, we obtain the static profile of the end-of-the-world brane 
in BTZ. When $LT<1$, we obtain (refer to Fig.~\ref{btzbranefig}):
\ba
r(X)=\frac{r_0 TL}{\s{1-T^2L^2}\sinh\frac{r_0 X}{L}}. \label{btzwakkf}
\ea
When $LT=1$ we have 
\ba
r(X)=2r_0 e^{-\frac{r_0 X}{L}}.
\ea
When $LT>1$ we find
\ba
r(X)=\frac{r_0 TL}{\s{T^2L^2-1}\cosh\frac{r_0 X}{L}}. \label{sufb}
\ea
Note that equally, the  translated profile $r(X-X_*)$ in the above solutions is also solution to  \eqref{bdycond}. In each case of $LT<1$, $LT=1$ and $LT>1$, we find the world volume is identical to AdS$_2$, R$_2$ and dS$_2$, respectively. If we set $p,q\neq 0$ in \eqref{solab}, then we get more profiles but they are time-dependent, which we will study elsewhere as a future problem.
Below, in this section, we focus on $LT<1$, which corresponds to a time-like boundary in a two dimensional CFT, to construct a basic example of wedge holography in BTZ. However, in section \ref{sec:space}, we will study the case $LT>1$ in Poincare AdS$_3$, which corresponds to a space-like boundary in a two dimensional CFT.

\subsection{Gravity action}

We would like to evaluate the gravity action for a wedge geometry in the Euclidean BTZ space 
(note the periodicity $t_E\sim t_E+\beta$ with $\beta=\frac{2\pi L}{r_0}$): 
\ba
ds^2=(r^2-r^2_0)dt_E^2+L^2\frac{dr^2}{r^2-r^2_0}+r^2dx^2,  \label{btzmm}
\ea
defined by the region surrounded by the surfaces of the profile \eqref{btzwakkf}
\ba
-\frac{r_0 TL}{\s{1-T^2L^2}\sinh\frac{r_0 x}{L}},<r(x)<\frac{r_0 TL}{\s{1-T^2L^2}\sinh\frac{r_0 x}{L}},
\ea
whose boundaries define $Q_1$ and $Q_2$. We assumed $0<TL<1$.
We put a UV cutoff as 
\be
r<r_\infty=\frac{1}{\ep},
\ee
 which introduces the surface
$\Sigma$, whose (infinitesimally small) width $\Delta x$ is estimated as
\ba
\Delta x\simeq \frac{2L^2 T}{\s{1-L^2T^2}}\ep+O(\ep^3).
\ea

We can also evaluate the trace of extrinsic curvature on $\Sigma$ as 
\ba
K|_{\Sigma}=\frac{2r_\infty^2-r_0^2}{Lr_\infty \s{r_\infty^2-r_0^2}}\simeq \frac{2}{L}+O\left(\frac{r_0^4}{r^4_\infty}\right).
\ea
Also the trace of the extrinsic curvature on $Q_1$ and $Q_2$ reads $K|_{Q}=2T$.
The induced metric on $Q$ looks like
\ba
ds^2&=&\frac{r_0^2\left[1-(1-L^2T^2)\cosh^2\frac{r_0 x}{L}\right]}{(1-L^2T^2)\sinh^2\frac{r_0 x}{L}}dt_E^2
\no
&&\ \ +\frac{L^2T^2r_0^2}{(1-L^2T^2)\sinh^2\frac{r_0 x}{L}\left[1-(1-L^2T^2)\cosh^2\frac{r_0 x}{L}\right]}dx^2.
\ea
We also introduce $x_*$ as the coordinate value at the intersection of $Q_2$ and the horizon $r=r_0$ which 
is the solution to
\ba
\sinh\frac{r_0x_*}{L}=\frac{LT}{\s{1-L^2T^2}}.
\ea 

Finally, the gravity action is estimated as follows:
\ba
&&I_W=\frac{1}{4\pi G_N L^2}\int_W\s{g}-\frac{T}{8\pi G_N}\int_{Q_1\cup Q_2} \s{h}-\frac{1}{4\pi G_N L}
\int_\Sigma \s{\gamma}\no
&&\ =\frac{\beta}{4\pi G_NL^2}\left[\Delta x \int^{1/\ep}_{r_0}L rdr+2\int^{x_*}_{\Delta x/2}dx\int^{r(x)}_{r_0}
Lrdr\right]\no
&&\ -\frac{T\beta}{4\pi G_N}\int^{x_*}_{\Delta x/2}\frac{LTr_0^2}{(1-L^2T^2)\sinh^2\frac{r_0 x}{L}}
-\frac{\beta\Delta x}{4\pi G_N L}r_\infty\s{r_\infty^2-r_0^2}\no
&&=-\frac{\beta LT}{4\pi G_N \ep\s{1-L^2T^2}}-\frac{\beta r_0^2 x_*}{4\pi G_N L}.
\ea
If we write $T=\frac{1}{L}\sinh\frac{\rho_*}{L}$ then we can rewrite the above result as 
\ba
I_W=-\frac{\beta\sinh\frac{\rho_*}{L}}{4\pi G_N\ep}-\frac{\rho_*}{2G_N}.
\ea
Notice that the finite term is equal to minus of the twice of the boundary entropy \cite{Affleck:1991tk} (for a surface with the tension $T=\frac{1}{L}\sinh\frac{\rho_*}{L}$) computed in \cite{AdSBCFT,AdSBCFT2,AdSBCFT3}:
\ba
S_{bdy}=\frac{\rho_*}{4G_N}.
\ea
This makes sense as $\Sigma$ can be regarded as a merger of two boundaries, each has the tension
$T=\frac{1}{L}\sinh\frac{\rho_*}{L}$. This result tells us that the holographic dual theory of the gravity on the wedge $W$  is one dimensional theory (quantum mechanics) which has the degree of freedom given by the above boundary entropy.

\section{Space-like boundaries in BCFT and gravity duals}
\label{sec:space}

We would like to explore gravity duals of space-like boundaries in two dimensional Lorentzian CFTs. 
This corresponds to the values of tension $|T|L>1$ in the solutions \eqref{solab} with the tensions\eqref{tensionp}. In the bulk gravity, they correspond to time-like end of world branes, whose induced metrics 
coincide with de Sitter spaces (refer to \cite{Karch:2020iit} for a recent application of such branes to brane world scenario). As we will see below, we can regard this as a new class of wedge holography in a Lorentzian AdS.

\subsection{Euclidean AdS/BCFT and entanglement entropy}

\begin{figure}[ttt]
  \centering
  \includegraphics[width=8cm]{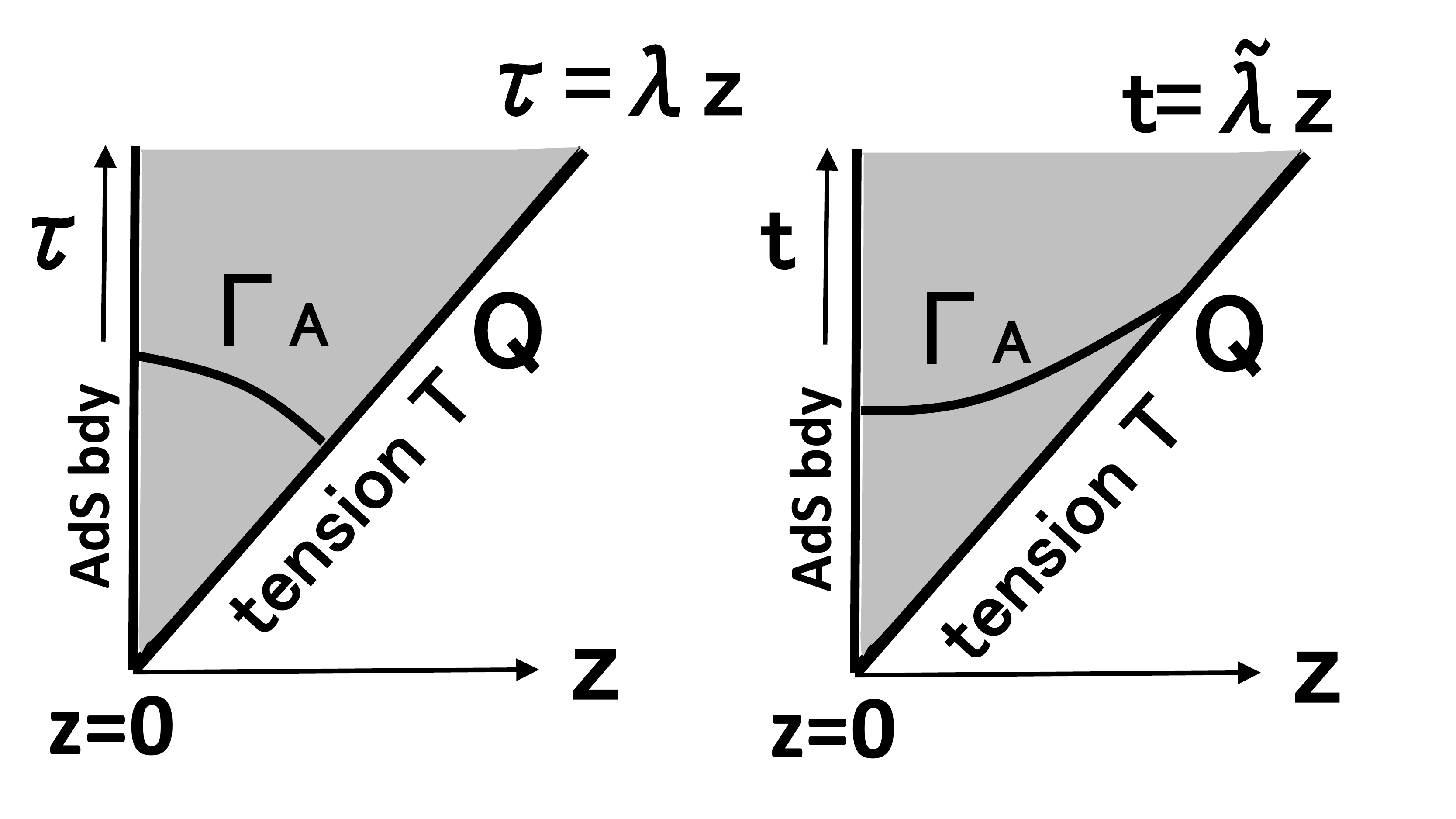}
  \caption{Small wedge regions of AdS/BCFT in the Euclidean (left) and Lorentzian (right) setup. The boundary surface $Q$ in the former and latter case has the tension $-1/L<T<0$ and $T< -1/L$, respectively.}
\label{ELsetupfig}
\end{figure}

\begin{figure}[ttt]
  \centering
  \includegraphics[width=8cm]{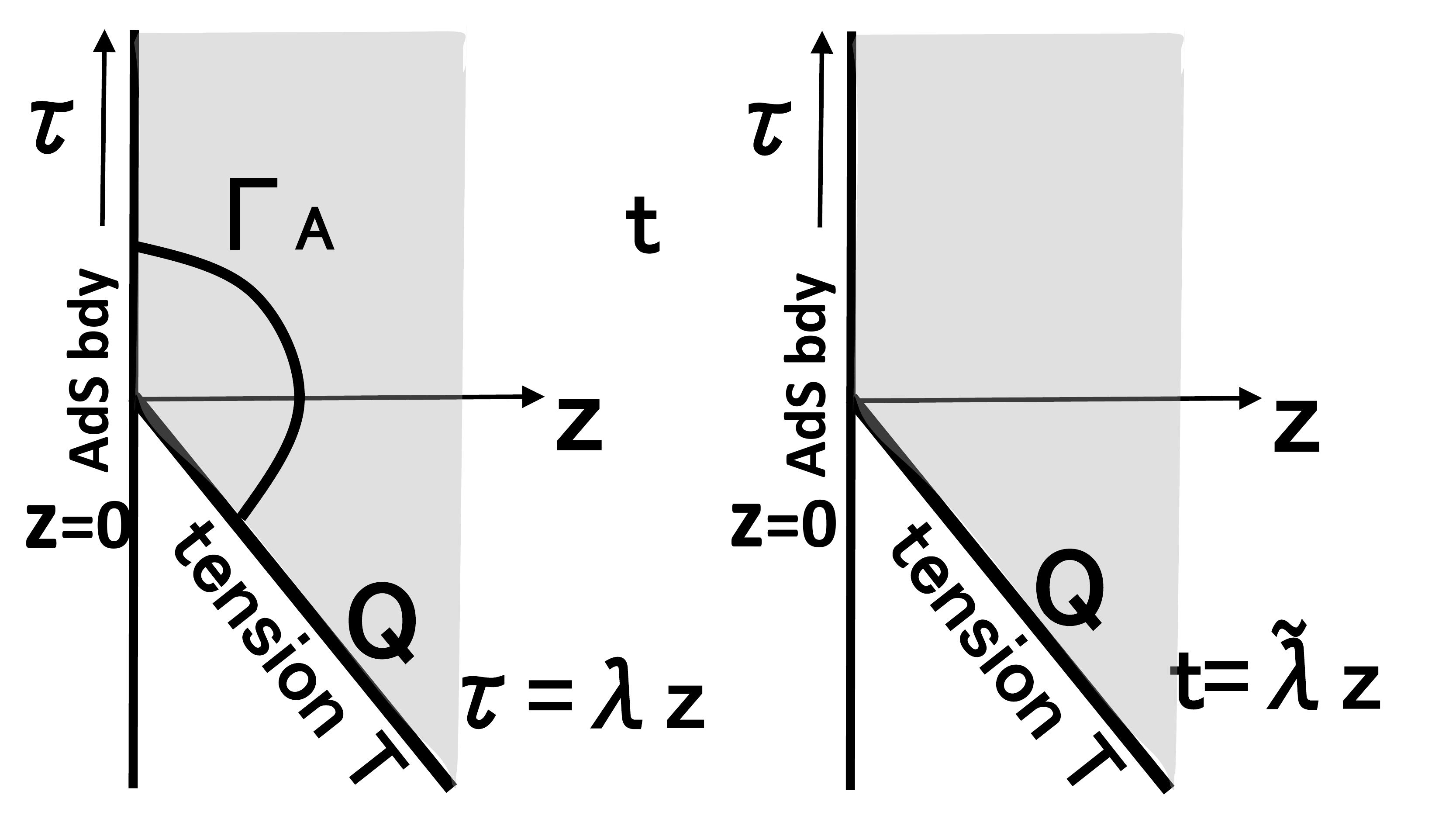}
  \caption{Large wedge regions of AdS/BCFT in the Euclidean (left) and Lorentzian (right) setup.  The boundary surface $Q$ in the former and latter case has the tension $0<T<1/L$ and $T>1/L$, respectively.}
\label{EELsetupfig}
\end{figure}

Let us first start with the familiar Euclidean AdS/BCFT setup where the gravity geometry is given by 
the following region in Euclidean Poincare AdS$_3$ $ds^2=\frac{L^2}{z^2}(d\tau^2+dx^2+dz^2)$:
\ba
\ \  z>\ep,\ \ \ \ \tau>\lambda z.  \label{wedgelo}
\ea
as depicted in the left picture of Fig.~\ref{ELsetupfig} ($\lambda>0$) 
and Fig.~\ref{EELsetupfig} ($\lambda<0$) .
The boundary $\tau=\lambda z$ defines the surface $Q$ (i.e. the end-of-the-world brane) and the parameter $\lambda$ is related to the tension as follows
\ba
\lambda=\sinh\frac{\rho_*}{L},\ \ \  \ TL=-\tanh\frac{\rho_*}{R}.     \label{ecgrh}
\ea
This is solved as a function of the tension $T$
\ba
TL=-\frac{\lambda}{\s{1+\lambda^2}}.
\ea
Therefore the tension takes the values in the range $|T|\leq 1/L$.
In this case, let us consider the geodesic which starts from  $(\tau,z)=(\tau_0,\ep)$ on the AdS boundary and ends at a point on $Q$.  The minimal geodesic length is realized when the latter point is at
$(\tau,z)=\left(\frac{\tau_0}{\s{1+\lambda^2}},\frac{\lambda\tau_0}{\s{\lambda^2+1}}\right)$ and its length leads to the contribution to HEE:
\ba
S_A=\frac{c}{6}\log\left[\frac{2\tau_0}{\ep}(\s{\lambda^2+1}-\lambda)\right].
\ea
At the initial time $\tau_0=O(\ep)\to 0$, the entanglement entropy vanishes $S_A$ as the boundary state 
does not have any real space entanglement up to the cutoff scale $\ep$ \cite{Miyaji:2014mca}.
Notice also that the two dimensional geometry of the boundary surface $Q$ is given by the hyperbolic space
$H_2$ (Euclidean AdS$_2$) with the AdS radius $\s{1+\lambda^2}L$.

\subsection{Wick rotation to Lorentzian AdS/BCFT}

Now we perform the Wick rotation $\tau=it$. This leads to the Lorentzian Poincare AdS$_3$ metic 
$ds^2=\frac{L^2}{z^2}(-dt^2+dx^2+dz^2)$. By introducing $\ti{\lambda}=i\lambda$, we find 
\ba
TL=-\frac{\ti{\lambda}}{\s{\ti{\lambda}^2-1}},
\ea
where we assume that the surface $Q$ given by $t=\ti{\lambda}z$ is time-like. Indeed, in gravity we usually do not admit space-like boundaries. Also notice that now the tension takes the values of 
\ba
|T|>1/L.  \label{tensrk}
\ea
For $\ti{\lambda}>0$ ($TL<-1$), we can find a space-like geodesic by analytical continuation from the previous Euclidean one  \eqref{ecgrh}, as in  the right of Fig.~\ref{ELsetupfig}.
 This leads to 
\ba
S_A=\frac{c}{6}\log\left[\frac{2t_0}{\ep}(\ti{\lambda}-\s{\ti{\lambda}^2-1})\right].\label{heelo}
\ea

For $\ti{\lambda}<0$ ($TL>1$) there is no space-like geodesic which ends on $Q$ from a point $(t,z)=(t_0,\ep)$ on the AdS boundary. This is depicted in the right of Fig.~\ref{EELsetupfig}.

Below we study further these two setups, mainly focusing on the first one. In these cases, the two 
dimensional geometry of the boundary surface $Q$ is given by the de Sitter  space ${\rm dS}_2$ with the dS radius $\s{\ti{\lambda}^2-1}L$.

\subsection{Space-like boundaries with \texorpdfstring{$\ti{\lambda}>0$}{lambda>0}}
\label{sec:ELSA}

Now let us focus on the space-like CFT boundary with  $\ti{\lambda}>0$ (i.e. the right of Fig.~\ref{ELsetupfig})
and discuss the interpretation of the entropy  (\ref{heelo}).  
We argue that the wedge Lorentzian geometry \eqref{wedgelo} is dual to a BCFT defined by a Lorentzian CFT with a space-like boundary at $t=0$. 

However, we should distinguish our setup from that given by a time evolution of a UV regulated boundary state:
\ba
|\Psi(t)\lb_{HH}={\cal N} e^{-iHt}e^{-\beta H/4}|B\lb,  \label{CCBo}
\ea
whose gravity dual is known to be the BTZ geometry with a time-like boundary surface given in \cite{Hartman:2013qma}.
As opposed to our setup, the boundary surface $Q$ in this geometry does not end on the Lorentzian AdS boundary and therefore does not correspond to a Lorentzian BCFT with a space-like boundary, though we might regard this as a regularized space-like boundary. Indeed the time evolution of entanglement entropy for 
(\ref{CCBo}) is given by the expression \cite{Hartman:2013qma}:
\ba
S_A=\frac{c}{6}\log \left[\frac{\beta}{\ep}\cosh\left(\frac{\pi t_0}{\beta}\right)\right],
\ea
which is totally different from (\ref{heelo}).

Instead our setup is expected to be dual to the time evolution of a non-regularized boundary state
\ba
|\Psi(t)\lb_{our}={\cal N} e^{-iHt}|B'\lb, \label{bdttim}
\ea
where $|B'\lb$ denotes a non-regulated boundary state. It is also useful to consider the behavior of boundary entropy \cite{Affleck:1991tk}
(or  $g$-function). In the Euclidean setup \eqref{ecgrh} the boundary entropy is found as 
\ba
S_{bdy}=\log g=\frac{c}{6}\log\left(\s{\lambda^2+1}-\lambda\right)=\frac{c}{6}\log\s{\frac{1-T}{1+T}}.
\label{bdyenthb}
\ea
In our Lorentzian continuation, this boundary entropy gets complex valued
\ba
 S_{bdy}&=&\log g=\frac{c}{6}\log\s{\frac{1-T}{1+T}}  \no
&=& \frac{c}{6}\log\left[-i\left(\ti{\lambda}-\s{\ti{\lambda}^2-1} \right)\right]
=\frac{c}{6}\log\left[\ti{\lambda}-\s{\ti{\lambda}^2-1}\right]-\frac{\pi c}{12}i.  \label{edyentrghw}
\ea
However, owing to this, the total entropy becomes positive valued as in \eqref{heelo}.
This suggests the dual Lorentzian BCFT assumes an exotic choice of boundary state. 
In subsection \ref{sec:bubble}, we will show how we obtain such a boundary state from a ordinary one 
via a Wick rotation of a given CFT.

Finally we would like to ask the precise meaning of the entropy \eqref{heelo}. 
One possibility is that we regard \eqref{heelo} as the holographic pseudo entropy \cite{Nakata:2020fjg},
where the initial state $|\psi_1\lb$ is chosen to (\ref{bdttim}), while the final state is the CFT vacuum 
$|\psi_2\lb=|0\lb$. In the replica calculation the pseudo Renyi entropy is evaluates as 
\ba
S^{(n)}_A=\frac{1}{1-n}\log \la 0 | \sigma_n\bar{\sigma}_n|B' \lb.
\ea
For example, in a massless free Dirac fermion CFT we can evaluate this explicitly (see the calculation in \cite{Takayanagi:2010wp})
\ba
S^{(n)}_A=\frac{1}{6}\left(1+\frac{1}{n}\right)\log \left(\frac{tx}
{\ep\s{t^2-x^2}}\right),
\ea
where the subsystem $A$ is chosen to be the interval $[0,x]$ at time $t$.   Note that at late time
$t\gg x$, we find the standard result $S_A=\frac{c}{3}\log\frac{x}{\ep}$.  On the other hand, 
when $t<x$ we have complex valued entropy and in the limit $t\to 0$ we find the behavior $S_A\sim \frac{c}{6}\log \frac{t}{\ep}
+\frac{\pi i}{12}$ which agrees with  (\ref{heelo}) up to the imaginary constant. This appearance of a complex 
value itself is not surprising because the pseudo entropy takes complex values in general.
However, we expect the imaginary part disappears in the large $c$ limit of holographic CFTs because the boundary entropy gets complex valued such that this cancels the imaginary contribution to the disconnected geodesic.

In addition to the above interpretation in terms of pseudo entropy, we also expect that  \eqref{heelo}
can also be interpreted as a standard entanglement entropy. This follows from the covariant holographic entanglement entropy \cite{HRT}, whose derivation was given in \cite{Dong:2016hjy}. Since the entanglement entropy gives a basic probe of time evolution of quantum states, it is useful to ask how the quantum state at time $t$ dual to our wedge geometry looks like. As the boundary surface $Q$ extends from the UV boundary toward the IR horizon under the time evolution, we expect that this state looks like adding layers from the UV layers in a scale invariant way which ends up with the full MERA tensor network \cite{Vidal:2007hda,Swingle:2009bg}. The time evolution of entanglement entropy for this simple model can qualitatively explain the logarithmic growth in  (\ref{heelo}).

\subsection{Space-like boundaries with \texorpdfstring{$\ti{\lambda}<0$}{lambda<0}}
\label{sec:ELSB}

Now let us focus on the space-like boundary  $\ti{\lambda}<0$ (i.e. the right of Fig.~\ref{EELsetupfig}). In this wedge geometry, which is much larger than the $\ti{\lambda}>0$ case, there is no space-like geodesic which connects a boundary point at $z=\ep$ and a point on the surface $Q$. Instead there is a time-like geodesic between the points.\footnote{For another interpretation of presence of a time-like geodesic as  a complex valued entanglement entropy, refer to 
\cite{Sato:2015tta, Narayan:2015oka, Narayan:2016xwq}.} Notice that the case $\ti{\lambda}<0$ and $\ti{\lambda}>0$ are separated by the case where the surface $Q$ is space-like and therefore it is natural that the interpretations of them are quite different. The brane world holography implies that the gravity on this large wedge region for   $\ti{\lambda}<0$   is dual to a CFT on an upper half plane plus a (quantum) gravity on dS$_2$. This is in contrast with the $\ti{\lambda}>0$ case where we argued that the gravity is dual to just a CFT on an upper half plane. In other words, the wedge holography for  $\ti{\lambda}<0$ has an additional degrees of freedom corresponding to the two dimensional dS gravity. In this sense, understandings of the physics in the $\ti{\lambda}<0$ case has a potential to answer the long standing problem of holography of dS gravity.

Consider the time evolution of quantum state in CFT in this large wedge spacetime. For $t<0$ we can regard the quantum state describes a time evolution as a MERA-like tensor network. Starting from the IR disentangled state, we gradually add each layer with quantum entanglement, and finally we realize the full MERA network which describes the CFT ground state. Therefore we have a genuine CFT vacuum $|0\lb$ at $t=0$. For later time, $t>0$ the time evolution is trivial because $e^{-iHt}|0\lb=|0\lb$ under the CFT hamiltonian $H$. This interpretation is consistent with the absence of 
space-like geodesic which connect the AdS boundary and $Q$. Indeed, for $t>0$ the correct entanglement entropy should be given by the familiar connected geodesic (i.e. a half circle shape) which leads to $S_A=\frac{c}{3}\log\frac{l}{\ep}$ and there is no need for a presence of disconnected geodesics. It will be an intriguing future problem to explore more on this and pursuit a possible connection to a de Sitter holography.

\subsection{Bubble nucleations in AdS}
\label{sec:bubble}

\begin{figure}
  \centering
  \includegraphics[width=7cm]{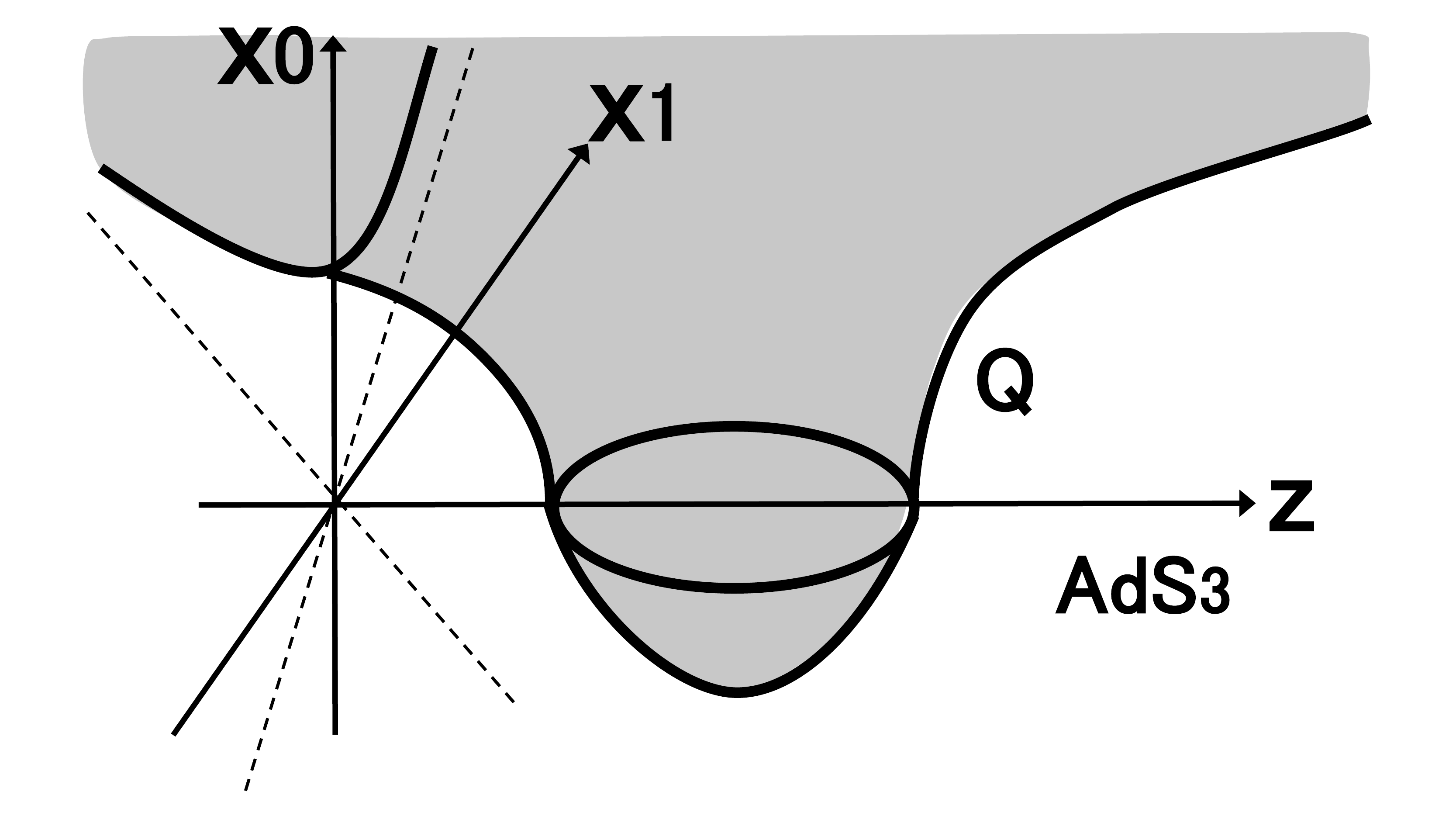}
 \includegraphics[width=7cm]{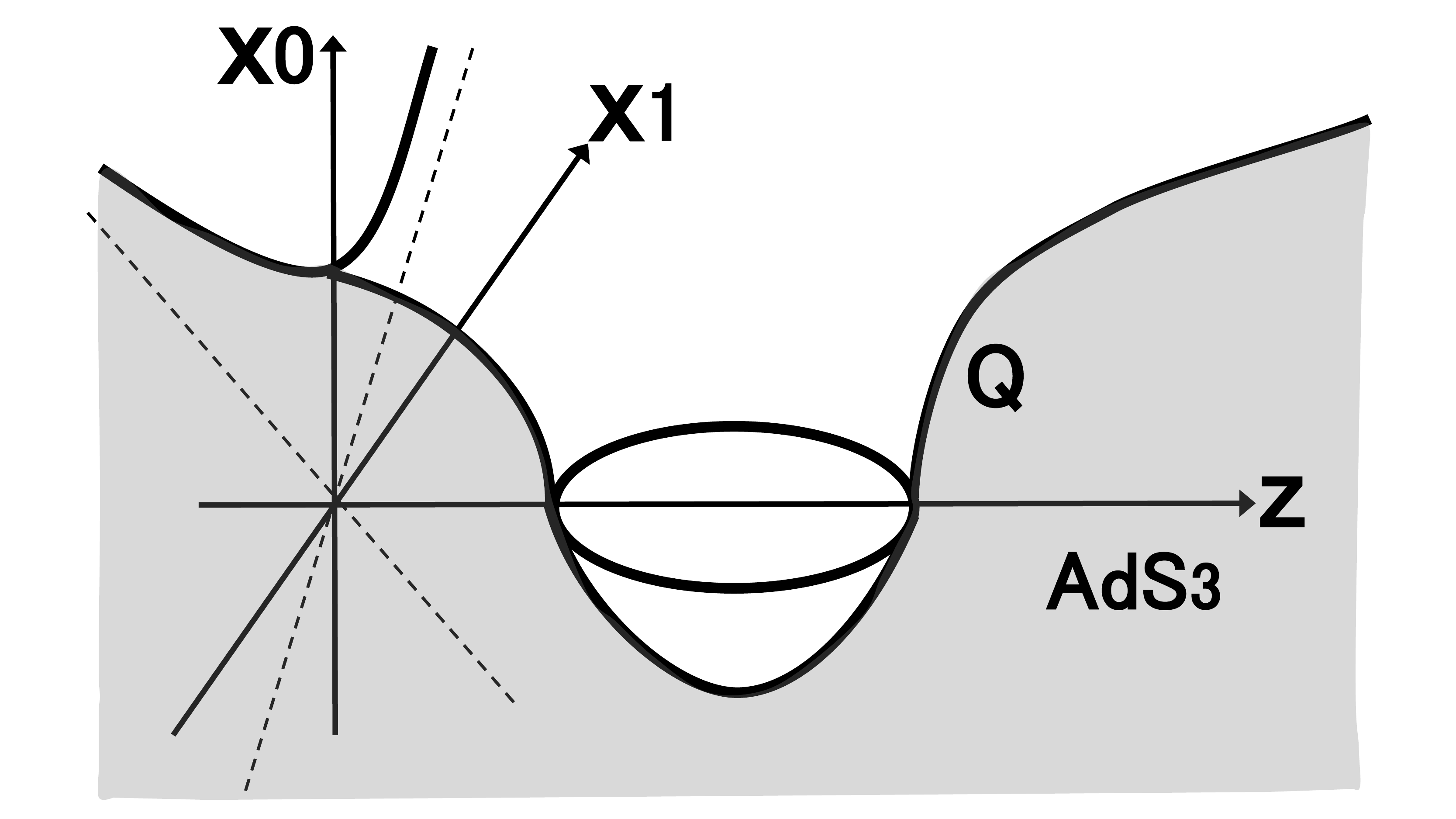}
  \caption{A bubble nucleation of universe (left) and a nucleation of bubble-of-nothing (right) in a Lorentzian AdS/BCFT setup, described by a boundary surface $Q$  with the tension $T\geq 1/L$ (left) and 
$T\leq -1/L$ (right). The colored regions describe physical spacetimes.}
\label{Bsetupfig}
\end{figure}

Let us get back to the general profile of the boundary surface $Q$  (\ref{solab}) in AdS/BCFT.
The spacetime boundary in BCFT corresponds to a time-like surface in AdS with 
the unusual values of the tension \eqref{tensrk}. This corresponds to the parameter region
$|\ap|>|\beta|$. In the Euclidean AdS, this boundary surface is floating in the bulk AdS without intersecting with the AdS boundary.
Its Lorentzian continuation  $x_0=\s{x_1^2-\beta^2+\ap^2}$ describes the bubble nucleation. The inside and outside of the Euclidean sphere $Q$ are time evolved as in  the left and right picture of Fig.~\ref{Bsetupfig}, which describes a bubble nucleation of universe and a nucleation of bubble-of-nothing, respectively. In this Lorentzian signature, the intersection between $Q$ and AdS boundary is eventually created as in Fig.~\ref{Bsetupfig}. 

In the Euclidean CFT, we can describe this Euclidean BCFT by the following disk with an imaginary radius
\ba
|w|^2=\beta^2-\ap^2<0.  \label{imdisk}
\ea
This imaginary radius corresponds to the fact that there is no actual intersection between $Q$ and the AdS boundary in the Euclidean setup. By taking Lorentzian analytical continuation we can see that the choice (\ref{imdisk}) in the AdS/BCFT leads to gravity side results which agree with the CFT calculations, as we will see below.

\subsubsection{BCFT description via analytical continuation}
In the following, we will formulate this nucleation of the bubble-of-nothing in AdS from the CFT viewpoint and evaluate the entanglement entropy in the bubble nucleation process.  We focus on the setup of right picture of Fig.~\ref{Bsetupfig}, which is related to the setup of subsection \ref{sec:ELSA} (i.e. $\ti{\lambda}>0$) via a global conformal transformation. On the other hand, the left one in  Fig.~\ref{Bsetupfig} is related to  subsection \ref{sec:ELSB} (i.e. $\ti{\lambda}<0$) and we can analyze this case in a similar way, though there is no disconnected geodesic which contributes the holographic entanglement entropy.
 
\begin{figure}[t]
 \begin{center}
  \includegraphics[width=7.0cm,clip]{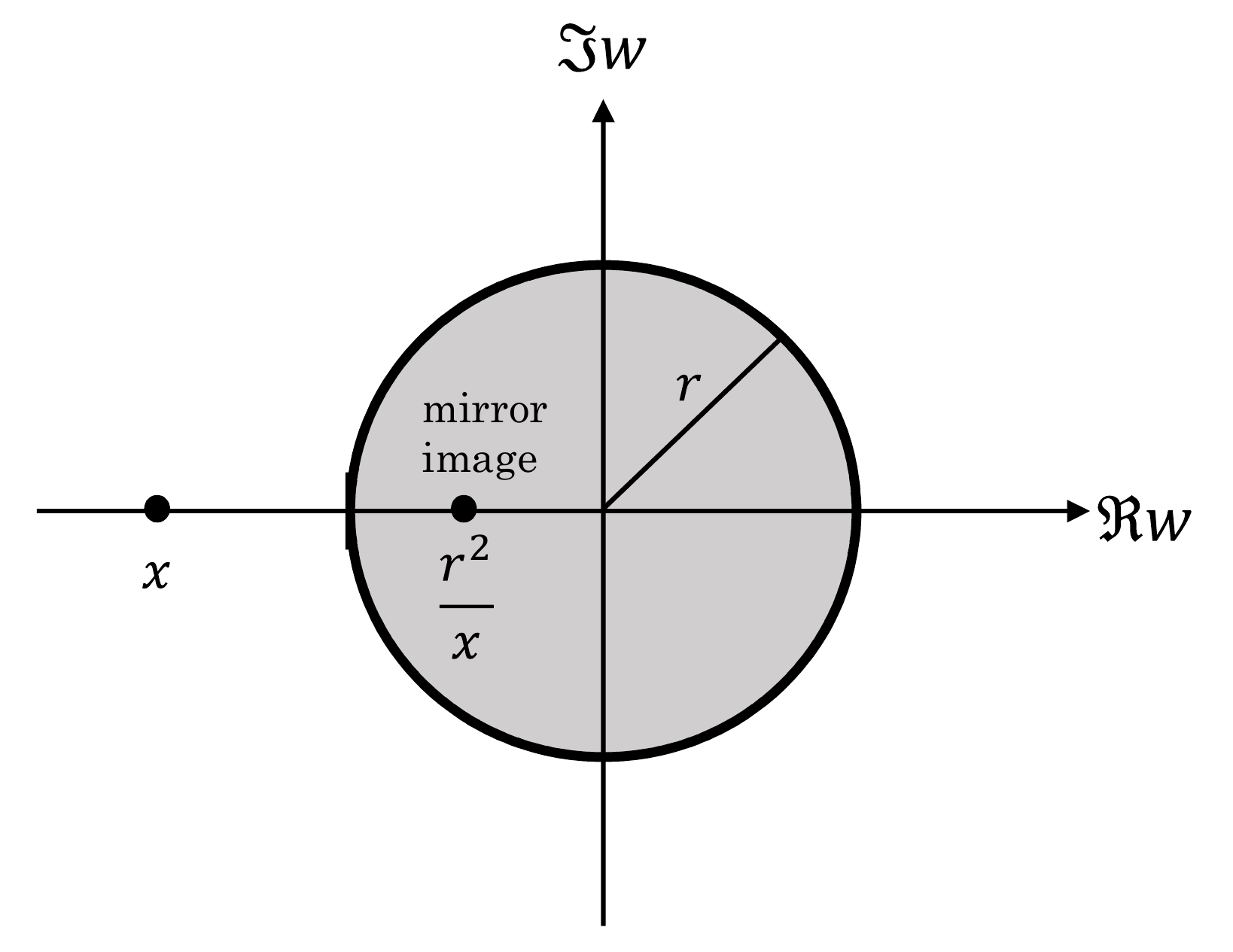}
  \includegraphics[width=7.0cm,clip]{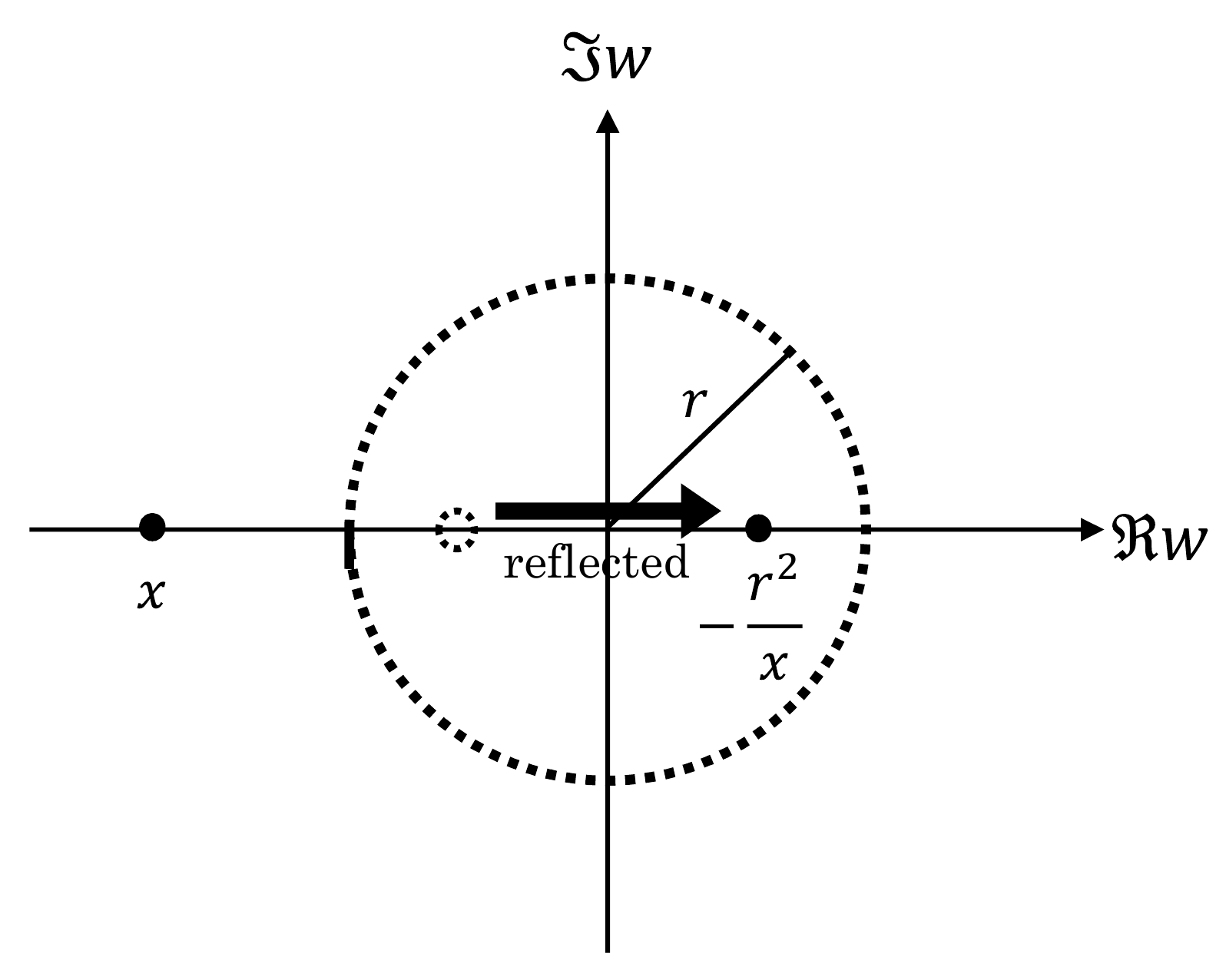}
 \end{center}
 \caption{The Euclidean CFT with the boundary $|w|^2 = r^2$. (Left) The doubling trick of the usual BCFT. The mirror image of the point $x$ is inserted onto the inversed point $r^2/x$. (Right) The doubling trick of the analytic continued BCFT. The analytic continuation can be interpreted as the reflection of the mirror point.}
 \label{fig:Mirror}
\end{figure}

Let us first consider the Euclidean BCFT on $|w|^2\geq r^2$ i.e. a region outside a radius $r$ disk. 
 To evaluate a correlator in this setup, we usually make use of the {\it doubling trick} \cite{Cardy2004}, as shown in the left of Fig.~\ref{fig:Mirror}. The kinematics of a BCFT $n$-point function is fixed by this $2n$-point function with {\it mirror points} on the full plane. Thus, we can evaluate a 1-point function with the circle boundary,
\begin{equation}
\braket{\phi(w)}_{\overline{\text{disk}}} \propto \pa{\frac{r}{(\abs{w}^2-r^2)}}^{2h},
\end{equation}
where $h$ is the conformal weight of $\phi$.
In general, the BCFT $n$-point function depends on the details of the boundary and thus has a non-universal and complicated form, which can be expressed as the sum of chiral conformal blocks with the coefficients related to the bulk-boundary 2-point functions \cite{Cardy:1991tv}.

To proceed our analysis, we focus on a particular boundary. Since we finally would like to justify our CFT calculation by comparing with a simple gravity calculation, we assume our CFT boundary to be dual to a purely gravitational end-of-the-world brane.
In that case, we can approximate the BCFT correlator by the chiral vacuum block in the semiclassical limit, like the usual prescription in the large $c$ CFT \cite{Hartman:2013mia}. In other words, we can think of the BCFT correlator as the chiral correlator (equivalently, the chiral block) on the full plane. This is the usual procedure to evaluate the BCFT correlator with the circle boundary in a holographic CFT.

Let us consider our analytic continuation $r^2 \to -r^2$. This unusual BCFT on a disk with an imaginary radius can also be treated by the doubling trick with a slight change. We put the mirror operator not on the inversed point but on the inversed + {\it reflected} point (see the right of Fig.~\ref{fig:Mirror}). Following the above prescription, this is also just the semiclassical correlator on the full plane, therefore, we expect this BCFT is reasonable. Note that we cannot find the boundary in this BCFT, that is for example, if one operator approaches the origin from outside the circle,  it needs an infinite energy to cross the circle barrier in the usual BCFT, on the other hand, in our BCFT, one finds no singularity even when it crosses the circle.

One may ask whether this CFT really leads a reasonable result. To answer this, we will evaluate the entanglement entropy in the setup (the right of Fig.~\ref{Bsetupfig}) following the above prescription.\footnote{
Even though there is a single boundary in this BCFT, we can regard \eqref{eq:2-point} as a regular entanglement entropy (or equally the pseudo entropy with the initial and final state identical) because at $t=0$ there is a (Euclidean) time reversal symmetry such that the upper path-integral is identical to the lower one.}  It is well-known that the entanglement entropy can be calculated by a correlator with twist operators. If we set a subsystem as $A = [a,b]$, then the entanglement entropy is given by
\begin{equation}\label{eq:2-point}
S_A(t) = \lim_{n \to 1}\frac{1}{1-n} \log \braket{\sigma_n(a,t) \bar{\sigma}_n(b,t)}_{\overline{\text{Im-disk}}},
\end{equation}
where we mean {\it Im
-disk} as the {\it imaginary} boundary (the right of Fig.~\ref{fig:Mirror}) with the radius $ir$ with
$r^2 = \alpha^2 - \beta^2 > 0 $.
For simplicity, we first calculate the entanglement entropy for a subsystem $A = [a, \infty)$. This is given by the 1-point function with the twist operator,
\begin{equation}
\braket{\sigma_n(w)}_{\overline{\text{disk}}} = \pa{\frac{r}{(|w|^2-r^2)}}^{2h_n},
\end{equation}
where $w = \tau + i a$ ($\tau = it$) and $h_n = \fr{c}{24}\pa{n - \fr{1}{n}}$ is the conformal weight of the twist operator. A naive analytic continuation to the imaginary radius leads to
\begin{equation}
\braket{\sigma_n(w)}_{\overline{\text{Im-disk}}} = \pa{\frac{ir}{(a^2+r^2-t^2)}}^{2h_n},
\end{equation}
and then we obtain the entanglement entropy as
\begin{equation}
S_A(t) =  \fr{c}{6} \log \fr{a^2 + r^2 - t^2}{r i \epsilon}.
\end{equation}
This {\it imaginary} entanglement entropy is problematic. To resolve this problem, we have to take into account the properties of the imaginary BCFT correlator more carefully.

\begin{figure}[ht]
 \begin{center}
  \includegraphics[width=12.0cm,clip]{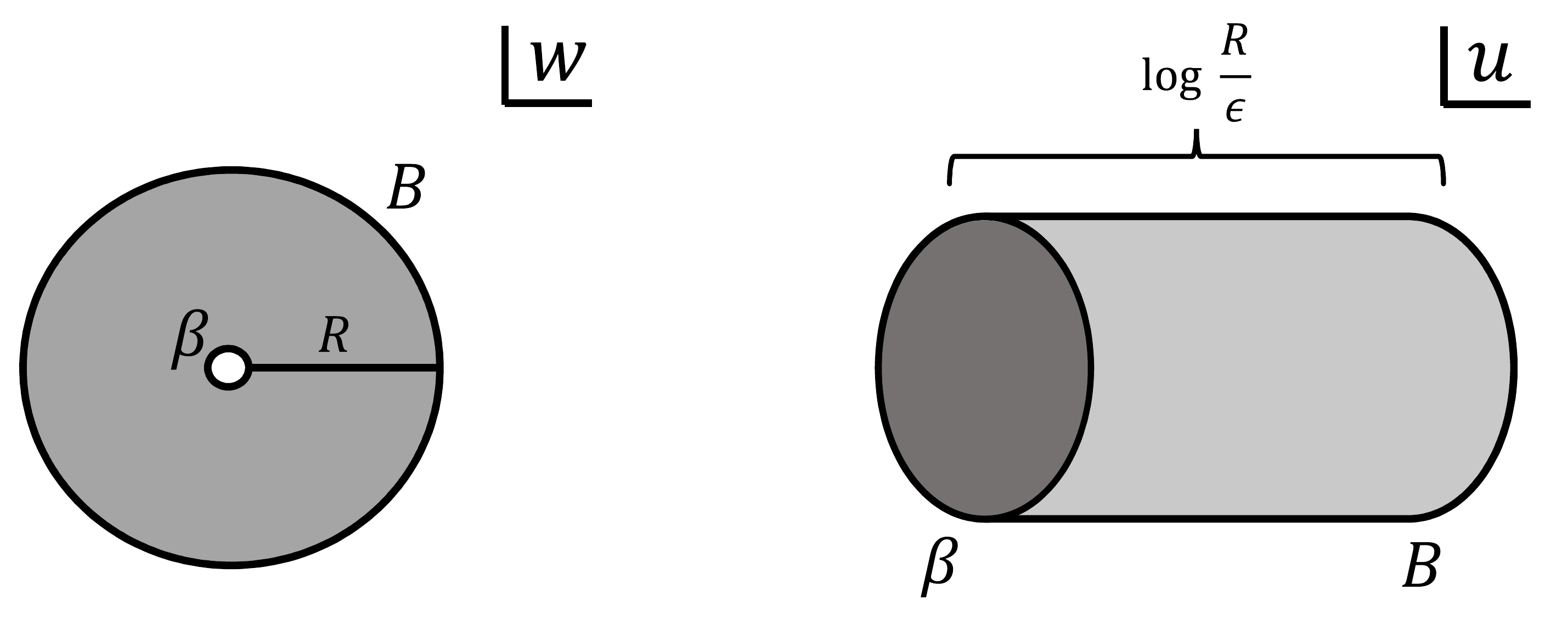}
 \end{center}
 \caption{The map from a disk (left) to a cylinder (right).}
 \label{fig:cylinder}
\end{figure}

For the purpose of the resolution, we follow the more rigorous evaluation of the entanglement entropy \cite{Cardy2016}.
From the viewpoint of the replica partition function, the twist operator is thought of as a specific boundary of radius $\epsilon$.
We describe the 1-point function on a unit dist by this partition function  $Z_n^{\beta, B}$ with two boundaries, the first $\beta$ is the boundary associated to the twist operator and the second $B$ is the original boundary,
\begin{equation}
\begin{aligned}
\braket{\sigma_n(0)}_{\text{unit-disk}}  = \braket{\sigma_n|B} = \frac{Z_n^{\beta, B}}{(Z_1^{\beta, B})^n}.
\end{aligned}
\end{equation}
From now on, we introduce a formal radius $R$ for the unit disk in order to take the effect of the analytic continuation $r^2 \to -r^2$ into account.
This replica partition function can be mapped into a cylinder (with coordinate $u$, see Fig.~\ref{fig:cylinder}) by the map,
\begin{equation}
\begin{aligned}
u = i \log (w).
\end{aligned}
\end{equation}
The length of this cylinder is $\log \fr{R}{\epsilon}$ and its circumference is $2\pi n$. This replica partition function can be written by
\begin{equation}
\begin{aligned}
Z_n^{\beta, B} = \braket{\beta| e^{-\fr{1}{2\pi n}\log \fr{R}{\epsilon}H }|B },
\end{aligned}
\end{equation}
where $H$ is the hamiltonian of our CFT. Since this cylinder is very long (because of the infinitesimal regulator $\epsilon$), we can approximate this amplitude as
\begin{equation}
Z_n^{\beta, B} =   e^{-\fr{1}{2\pi n}\log \fr{R}{\epsilon}E_0 } \braket{\beta|0 } \braket{0|B },
\end{equation}
where $E_0 = -\fr{c \pi }{6}$ is the vacuum energy. Thus, we obtain
\begin{equation}
\braket{\sigma_n|B} = \pa{\fr{R}{\epsilon}}^{-2h_n} \pa{\braket{\beta|0 } \braket{0|B }}^{1-n}.
\end{equation}
From this result, we find that the analytic continuation leads to
\begin{equation}
\braket{\sigma_n|\text{Im-}B} = \pa{\fr{i}{\epsilon}}^{-2h_n} \pa{\braket{\beta|0 } \braket{0|B }}^{1-n},
\end{equation}
where $\text{Im-}B$ is the imaginary boundary state. This term is known as the boundary entropy \cite{Affleck:1991tk}. Taking this term into account, our calculation is improved as
\begin{equation}
\braket{\sigma_n(w)}_{\overline{\text{Im-disk}}} = \braket{\sigma_n|\text{Im-}B} \pa{\frac{i r}{(a^2 + r^2-t^2)}}^{2h_n}
=\pa{\braket{\beta|0 } \braket{0|B }}^{1-n} \pa{\frac{\epsilon r}{(a^2 + r^2-t^2)}}^{2h_n},
\end{equation}
and the entanglement entropy is improved by
\begin{equation}
S_A(t) =  \fr{c}{6} \log \fr{a^2 + r^2 - t^2}{r \epsilon}  + g_B,  \label{eq:bubbleEECFT}
\end{equation}
where $g_B$ is the boundary entropy $g_B = \log \braket{0|B}$ and the contribution $\braket{\beta|0 } $ is absorbed by the redefinition of the regulator $\epsilon$ as usual. Consequently, we find the ``cancellation'' between the imaginary factor of the leading part of the entanglement entropy and the imaginary factor of the boundary entropy, and then the imaginary problem is perfectly resolved, therefore, we argue that our formulation of this imaginary BCFT is reasonable in this sense.

\begin{figure}[t]
 \begin{center}
  \includegraphics[width=13.0cm,clip]{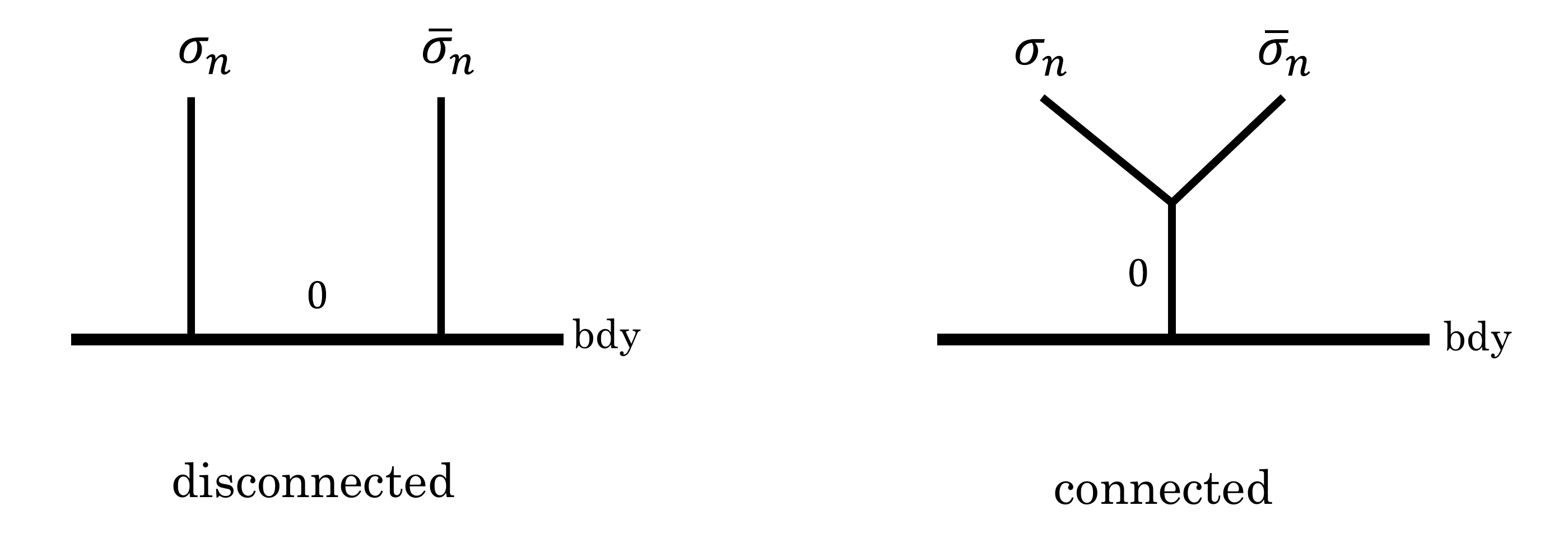}
 \end{center}
 \caption{The vacuum block approximation of a two-point function in the holographic BCFT. There are two possibilities, like the s- and t-channel vacuum block approximation of a full plane four-point function with same operators.}
 \label{fig:block}
\end{figure}

Let us move on to the more general case where the subsystem is given by an interval $A = [a,b]$  with $a < -r < r < b$.
If we assume that in the gravity dual, the end-of-the-world brane is only coupled to the metric, we can approximate the two-point function (\ref{eq:2-point})  as \cite{Sully:2020pza}
\begin{equation}
\begin{aligned}
\braket{\sigma_n(a,t) \bar{\sigma}_n(b,t)}_{\overline{\text{Im-disk}}}
\ar{n \to 1}&\left\{
    \begin{array}{ll}
     \braket{0|B}^{2(n-1)}\pa{\frac{r\epsilon}{(a^2 + r^2 - t^2)}}^{2h_n} \pa{\frac{r\epsilon}{(b^2 + r^2 -t^2)}}^{2h_n}   ,& \text{disconnected}   ,\\
      \pa{\frac{\epsilon}{(b-a)}}^{4h_n} ,& \text{connected}   .\\
    \end{array}
  \right.\\
\end{aligned}
\end{equation}
Here there are two possibilities of the vacuum block approximation. One is the {\it disconnected} channel approximation (the left of Fig.~\ref{fig:block}) and the other is the {\it connected} channel approximation (the right of Fig.~\ref{fig:block}). As a result, we obtain
\begin{equation}\label{eq:SA}
\begin{aligned}
S_A(t) &=\left\{
    \begin{array}{ll}
   \fr{c}{6} \log \fr{a^2+r^2-t^2}{r \epsilon} + \fr{c}{6} \log \fr{b^2+r^2-t^2}{r \epsilon} + 2g_B   ,& \text{disconnected}    ,\\
   \fr{c}{3} \log \fr{b-a}{\epsilon}    ,& \text{connected},\\
    \end{array}
  \right.\\
\end{aligned}
\end{equation}
where we remind $r = \sqrt{\alpha^2 - \beta^2} > 0$.
If we neglect the contribution from the boundary degrees of freedom i.e. $g_B=0$, for simplicity,
the transition time $t_*$ from the connected geodesic to disconnected one is
\begin{equation}
\begin{aligned}
t_* = \sqrt{\fr{(a-r)^2 +(b-r)^2}{2}}.
\end{aligned}
\end{equation}
In particular, if we set $a = -b$, then we have $t_* = b - r$.

Note that if the time $t$ approaches $t = \sqrt{b^2 + r^2}$ (more formally, $\abs{t-\sqrt{b^2 + r^2}}\to \epsilon$) with $a = -b$, then the entanglement entropy vanishes as one can see in (\ref{eq:SA}). This is natural because, in that time, the subsystem $A$ coincides with the space-like boundary where the two dimensional spacetime of the CFT ends.

\subsubsection{Gravity dual description}

In the former subsection, we performed CFT computations via a doubling trick with a Wick rotated boundary to compute the entanglement entropy in the setup given by the right picture of Fig.~\ref{Bsetupfig}. Here in this subsection, we will perform a holographic calculation on the gravity side and confirm that the two results indeed match with each other. 

Let us first consider the subsystem $A=[a,\infty)$. The holographic entanglement entropy is given by
\begin{align}
    &S_A = S_A^{dis} = \frac{L_{a}}{4G_N},
\end{align}
where $L_a$ is the length of the so-called disconnected geodesic which is the spacelike geodesic extending from $(t,x,z) = (t,a,\ep)$ and ending on the end-of-the-world brane $Q$. This calculation of entanglement entropy 
in AdS/BCFT \cite{AdSBCFT,AdSBCFT2} is similar to  those for the joining/splitting local quenches 
\cite{Calabrese:2007mtj,Ugajin:2013xxa,Shimaji:2018czt}  (see also \cite{Caputa:2019avh,Mezei:2019zyt,Cavalcanti:2020rsp}). Note that the Newton constant $G_N$ in AdS$_3$ is related to the central charge $c$ in CFT$_2$ by $1/(4G_N) = c/6$. 

\begin{figure}[ht]
  \centering
  \includegraphics[width=7cm]{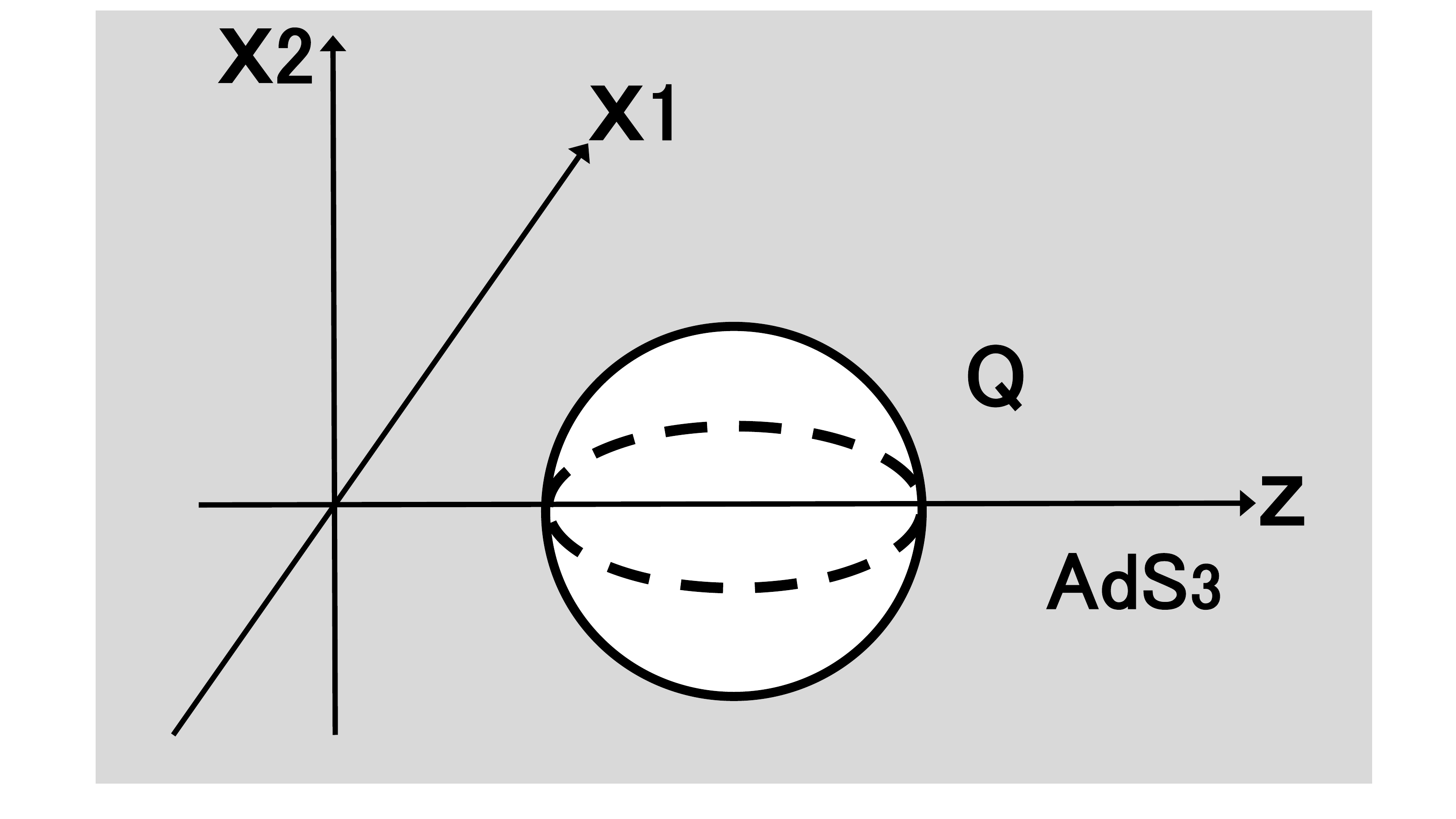}
  \caption{The Euclidean setup corresponding to the right picture of Fig.~\ref{Bsetupfig} with $x_2=ix_0$. The end-of-the-world brane $Q$ is given by $x_1^2 + x_2^2 + (z-\alpha)^2 = \beta^2$.}
\label{BsetupfigE}
\end{figure}

Consider the corresponding Euclidean setup of the right picture in Fig.~\ref{Bsetupfig} with $x_2 = ix_0$. The bulk is given by the Poincare AdS outside of the bubble: 
\begin{align}
    x_1^2 + x_2^2 + (z-\alpha)^2 \geq \beta^2. 
\end{align}
This is shown in Fig.~\ref{BsetupfigE}. 

\begin{figure}[t]
  \centering
  \includegraphics[width=12cm]{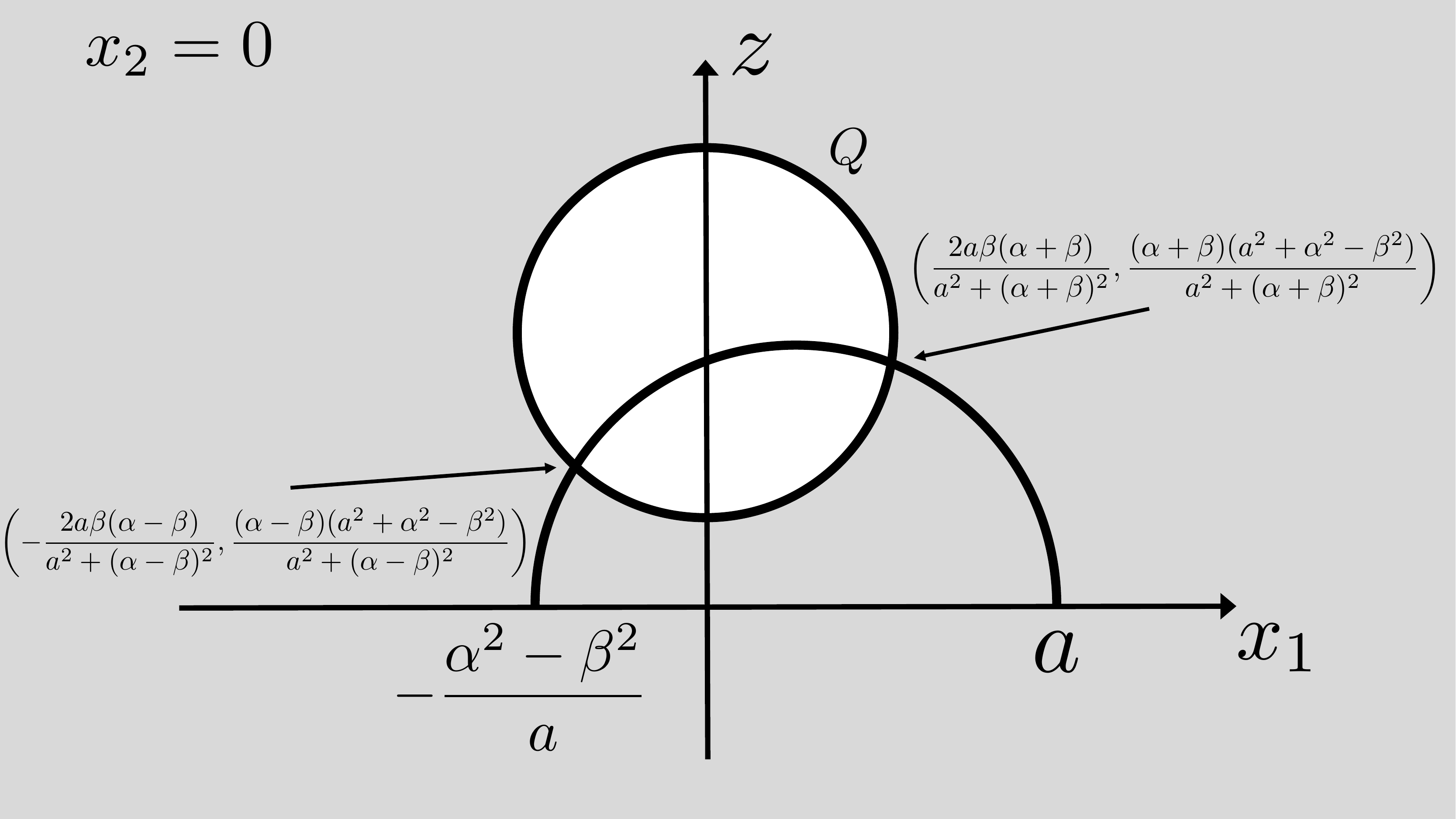}
  \caption{The $x_2=0$ slice. It is shown how the disconnected geodesic (and its extension) intersects with the bubble and the AdS boundary $z=0$.}
\label{bubbleslice}
\end{figure}

Now, let us find out the disconnected geodesic between $(x_2,x_1,z) = (0,a,\ep)$ and the end-of-the-world brane. The disconnected geodesic should lie on the $x_2=0$ slice due to the $x_2\leftrightarrow-x_2$ symmetry in our setup. Therefore, the geodesic's ending point on the brane can be denoted as $(x_2,x_1,z) = (0,\xi,\zeta)$. The distance between $(x_2,x_1,z) = (0,a,\ep),(0,\xi,\zeta)$ is given by
\begin{align}
    \cosh^{-1}\left(\frac{(a-\xi)^2+\zeta^2+\ep^2}{2\epsilon\zeta}\right) \simeq \log \left(\frac{(a-\xi)^2+\zeta^2}{\epsilon\zeta}\right).
    \label{eq:bubblelength}
\end{align}
Varying $(\xi,\zeta)$ under the restriction $\xi^2+(\zeta-\alpha)^2 = \beta^2$, we can see that (\ref{eq:bubblelength}) takes extremal value at 
\begin{align}
    (\xi,\zeta)= &\left(-\frac{2a\beta(\alpha-\beta)}{a^2+(\alpha-\beta)^2},\frac{(\alpha-\beta)(a^2+\alpha^2-\beta^2)}{a^2+(\alpha-\beta)^2}\right), \label{dismax}\\
    ~&\left(\frac{2a\beta(\alpha+\beta)}{a^2+(\alpha+\beta)^2},\frac{(\alpha+\beta)(a^2+\alpha^2-\beta^2)}{a^2+(\alpha+\beta)^2}\right). \label{dismin}
\end{align}
If we carefully look at the $x_2=0$ slice, we can see from Fig.~\ref{bubbleslice} that (\ref{dismin}) is the point where the disconnected geodesic ends.  
Plugging this back into (\ref{eq:bubblelength}), we can get the length of the disconnected geodesic
\begin{align}
    L_a = \log\left(\frac{a^2+\alpha^2-\beta^2}{(\alpha+\beta)\ep}\right). 
\end{align}
The geodesic itself is given by the following equation
\begin{align}
  \begin{cases}
    z^2 + (x_1-a)\left(x_1+\frac{\alpha^2-\beta^2}{a}\right) = 0, \\
    x_2 = 0.
  \end{cases}
\end{align}
Here note that $(x_2,x_1) = (0, (\alpha^2-\beta^2)/a)$ is nothing but the inversed + reflected point of $(x_2,x_1) = (0, a)$ shown in the right of Fig.~\ref{fig:Mirror}. 

Thanks to the rotation symmetry of the Euclidean setup, these results can be easily extended to the disconnected geodesic starting from $(x_2,x_1,z) = (\tau,a,\ep)$. In this case, the length is
\begin{align}
    L_a = \log\left(\frac{a^2+\tau^2+\alpha^2-\beta^2}{(\alpha+\beta)\ep}\right), 
\end{align}
and the geodesic follows 
\begin{align}
  \begin{cases}
    z^2 + (\frac{a}{\sqrt{a^2+\tau^2}}x_1+\frac{\tau}{\sqrt{a^2+\tau^2}}x_2-\sqrt{a^2+\tau^2})\left(\frac{a}{\sqrt{a^2+\tau^2}}x_1+\frac{\tau}{\sqrt{a^2+\tau^2}}x_2+\frac{\alpha^2-\beta^2}{\sqrt{a^2+\tau^2}}\right) = 0, \\
    a x_2 - \tau x_1= 0.
  \end{cases}
\end{align}
Performing analytic continuation $x_2 = ix_0$, $\tau=it$, 
\begin{align}
    L_a = \log\left(\frac{a^2-t^2+\alpha^2-\beta^2}{(\alpha+\beta)\ep}\right),
\end{align}
and the geodesic follows
\begin{align}
  \begin{cases}
    z^2=\frac{(t-x_0)\left((a^2-t^2)x_0+(\alpha^2-\beta^2)t\right)}{t^2}, \\
    x_1 = \frac{a}{t} x_0.
  \end{cases}
\end{align}
Here we can explicitly write down the tangent vector of this geodesic and check it is spacelike. Accordingly, the holographic entanglement entropy of subsystem $A=[a,\infty)$ is given by
\begin{align}
    S_A = \frac{L_a}{4G_N} = \frac{c}{6}\log\left(\frac{a^2-t^2+\alpha^2-\beta^2}{(\alpha+\beta)\ep}\right).
\end{align}
This perfectly matches with (\ref{eq:bubbleEECFT}) with the boundary entropy identified as
\begin{align}
    g_B = \frac{c}{6}\log\left(\frac{\sqrt{\alpha^2-\beta^2}}{\alpha+\beta}\right) = \frac{c}{6}\log\left(\sqrt{\frac{{\alpha-\beta}}{\alpha+\beta}}\right) .
\end{align}
The assumption $0\leq\beta<\alpha$ in our setup implies that the boundary entropy $g_B$ should always be non-positive. This value of $g_B$ agrees with the real part of our earlier result \eqref{edyentrghw}
for the plane shape boundary surface.

We can also consider a more general case where the subsystem is given by $A=[a,b]$. In this case, the holographic entanglement entropy is given by
\begin{align}
    &S_A = \min \{S_A^{con}, S_A^{dis}\}, \\
    &S_A^{con} = \frac{L_{ab}}{4G_N},~  S_A^{dis} = \frac{L_{a}+L_b}{4G_N}.
\end{align}
where $L_{ab}$ is the length of the so-called connected geodesic which is spacelike and connects $(x_2,x_1,z) = (t,a,\ep)$ and $(x_2,x_1,z) = (t,b,\ep)$. The connected geodesic and the disconnected geodesic correspond to the connected channel and the disconnected channel on the CFT side shown in Fig.~\ref{fig:block} respectively. This holographic formula gives us the same results with (\ref{eq:SA}) and hence we would like to omit the discussion here.

\section{Conclusion and discussion}
\label{sec:concs}

In this paper, we have proposed a codimension two holography, called wedge holography, 
between gravity on a $d+1$ dimensional wedge region $W_{d+1}$ in AdS, surrounded by two end-of-the-world branes and a $d-1$ dimensional CFT on its corner $\Sigma_{d-1}$. 
We can derive our wedge holography via two different routes. One is to employ the brane world holography and apply holography twice. The other one is to take a limit of the AdS/BCFT formulation. After we have formulated this holography, we have calculated the total free energy, holographic entanglement entropy and spectrum of conformal dimensions. In particular, we have studied the $d=2$ and $d=3$ cases in detail. We have shown that the expected correlation functions in $d-1$ dimensional CFT can be recovered from a bulk scalar field on the $d+1$ dimensional wedge space. The forms of free energy and entanglement entropy agree with general expectations for $d-1$ dimensional CFTs. 
 
In $d=3$, we have computed the central charge from the conformal anomaly. We have independently evaluated the holographic entanglement entropy and shown that the result perfectly matches with the known result  of entanglement entropy in two dimensional CFTs.  The central charge computed from the holographic entanglement entropy agrees with that from the conformal anomaly. Moreover, we have confirmed that our wedge holography leads to phase transition behaviors of entanglement entropy for disconnected subsystems 
which are expected from holographic CFTs. These provide non-trivial tests of our wedge holography. 

In $d=2$, the dual theory is supposed to be a quantum mechanics. However, our gravity calculation shows that no Schwarzian action appears. Moreover, we find that the free energy at finite temperature is given by a universal quantity analogous to the boundary entropy (or $g$ function), which is expected to describe the degrees of freedom of our conformal quantum mechanics. Since our wedge geometry is a part of AdS$_3$, we expect that the ground state preserves the $SL(2,R)$ conformal symmetry as opposed to the  Schwarzian quantum mechanics and its dual JT gravity. Though the dual theory can be a genuine conformal quantum mechanics, we need to remember that there is a Kaluza-Klein tower of primary operators as the gravity is actually three dimensional. It will be an interesting future problem to explore this more and identify the dual theory.

If we impose the Neumann boundary condition on $\Sigma$ instead of the Dirichlet condition, we find a third interpretation of our wedge setup in terms of two $d-1$ dimensional CFTs on $\Sigma$ which are interacting via $d-1$ dimensional gravity. It is an interesting future problem to explore this gravity dual of interacting two CFTs obtained by gluing two AdS$_d$ geometries in more depth.

Finally, we have studied another class of wedge geometries, which is expected to be dual to a space-like boundary in a Lorentzian CFT.  Even though we have analyzed the AdS$_3$ case, our studies can be generalized to higher dimensions in a straightforward way.  In this class of examples, the end-of-the-world brane takes the form of  de Sitter spacetime and the tension takes unusual values $|T|L>1$. We have argued that the wedge geometry with $T<-1/L$ is dual to  a BCFT with a space-like boundary, while that with $T>1/L$ is dual to a combined system of a BCFT with a space-like boundary and a gravity on de Sitter space. Our result of holographic  entanglement entropy is consistent with this interpretation. We also find that more general profiles of end-of-the-world branes, obtained from global conformal maps, can describe either a bubble nucleation of universe or a nucleation of a bubble-of-nothing. These provide new setups of AdS/BCFT going beyond the standard 
ones with the values of tension $|T|L<1$. Notice that in them,  boundaries of the Lorentzian spacetimes where CFTs are defined, are space-like, while the end-of-the-world branes in the bulk are time-like. We have also given a CFT description of these new setups which we originally found from the gravity viewpoint.
Interestingly, we encounter unusual analytical continuation of the standard Euclidean CFT such that the radius of disk is imaginary. Though this imaginary continuation gives a complex value for the boundary entropy, the final values of entanglement entropy turn out to be real, matching with the holographic result. It will be an intriguing future direction to further explore this class of boundaries in CFTs.

\section*{Acknowledgements}

We are grateful to Pawel Caputa, Kotaro Tamaoka and Tomonori Ugajin for useful discussions.
We would like to thank Raphael Bousso, Hao Geng and Andreas Karch very much for valuable comments on this paper. 
IA is supported by the Japan Society for the
Promotion of Science (JSPS) and the Alexander von Humboldt (AvH) foundation. 
IA and TT are supported by Grant-in-Aid for JSPS Fellows No.~19F19813.
YK and ZW are supported by the JSPS fellowship.
YK is supported by Grant-in-Aid for JSPS Fellows No.~18J22495.
TT is supported by the Simons Foundation through the ``It from Qubit'' collaboration.  
TT is supported by Inamori Research Institute for Science and 
World Premier International Research Center Initiative (WPI Initiative) 
from the Japan Ministry of Education, Culture, Sports, Science and Technology (MEXT). 
TT is supported by JSPS Grant-in-Aid for Scientific Research (A) No.~16H02182 and 
by JSPS Grant-in-Aid for Challenging Research (Exploratory) 18K18766.
ZW is supported by the ANRI Fellowship and Grant-in-Aid for JSPS Fellows No.~20J23116.

\appendix

\section{A brief review of the Schwarzian action}
\label{sec:schac}

The Schwarzian derivative is defined by ($\dot{t}=\frac{dt}{du}$)
\ba
\mbox{Sch}(t,u)=\dot{t}^{-2}\left(\dddot{t}~\dot{t}-\frac{3}{2}(\ddot{t})^2\right).
\ea
It satisfies the following relations under coordinate transformations
\ba
&& \mbox{Sch}(t,u)=\mbox{Sch}(\ti{t},u)+\left(\frac{d\ti{t}}{dt}\right)^2\cdot \mbox{Sch}(t,\ti{t}),\no
&& \mbox{Sch}(t,u)=\mbox{Sch}(\ti{u},u)+\left(\frac{d\ti{u}}{du}\right)^2\cdot \mbox{Sch}(t,\ti{u}),\no
&& \mbox{Sch}(t,u)=-\left(\frac{dt}{du}\right)^4\mbox{Sch}(u,t).\no
\ea

We have $\mbox{Sch}(t,u)=0$ iff 
\ba
t(u)=\frac{au+b}{cu+d}.\label{slt}
\ea
Also the equation of motion for the Schwarzian action, i.e. $I_{Sch}=\int du\ \mbox{Sch}(t,u)$, is 
proportional to $\dot{t}^{-1}\frac{d}{du}\mbox{Sch}(t,u)$, which means that $\mbox{Sch}(t,u)$ is a constant.
Thus, the solutions to this equation of motion are either  \eqref{slt} or 
\ba
t(u)=p+q\tan(ru+s).
\ea
If we set 
\ba
\frac{dt}{du}=e^{-\vp(u)},
\ea
then we find
\ba
\mbox{Sch}(t,u)=-\frac{1}{2}\dot{\vp}^2-\ddot{\vp}
=e^{-2\vp}\left(-\vp''+\frac{1}{2}(\vp')^2\right),
\ea
where note that $\vp'=\de_t\vp=e^\vp \dot{\vp}$. In this case, the Schwarzian action can be expressed as
\ba
\int du\ \mbox{Sch}(t,u)=\int dt e^{-\vp}\left(-\vp''+\frac{1}{2}(\vp')^2\right).\label{schg}
\ea

\section{Gravity action for \texorpdfstring{$d=2$}{d=2} with another UV cutoff}
\label{sec:another}

\begin{figure}[h!]
  \centering
  \includegraphics[width=6cm]{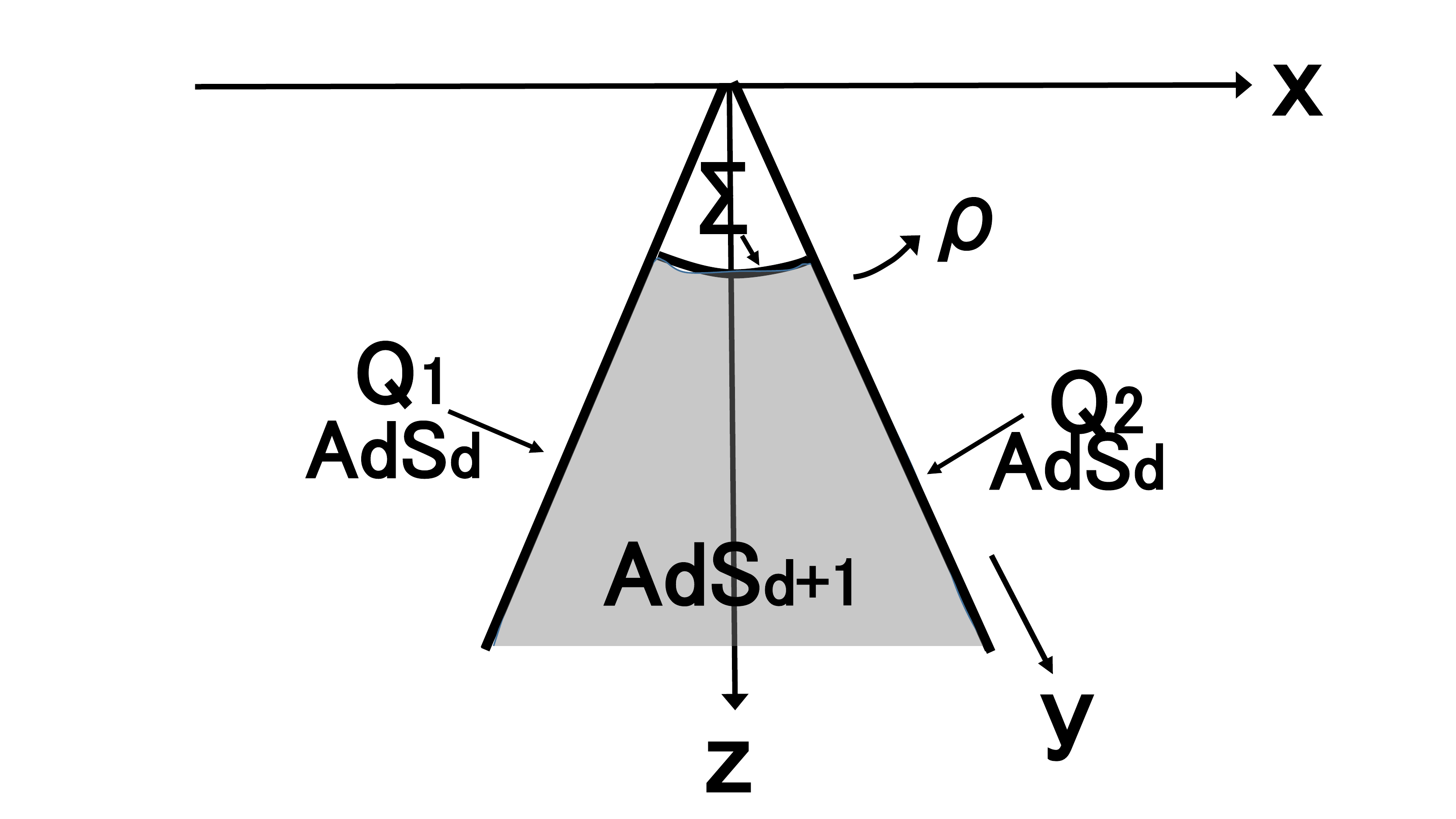}
  \caption{A sketch of $d=2$ wedge holography with another UV cutoff.}
\label{anothersetupfig}
\end{figure}

Here, we evaluate the gravity action (in Lorentzian signature) for the choice of the UV cutoff 
given by (refer to Fig.~\ref{anothersetupfig})
\ba
y\geq f(t).
\ea
Notice that this is different from the regularization we employed in section \ref{whdt}.
The total gravity action looks like
\ba
I_G=\frac{1}{16\pi G_N}\int_{Wedge} 
\s{-g}(R-2\Lambda)+\frac{1}{8\pi G_N}\int_{Q_1\cup Q_2}\s{-\gamma}(K-T)
+\frac{1}{8\pi G_N}\int_{\Sigma}\s{-\gamma}K.\no \label{actiong}
\ea
Note that in our setup we have
\ba
R=-\frac{6}{L^2},\ \ \ \Lambda=-\frac{1}{L^2},\ \ \ T=\frac{1}{L}\tanh\frac{\rho_*}{L}.
\ea
We find
\ba
&& \frac{1}{16\pi G_N}\int_{Wedge}\s{-g}(R-2\Lambda)=-\frac{1}{4\pi G_N}\left(\rho_*+L
\sinh\frac{\rho_*}{L}\cosh\frac{\rho_*}{L}\right)\int \frac{dt}{f(t)},\no 
&& \frac{1}{8\pi G_N}\int_{Q_1\cup Q_2}\s{-\gamma}(K-T)=\frac{L}{4\pi G_N}\sinh\frac{\rho_*}{L}\cosh\frac{\rho_*}{L}\int\frac{dt}{f(t)}.
\ea
Thus, the sum of these two contributions simplifies as
\ba
 \frac{1}{16\pi G_N}\int_{Wedge}\s{-g}(R-2\Lambda)+
\frac{1}{8\pi G_N}\int_{Q_1\cup Q_2}\s{-\gamma}(K-T)= \frac{- \rho_*}{4\pi G_N}\int \frac{dt}{f(t)}.
\label{bulkc}
\ea

To calculate the contribution from the $\Sigma$ surface, we need to calculate the extrinsic curvature 
$K_{ab}=(\nabla_a N_b)_{\Sigma}$. The out-going (unit normalized) normal vector is given by 
\ba
(N^{t},N^{y},N^\rho)=\frac{f(t)}{L\cosh\frac{\rho}{R}\s{1-f'(t)^2}}(-f'(t),-1,0).
\ea 
The trace of extrinsic curvature is computed as 
\ba
K=\frac{1}{L\cosh\frac{\rho}{L}}\cdot \frac{1-f'^2-ff''}{(1-f'^2)^{3/2}}.
\ea
Therefore, we find
\ba
\frac{1}{8\pi G_N}\int_{\Sigma}\s{-\gamma}K =\frac{\rho_*}{4\pi G_N}\int dt \frac{\s{1-f'^2}}{f}\cdot \frac{1-f'^2-ff''}{(1-f'^2)^{3/2}}. \label{gkw}
\ea
Since the induced metric on $\Sigma$ looks like
\ba
ds^2=-L^2\frac{1-f'^2}{f^2}dt^2,
\ea
we introduce the Weyl scaling factor as $e^{2\vp}=\frac{1-f'^2}{f^2}$.
Assuming the usual UV cutoff property $f\ll 1$, we can make the expansion
\ba
f(t)\simeq e^{-\vp(t)}+O(e^{-3\vp}).
\ea
This leads to 
\ba
\frac{1}{8\pi G_N}\int_{\Sigma}\s{-\gamma}K   \simeq \frac{\rho_*}{4\pi G_N}\int dt e^{-\vp}\left(e^{2\vp}-\frac{1}{2}(\vp')^2+\vp''\right). \label{vpex}
\ea
By using the expression for the Schwarzian action \eqref{schg}, the latter action \eqref{vpex} can be rewritten as\footnote{Actually, here we have rescaled $\ep e^\vp\to e^\vp$.}
\ba
 \frac{1}{8\pi G_N}\int_{\Sigma}\s{-\gamma}K  \simeq \frac{\rho_*}{4\pi G_N}\int dt e^{\vp}\left(1- \mbox{Sch}(t,u)\right). \label{vpxxx}
\ea

However, if we add the bulk contribution \eqref{bulkc} and evaluate the total gravity action on this wedge, then the  Schwarzian action above is canceled, i.e.
\ba
I_G\simeq  \frac{\rho_*}{4\pi G_N}\int dt e^{-\vp}(-(\vp')^2+\vp'')= 0,
\ea
where we have performed partial integration.

So far we ignored the Hayward term. Indeed, this gives the trivial contribution as the angle between $Q_{1,2}$ and $\Sigma$ is $\theta=\frac{\pi}{2}$:
\ba
I_H=\frac{1}{4\pi G_N}\int_\Sigma (\pi-\theta)\s{\gamma}=\frac{1}{8G_N}\int dt \frac{\s{1-\dot{g}^2}}{g}=\frac{1}{8G_N}\int dt  e^{\vp}.
\ea
In this way, we again find that there is no Schwarzian term in the total gravity action as long as we impose the Neumann boundary condition on $Q_{1}$ and $Q_2$.

On the other hand, if we assume the Dirichlet boundary condition on both $Q_1$ and $Q_2$, then we do not need to 
add the term $-\frac{1}{8\pi G_N}\int \s{-\gamma}T$ in \eqref{actiong}. Then, the bulk contribution vanishes 
and we find that the Schwarzian term appears as 
\ba
I_G\simeq \frac{\rho_*}{4\pi G_N}\int dt e^{\vp}\left(1- \mbox{Sch}(t,u)\right),  \label{scdef}
\ea
where the coordinate $u$ is defined such that $ds^2=-L^2\left(\frac{1-f'^2}{f^2}\right)dt^2=-L^2 du^2$.

\bibliographystyle{JHEP}
\bibliography{JoinCFTs_bib}

\end{document}